\documentclass{article}

\usepackage{arxiv}

\usepackage[utf8]{inputenc} 
\usepackage[T1]{fontenc}    
\usepackage{hyperref}       
\usepackage{url}            
\usepackage{booktabs}       
\usepackage{amsfonts}       
\usepackage{nicefrac}       
\usepackage{microtype}      
\usepackage{graphicx}       
\usepackage{physics}        
\usepackage{amssymb}
\usepackage{enumitem}

\usepackage{placeins}
\usepackage{float}
\usepackage{mathtools}
\DeclarePairedDelimiter{\nint}\lfloor\rceil

\usepackage[ruled,linesnumbered]{algorithm2e}
\makeatletter
\newcommand{\nosemic}{\renewcommand{\@endalgocfline}{\relax}}
\newcommand{\dosemic}{\renewcommand{\@endalgocfline}{\algocf@endline}}
\newcommand{\pushline}{\Indp}
\newcommand{\popline}{\Indm\dosemic}
\let\oldnl\nl
\newcommand{\nonl}{\renewcommand{\nl}{\let\nl\oldnl}}
\makeatother

\usepackage[style=numeric-comp, sorting=none]{biblatex}
\title{Resolving gas bubbles ascending in liquid metal \\ from low-SNR neutron radiography images}

\author{
  Mihails Birjukovs\\
  Institute of Numerical Modelling\\
  University of Latvia\\
  Riga, Latvia, Jelgavas 3, 1004 \\
  \texttt{mihails.birjukovs@lu.lv} \\
  \And
 Pavel Trtik \\
  Research with Neutrons and Muons\\
  Paul Scherrer Institut\\
  Villigen, Switzerland, Forschungsstrasse 111, 5232 \\
  \texttt{pavel.trtik@psi.ch} \\
  \And
 Anders Kaestner \\
  Research with Neutrons and Muons\\
  Paul Scherrer Institut\\
  Villigen, Switzerland, Forschungsstrasse 111, 5232 \\
  \texttt{anders.kaestner@psi.ch} \\
  \And
 Jan Hovind \\
  Research with Neutrons and Muons\\
  Paul Scherrer Institut\\
  Villigen, Switzerland, Forschungsstrasse 111, 5232 \\
  \texttt{jan.hovind@psi.ch} \\
  \And
  Martins Klevs \\
  Institute of Numerical Modelling\\
  University of Latvia\\
  Riga, Latvia, Jelgavas 3, 1004 \\
  \texttt{martins.klevs@lu.lv} \\
  \And
  Dariusz Jakub Gawryluk \\
  Research with Neutrons and Muons\\
  Paul Scherrer Institut\\
  Villigen, Switzerland, Forschungsstrasse 111, 5232 \\
  \texttt{dariusz.gawryluk@psi.ch} \\
  \And
 Knud Thomsen \\
  Research with Neutrons and Muons\\
  Paul Scherrer Institut\\
  Villigen, Switzerland, Forschungsstrasse 111, 5232 \\
  \texttt{knud.thomsen@psi.ch} \\
  \And
 Andris Jakovics\\
Institute of Numerical Modelling\\    
  University of Latvia\\
  Riga, Latvia, Jelgavas 3, 1004 \\
  \texttt{andris.jakovics@lu.lv} \\
  \And
}

\begin{document}

\maketitle

\begin{abstract}

We demonstrate a new image processing methodology for resolving gas bubbles travelling through liquid metal from dynamic neutron radiography images with intrinsically low signal-to-noise ratio. Image pre-processing, denoising and bubble segmentation are described in detail, with practical recommendations. Experimental validation is presented -- stationary and moving reference bodies with neutron-transparent cavities are radiographed with imaging conditions similar to the cases with bubbles in liquid metal. The new methods are applied to our experimental data from previous and recent imaging campaigns, and the performance of the methods proposed in this paper is compared against our previously developed methods. Significant improvements are observed as well as the capacity to reliably extract physically meaningful information from measurements performed under highly adverse imaging conditions. The showcased image processing solution and separate elements thereof are readily extendable beyond the present application, and have been made open-source.

\end{abstract}


\clearpage

\section{Introduction}

Gas bubble flow in liquid metal is encountered in a variety of industrial processes. Examples include liquid metal stirring, purification and continuous casting in metallurgy, liquid metal-based chemical reactors, and more. Controlling the output of these processes is essential and it was proposed that this can be (and in some cases already is) achieved via applied magnetic field (MF) \cite{baakeNeutronRadiographyVisualization2017, casting-euler-musig, embr-experiment, embr-experiment-2, limmcast, embr-visualized, embr-cift, birjukovsPhaseBoundaryDynamics2020}.

Single-bubble flow with and without applied MF has been systematically studied using ultrasound Doppler velocimetry (UDV) \cite{zhang-thesis, udv-review-article, udv-longitudinal-field, udv-transverse-field}, ultrasound transit time technique \cite{uttt-path-instability, uttt-x-ray-single-bubble}, X-ray imaging \cite{uttt-x-ray-single-bubble} and numerical simulations \cite{hzdr-ibm-bubbles-thesis, dns-longitudinal-field, imb-transverse-field, zhang-mf-vertical, zhang-mf-simulations, gaudlitz-shape-wake-variations-1-bubble, hele-shaw-bubbles-vof, hele-shaw-bubbles-experiment}. The most important features and physical mechanisms involved in single-bubble flow are presently rather well known \cite{prl-path-instability, natcomms-shape-dynamics, spiral-to-zigzag-explained, shape-and-wake-simulations, spiral-to-zigzag-explained-2, in-depth-study-of-ellipsoid-kinematics}, but many aspects of bubble collective dynamics, especially in presence of MF, are not fully understood or have not been studied in-depth \cite{x-ray-bubble-breakup, x-ray-bubble-coalescence, optics-collective-dynamics, 2021-review-article-bubbles-in-liquid-metal}. Clearly, this is a problem from the process engineering and optimization perspective, and also from the point of view of computational fluid dynamics (CFD) where it is of interest to improve effective models for bubble flow (Euler-Euler and Lagrangian) \cite{casting-euler-les, casting-euler-musig, casting-lagrange-bubbles, casting-new-collective-dynamics-models, taborda-les-euler-lagrange}.

Through recent efforts and the advent of dynamic x-ray and neutron radiography of two-phase liquid metal flow \cite{megumi-x-rays, saito-neutrons-1, saito-neutrons-2, neutrons-particles-lappan, neutrons-particles-stirrer-scepanskis, neutrons-particles-stirrer-scepanskis-2, neutrons-simulations-stirrer-valters}, fundamental investigation of bubble chain systems mimicking industrially relevant flow conditions is underway \cite{hzdr-ibm-bubbles-thesis, birjukovsArgonBubbleFlow2020, birjukovsPhaseBoundaryDynamics2020, baakeNeutronRadiographyVisualization2017, x-ray-bubble-chain-simulate, x-ray-prime-code, x-ray-bubble-breakup, x-ray-bubble-coalescence, x-ray-validation, megumi-x-rays}. In bubble chain flow, bubbles are released into a liquid metal system one-by-one with a uniform time delay between each, at a certain gas flow rate, and ascend to the free surface of liquid metal. Such systems are usually rectangular vessels filled with gallium \cite{birjukovsArgonBubbleFlow2020, birjukovsPhaseBoundaryDynamics2020, baakeNeutronRadiographyVisualization2017} or an eutectic gallium-indium-tin alloy \cite{x-ray-bubble-chain-simulate, x-ray-bubble-breakup, x-ray-bubble-coalescence, x-ray-validation, megumi-x-rays} where bubbles are introduced via horizontal \cite{birjukovsArgonBubbleFlow2020,birjukovsPhaseBoundaryDynamics2020} or vertical \cite{baakeNeutronRadiographyVisualization2017, x-ray-bubble-chain-simulate, x-ray-bubble-breakup, x-ray-bubble-coalescence, x-ray-validation} nozzles at the bottom of the vessel, or top-submerged vertical \cite{megumi-x-rays} nozzles. Bubble chains flow systems are the next logical step from single-bubble flow investigations, since single-bubble flow, while very informative of the bubble wake flow dynamics and characteristic trajectories without and with applied MF, is not representative of the actual flow conditions typical for the above mentioned industrial processes where one has columns and swarms with a high number density of deformable bubbles \cite{stirring-review, simulations-inclusion-removal, simulations-experiments-dual-jets, stirring-cfd-srm}.

 Bubble chains are still simple enough to enable experimentation with compact systems \cite{birjukovsArgonBubbleFlow2020, birjukovsPhaseBoundaryDynamics2020, x-ray-prime-code, x-ray-validation, baakeNeutronRadiographyVisualization2017} and contain computationally manageable numbers of bubbles within the liquid metal volume \cite{hzdr-ibm-bubbles-thesis, x-ray-prime-code, birjukovsPhaseBoundaryDynamics2020, birjukovsArgonBubbleFlow2020}. Meanwhile, they already exhibit collective dynamics between leading and trailing bubbles \cite{hzdr-ibm-bubbles-thesis, x-ray-prime-code, birjukovsPhaseBoundaryDynamics2020, birjukovsArgonBubbleFlow2020, x-ray-bubble-chain-simulate} and, depending on the system geometry and flow rate, bubble agglomeration, coalescence and breakup can occur \cite{x-ray-bubble-coalescence, x-ray-bubble-breakup, megumi-x-rays}.

Therefore these systems are a crucial milestone in a transition from studying single-bubble flow to investigations of many-bubble systems that are very close to their actual industrial counterparts. However, despite the \textit{relative simplicity}, dynamics exhibited by bubble chain flow in liquid metal without or with applied MF are still very complex. Depending on the gas flow rate, bubbles produce unstable elongated wake flow regions where periodic vortex detachment occurs and turbulent pulsations are generated -- shed vortices and turbulent wakes of leading bubbles strongly affect the trailing bubbles, leading to bubble pair coupling across the ascending chain \cite{hzdr-ibm-bubbles-thesis, x-ray-prime-code, birjukovsPhaseBoundaryDynamics2020, birjukovsArgonBubbleFlow2020, kusuno-bubble-pair-coupling-2021, bubbles-side-by-side, bubbles-in-line-2020, hele-shaw-bubbles-coupling}. There exists a feedback loop involving combined perturbations of bubble shapes and within the bubble chain, surrounding liquid metal flow, and the influence of the free surface at the top of the metal vessels with instabilities and oscillations in the bubble chain shape \cite{hzdr-ibm-bubbles-thesis, x-ray-prime-code, birjukovsPhaseBoundaryDynamics2020, birjukovsArgonBubbleFlow2020, hele-shaw-bubbles-coupling}. Recently, dynamic mode decomposition (DMD) has been applied to the output of the MHD bubble chain flow simulation to study both large-scale flow structures and bubble wake flow in the bubble reference frame \cite{klevs2021dynamic}. It was demonstrated that DMD is a viable tool for an in-depth analysis of the complex dynamics mentioned above, and it was shown that there exists a very complex interplay of bubble wake flow and large-scale flow modes with a wide range of spatial and temporal scales.

It has become clear that specialized and rather advanced image processing methods and tools are required to extract physically meaningful data from data sets acquired via dynamic neutron and/or X-ray imaging \cite{baakeNeutronRadiographyVisualization2017, birjukovsArgonBubbleFlow2020, birjukovsPhaseBoundaryDynamics2020}. This is mainly due to the low signal-to-noise ratio (SNR) associated with imaging thick ($> 20$-$30~ mm$) layers of liquid metal at frame rates $\gtrsim 100$ frames per second (FPS) and the need to resolve many often closely packed interacting objects. High frame rates are a requirement to enable capturing fast bubbles, drops and particles flowing in liquid metal and to avoid motion blur \cite{baakeNeutronRadiographyVisualization2017, birjukovsArgonBubbleFlow2020, birjukovsPhaseBoundaryDynamics2020}. Meanwhile, neutron flux that can be used in experiments is limited by both the utilized neutron source and the rapid activation of model liquid metals such as gallium.

Here we describe and demonstrate our new image processing methodology developed over the course of our dynamic neutron imaging experiments with bubble flow in liquid metal. We show that the implemented code is robust and can operate reliably at very low SNR in presence of image artefacts. It is also shown that this version clearly outperforms our previous solution by resolving/avoiding known issues. In addition to demonstrating performance for dynamic neutron imaging data sets from our measurements with a rectangular gallium vessel with bubble chain flow, we have also performed direct experimental validation of the code by imaging a reference spherical body, both stationary and in motion, and have quantified the shape detection errors.

\clearpage

\section{Image \& bubble characterization}

\subsection{Image acquisition}
\label{sec:image-acquisition}

Bubble flow imaging with and without applied magnetic field was conducted at the thermal neutron imaging beamline NEUTRA (SINQ, PSI, $20~ mm$ aperture, $10^7~ n~ cm^{-2}~ s^{-1}~ mA^{-1}$ flux) at the Paul Scherrer Institute (PSI) using a medium spatial resolution set-up (MIDI) \cite{neutra-props}. Experiments were performed with both horizontally and vertically oriented inlet nozzles. The setup with the horizontal inlet is described in \cite{birjukovsArgonBubbleFlow2020, birjukovsPhaseBoundaryDynamics2020}, whereas a modified version of that model gallium/argon system was designed for the new experiments with the vertical inlet -- the latter will be described in a follow-up publication. A thin-walled ($4~mm$) rectangular $150~ mm \times 90~ mm~ (95~ mm) \times 30~ mm$ (interior dimensions) glass (boron-free) vessel filled with liquid gallium up to the $130$- to $140$-$mm$ mark was used. Neutron flux was parallel to the $30$-$mm$ dimension of the vessel. A square field of view (FOV, $112.8$- or $123.125$-$mm$ side) above the gas inlet was imaged at 100 FPS. The distance between the liquid metal layer and the scintillator varied depending on the setup (magnetic field system used, if any) and was $[ 4 ; 32 ]~ mm$. A sCMOS \textit{ORCA Flash 4.0} camera was used to collect the scintillator ($200 ~ \mu m$ thick ${}^6$\textit{LiF/ZnS} screen).

Reference experiments were performed at the cold neutron beamline ICON (SINQ, PSI, $20~ mm$ aperture, $\sim 1.3 \times$ NEUTRA flux) to validate the developed image processing methodology \cite{icon-props}. A $20~ mm \times 20~ mm \times 30~ mm$ rectangular brass reference body (stationary and moving) with a central spherical cavity ($5~mm$ radius) was imaged at 100 FPS within a $120.06~mm$ square FOV (a sCMOS \textit{ORCA Flash 4.0 V2} camera with a $200 ~ \mu m$ ${}^6$\textit{LiF/ZnS} scintillator) to reproduce the imaging conditions for argon bubbles in liquid gallium. The reference body was imaged with neutron flux directed along its shorter or longer axes. In addition, the distance between the scintillator and the body was either $0$- (static body) / $2~ mm$ (moving body) or either with an extra $1~cm$. These adjustments to the test conditions were meant to determine how the reference shape acquisition error depends on SNR which these conditions modify.


For each image sequence recording at NEUTRA and ICON camera dark current signal and neutron beam profile signals were recorded to be used for subsequent image normalization during pre-processing.

\subsection{Image properties}
\label{sec:image-props}

The acquired images are 16-bit 1-channel TIFFs with a $1024 \times 1024$ pixel resolution ($2 \times 2$ average-binned $2048 \times 2048$ frames) with a $~ 0.11$-$0.12~ mm$ pixel size ($0.12~mm$ for the reference experiments). Short exposure times ($10~ ms$) result in strong Poisson (multiplicative) noise from neutrons and converted photons, and salt-and-pepper noise of varying density is present due to overexposed (gamma ray noise) or "dead" camera pixels. The neutron beam flux over the FOV is non-uniform with a fall-off near the edges of the acquired images.

\begin{figure}[htbp]
\centering
\includegraphics[width=0.80\textwidth]{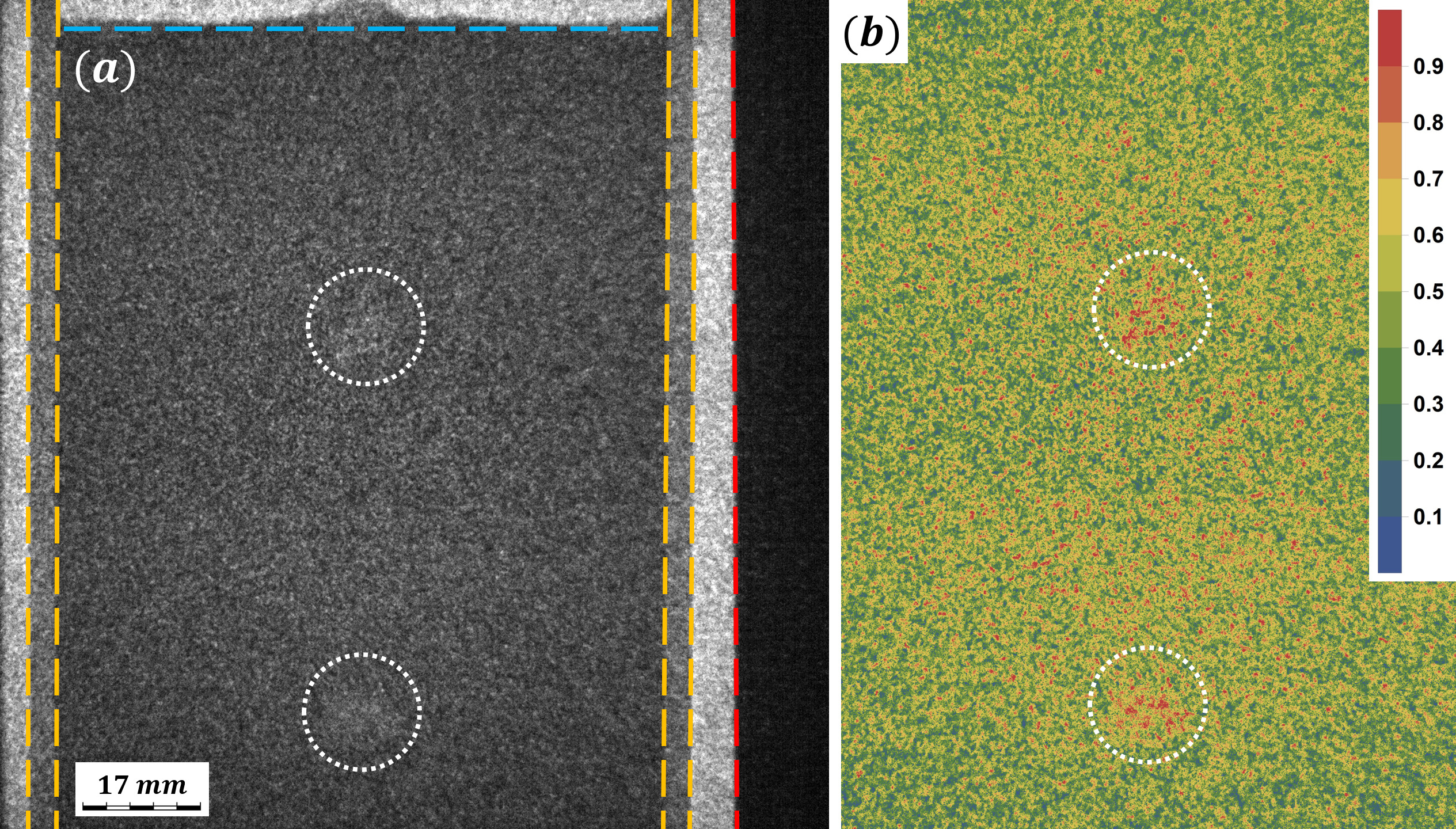}
\caption{
(a) Original captured FOV after outlier removal and luminance normalization with marked container walls (orange dashed lines), the mean gallium free surface level (light blue), the neutron flux shielding and bubble locations within the FOV (white). Note the scale bar in the bottom-left corner. (b) FOV in false color (color bar on the right) after cropping to the container walls and the metal free surface, dark current and flat field corrections, and normalization.
}
\label{fig:og-fov-and-crop}
\end{figure}

Figure \ref{fig:og-fov-and-crop}a is an example of an acquired raw image. Here gas flow rate was $120~sccm$ (standard cubic centimeters per minute) and static vertical magnetic field of $125~ mT$ was applied to the bubble chain region. Typically one can see the following captured in a full FOV image: the liquid gallium volume within the glass container (delimited by the interior orange lines and the light blue line), the container walls (orange lines), the free surface of liquid metal (the light blue line), the surrounding air (regions outside the walls and above the metal free surface) and the neutron flux shielding (the red line) for the magnetic field system. However, not all of the FOV is of interest -- Figure \ref{fig:og-fov-and-crop}b shows the FOV cropped to the liquid metal volume in false color after pre-processing (Section \ref{sec:pre-processing}). Note also the bubble regions highlighted in Figures \ref{fig:og-fov-and-crop}a and b -- these are the objects of interest that must be segmented and their properties such as centroids, projection areas, tilt angles, aspect ratios, etc. measured. Note that the same color scheme and normalized luminance scale as in Figure \ref{fig:og-fov-and-crop}b are used in all other figures beyond this point unless stated otherwise.

\begin{figure}[htbp]
\centering
\includegraphics[width=0.9\textwidth]{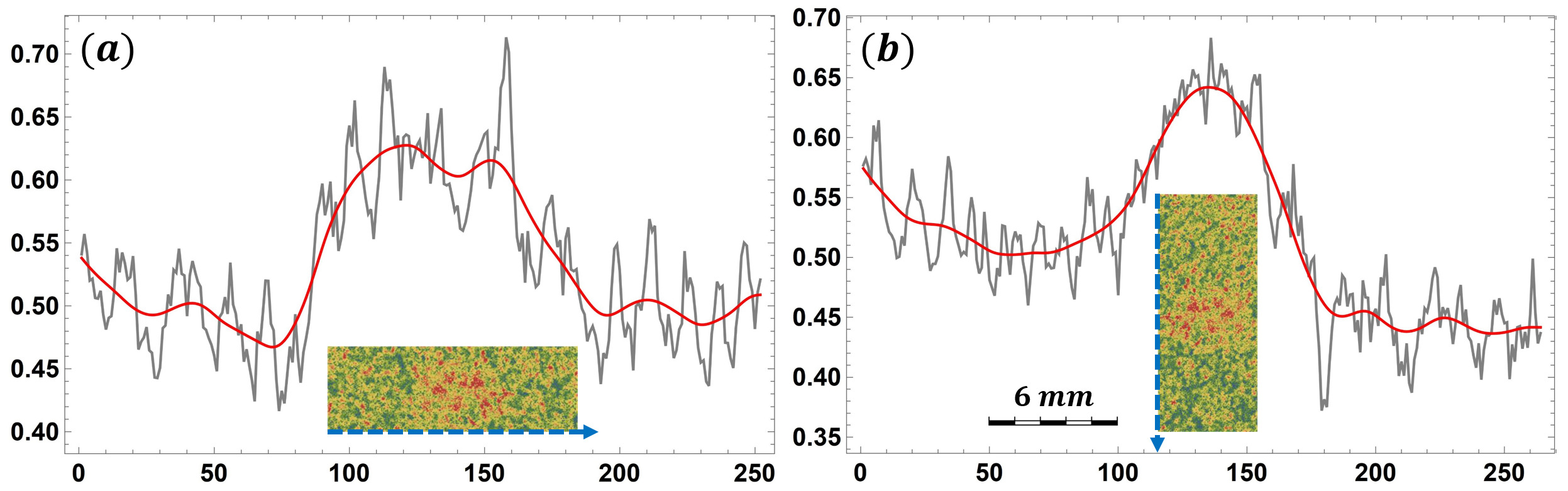}
\caption{
Bubble neighborhood analysis: width mean of luminance (gray) over the length of (a) horizontal and (b) vertical patches roughly fitted to the bubble region within the FOV. Normalized luminance versus pixel coordinates is shown in both cases. Note the pixel-to-$mm$ scale bar in (b). Bubble neighborhood patches shown in (a,b) are not to scale and are normalized as in Figure \ref{fig:og-fov-and-crop}. Scan directions are indicated by the dashed blue arrows and the red curve is the total variation filtered (Gaussian, regularization parameter equal to 1 \cite{total-variation-rof-model}) gray curve.
}
\label{fig:bubble-profiles}
\end{figure}

\begin{figure}[htbp]
\centering
\includegraphics[width=0.9\textwidth]{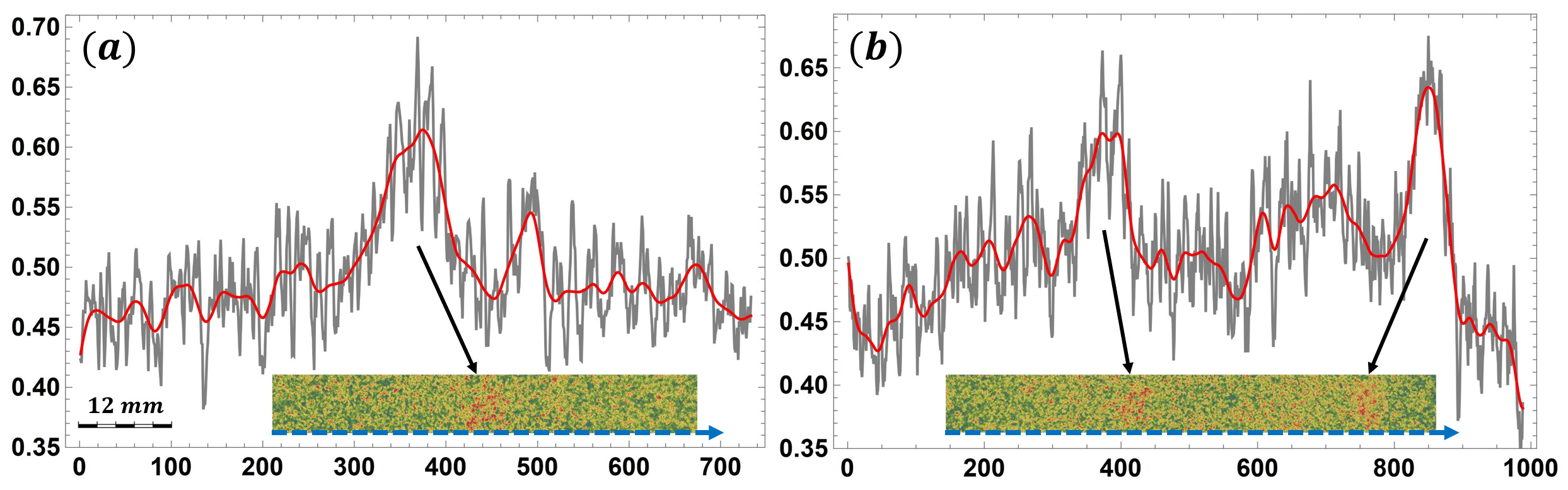}
\caption{Bubble signal analysis for the cropped FOV: width mean of luminance (gray) over the length of (a) horizontal and (b) vertical patches roughly fitted to the bubble dimensions, spanning the FOV. Note the scale bar in (a).}
\label{fig:bubble-profiles-fov}
\end{figure}

To illustrate the image processing challenge, Figures \ref{fig:bubble-profiles} and \ref{fig:bubble-profiles-fov} show examples of image noise and neutron flux transmission signal for bubbles \textit{after} pre-processing (Section \ref{sec:pre-processing}). Figure \ref{fig:bubble-profiles} shows a typical luminance distribution about a bubble region: note that here moving averages (window width in the scan direction is equal to one pixel) over (a) horizontal and (b) vertical image patches are shown, since plots over any given single pixel line would be unintelligible. Estimates from a control set of bubble regions for different frames indicate many of the images have SNR $\sim 1.5$-$1.6$, and there are cases with SNR as low as $\sim 1.2$. Similar plots for (a) horizontal and (b) vertical patches for the entire width/length of the cropped FOV are shown in Figure \ref{fig:bubble-profiles-fov}.

While visible in (a), it is especially evident in (b) that despite the compensation for neutron flux non-uniformity over the FOV, the background "mean" is still not uniform/flat. One also has rather densely packed noise peaks with luminance values comparable to the values associated with the bubbles.

\subsection{Bubble flow properties}
\label{sec:bubble-props}

Estimates indicate that Eötvös number is $Eo \in [2.1;4.1]$ and the Reynolds number near the bubbles is within $Re \in [10^3;10^4]$. This corresponds to a flow regime wherein bubbles have oscillating elliptic shapes \cite{clift-bubbles, natcomms-shape-dynamics, birjukovsPhaseBoundaryDynamics2020, birjukovsArgonBubbleFlow2020}. Bubbles have equivalent diameters of $d_\text{b} \in [6;8]~mm$  and travel with varying velocities usually in the $[20;40]~ cm/s$ range \cite{birjukovsArgonBubbleFlow2020}. Such velocities and dynamic phase boundaries dictate the high acquisition frame rate required for physical analysis. 100 FPS was considered the optimal trade-off frame rate for the experiments covered herein in that one already avoids significant motion blur about the bubbles while still maintaining manageable image SNR.

Bubbles ascend in vertical chains from the bottom of the gallium vessel to the free surface where they exit the liquid metal volume. Bubbles generally exhibit non-uniformly accelerated motion depending on the applied magnetic field and flow rate. Bubble trajectories are mostly planar zigzags (plane parallel to the largest container face) with slight out-of-plane perturbations that intensify with gas flow rate. Bubble collisions, coalescence or breakup do not occur for image sequences considered in this paper, but rather chains with closely packed bubbles manifest for higher flow rates. In the future, the methodology developed herein will also be applied for even higher flow rate cases where bubble collisions do take place.

\section{Image Processing}
\label{sec:image-processing}

\subsection{Assumptions \&  considerations for algorithm development}

The results obtained via the previous version of our image processing code \cite{birjukovsPhaseBoundaryDynamics2020, birjukovsArgonBubbleFlow2020} determined the objectives that must be met here:

\begin{enumerate}[noitemsep,topsep=0pt]

    \item Improved bubble edge detection stability -- more reliable detection, less artefacts and false positives
    
    \item Greater bubble shape detection precision
    
    \item Increased bubble detection rates at the bottom of the FOV
\end{enumerate}

The following assumptions regarding the bubbles are in effect:

\begin{itemize}[noitemsep,topsep=0pt]

    \item Bubbles have perturbed elliptic/circular shapes, i.e. not necessarily convex.
    
    \item Bubble phase boundaries do not exhibit local curvature radii less that a pre-defined fraction of their mean curvature, i.e. smoothness is assumed below a certain length scale due to surface tension.
    
    \item No coalescence or breakup of bubbles is expected at flow rates $< 500~ sccm$ considered herein.
    
\end{itemize}

In addition, the development of the methodology outlined herein was subject to the following considerations:

\begin{itemize}[noitemsep,topsep=0pt]

    \item Given the image properties/quality as described in Section \ref{sec:image-props}, we do not attempt to perform filtering \& segmentation such that argon/gallium volume fraction can be recovered, since low SNR will translate to significant errors in the output if both volume fraction field and shape recovery are attempted.
    
    \item Since the 100 FPS frame rate is the lower limit for prevention of considerable motion blur, we do not employ methods that perform spatio-temporal noise filtering -- only spatial denoising is used.
    
    \item Furthermore, given the above and to make our methods more general, we treat image sequence frames \textit{separately} at all stages of image processing. As such, it is possible to implement parallelization on the frame level with simultaneous multi-threading on to speed up the processing.
    
\end{itemize}

To achieve the above goals, it was decided to separate the bubble segmentation into two stages: \textit{global} and \textit{local} filtering/segmentation. That is, one starts by obtaining first estimates for bubble segments within images (the entire cropped FOV), and then uses a separate routine for local filtering/segmentation about the preliminary segments to improve shape detection precision at local scales (i.e. lower-wavelength corrections to first shape estimates) and to resolve false positives. 

The global filtering/segmentation routine, which is essentially a completely overhauled version of our previous image processing approach \cite{birjukovsArgonBubbleFlow2020, birjukovsPhaseBoundaryDynamics2020}, was designed and adjusted to maximize bubble detection rates at the cost of reduced shape detection precision and higher false positive rates. Further, to improve edge detection stability, we have opted for \textit{implicit} edge detection. Global noise filtering is performed in multiple stages, each targeting a certain noise type and/or wavelength range. 

As for the local filtering/segmentation routine, a recursive multi-scale analysis algorithm was implemented that is shown to perform well even for images with especially low SNR and recovers bubble shapes in cases where the first segment estimation outright fails to capture initial bubbles shapes within an acceptable margin of error. The overall structure of the proposed image processing solution is outlined in Algorithm \ref{alg:new-pipeline}.

\begin{algorithm}

    \nonl \textbf{Input}: Raw image sequence
    
    Pre-process images (Algorithm \ref{alg:pre-processing})
    
    Perform \textit{global} filtering for all images (Algorithm \ref{alg:global-filter})
    
    Process images with the \textit{multi-scale recursive interrogation filter} (MRIF, Algorithm \ref{alg:mrif-core})
    
    Apply the luminance map-based false positive filter to the MRIF output (Algorithm \ref{alg:false-positive-luminance-filter})
    
    Apply logical filters
    
    \nonl \textbf{Output:} Centroids and shape parameters for bubbles detected in each image

\caption{Overall structure of the new image processing pipeline}
\label{alg:new-pipeline}
\end{algorithm}

\FloatBarrier

\subsection{Pre-processing}
\label{sec:pre-processing}

Image pre-processing is performed in \textit{ImageJ} as shown in Algorithm \ref{alg:pre-processing}).

\begin{algorithm}

    \nonl \textbf{Input}: Raw image sequence: $16$-bit 1-channel TIFFs, $1024 \times 1024$ pixels
    
    Crop the FOV to the liquid metal domain (Figure \ref{fig:og-fov-and-crop})
    
    Compensate the images for camera dark current (mean)
    
    Compensate the flat field images for dark current (mean)
    
    Perform flat field correction (32-bit precision)
    
    Convert to 16-bit
    
    Remove outlying bright luminance values (median thresholding)
    
    Normalize the resulting images (pixel luminance re-scaled to $[0;1]$)
    
    \nonl \textbf{Output:} Dark current and flat-field corrected normalized images

\caption{Pre-processing for raw images}
\label{alg:pre-processing}
\end{algorithm}

 Images as cropped as indicated in Figures \ref{fig:og-fov-and-crop}a and b. Dark current compensation is performed by subtracting the mean projection of 5K-10K recorded dark current noise images from all images in the raw image sequence (typically 3K-9K images). Then the dark current-compensated flat field is computed from the mean projection of 5K-10K neutron beam flux distribution images. Afterwards the dark current-compensated FOV images are normalized with respect to the flat field compensated for the dark current. These corrections for the cropped raw FOV images can be expressed as

\begin{equation}
    I = \frac{I_0 - \left< I_{dark} \right>}{ \left< I_{beam} \right> - \left< I_{dark} \right>}
\label{eq:image-correction-dark-beam}    
\end{equation}
where $I$ are the luminance maps of the corrected images, $I_0$ are the cropped raw images, and $I_{dark}$ and $I_{beam}$ are the dark current and neutron beam flux images. This correction results in the spatial dependence of the SNR as a consequence of less neutrons detected behind the sample compared to the open beam.

Afterwards, bright outliers are removed using median thresholding with a 1- to 2-pixel radius. The threshold is chosen such that outlier removal modifies only the pixels where luminance by far exceeds the local median (within the designated radius). Pixels with luminance above the threshold are then assigned the local median values. Finally, the images are normalized and saved, then passed to \textit{Wolfram Mathematica} for further processing.

\subsection{First segment estimates}
\label{sec:first-segment-estimates}

First segment estimates are obtained using an algorithm referred to in this article as the \textit{global filter}, which is outlined in Algorithm \ref{alg:global-filter}. Segment estimation is performed in three stages -- noise filtering \& background removal, implicit edge detection and segment filling \& cleanup.

\begin{algorithm}

    \nonl \textbf{Input:} Normalized pre-processed images (Algorithm \ref{alg:pre-processing})
    
    \nonl \underline{\textit{Noise filtering \& background removal}}
    
    \pushline Perona-Malik (PM) filter (\ref{eq:perona-malik})
    
    Total variation (TV) filtering, Poisson model (\ref{eq:total-variation})
    
    Self-snakes curvature flow (SSCF) filter (\ref{eq:self-snakes}); \textit{memoize output}
    
    Soft color tone map masking (SCTMM) (\ref{eq:ctm-correction}); \textit{memoize output}
    
    \nonl \popline \underline{\textit{Implicit edge detection}}
    
    \pushline Compute the luminance using the gradient filter (Gaussian regularization + Bessel derivative kernel)
    
    2-threshold hysteresis binarization
    
    (Optional) Morphological erosion
    
    Thinning transform
    
    \nonl \popline \underline{\textit{Segment filling \& cleanup}}
    
    \pushline Filling transform
    
    Mean filtering (small radius)
    
    Otsu binarization
    
    Remove border components

    \nonl \popline \textbf{Output:} 
    \begin{itemize}[noitemsep,topsep=0pt]
    \popline    \item Image mask with first segment estimates
        \item Memoized SSCF-filtered image for later use in Algorithm \ref{alg:mrif-core}
        \item Memoized SCTMM-filtered image for later use in Algorithm \ref{alg:false-positive-luminance-filter}
    \end{itemize}
    
\caption{Global first segment estimator for pre-processed FOV images}
\label{alg:global-filter}
\end{algorithm}

Noise filtering is done as follows. First, Perona-Malik (PM) filtering is performed on an input image whereby the following non-linear diffusion partial differential equation (PDE) is solved over the image luminance map $I$ over virtual time $t$ with Neumann boundary conditions (BCs) \cite{perona-malik}:

\begin{equation}
    \pdv{I}{t} = \nabla \cdot \left( f(I,\alpha) \cdot \nabla I \right); ~~ f(I,\alpha) = \exp ( - { \abs{ \nabla_\sigma I } }^2 / \alpha^2 )
\label{eq:perona-malik}
\end{equation}
where $I = I(\vec{r},t)$, $f(I,\alpha)$ is the diffusion coefficient and $\alpha$ is its control parameter, and $I_0 (\vec{r}) = I (\vec{r},0)$ is the input image. $f(I,\alpha)$ prescribes anisotropic diffusion by restricting luminance diffusion across sharp edges. The number of PM iterations, $\alpha$ and the gradient Gaussian regularization width $\sigma$ that controls the sensitivity of $f(I,\alpha)$ to noise are chosen such that the PM process only acts on very small-scale noise structures -- Equation (\ref{eq:perona-malik}) is edge-preserving -- and it is mainly aimed at removing the remainder of the salt and pepper noise due to bright and dark outliers that survived pre-processing (Algorithm \ref{alg:pre-processing}). Note that

\begin{equation}
    \nabla_\sigma I = \nabla \left( K_\sigma \circledast I \right)
\label{eq:gaussian-convolution}
\end{equation}
where $K_\sigma$ is a Gaussian kernel with its standard deviation $\sigma$. In the case of our images, we found that is productive to set $\alpha$ to the $0.5$ quantile of $\abs{\nabla I}$ in $I_0 (\vec{r})$, and to set $\sigma = 1$ and the number of PM iterations to 5 \cite{weickert-nonlinear-aniso-diff-schemes, wolfram-perona-malik}.

Next, the total variation (TV) filter is applied assuming Poisson noise (typical for low-SNR underexposed images as in this case) by solving the following PDE with Neumann BCs \cite{total-variation-poisson}:

\begin{equation}
    \pdv{I}{t} = 
    \underbrace{
    \nabla \left( \frac{\nabla I}{ \abs{\nabla I } } \right)
    }_{\text{Noise filtering}}  + 
    \underbrace{
    \frac{1}{\beta I} \left( I_0 - I \right)
    }_{\text{Input/output similarity}}
\label{eq:total-variation}
\end{equation}
where the regularization parameter $\beta$ determines the balance between noise filtering (pixel value variation minimization) and input/output similarity preservation. Solving (\ref{eq:total-variation}) over virtual time $t$ asymptotically transforms the input image into a stationary (with respect to $t$) filtered image. The TV filter is set up such that it eliminates noise structures up to a fraction of the length scale of bubbles on the order of bubbles while avoiding overly distorting the luminance map. In our case, we find that setting $\beta \in [0.8,1]$ and limiting the number of TV iterations to $100$ \cite{wolfram-total-variation} yields better results.

The third stage is the application of the self-snakes curvature flow (SSCF) filter. Here the following PDE is solved with Neumann BCs \cite{self-snakes-curvature-flow}:

\begin{equation}
    \pdv{I}{t} = \abs{\nabla I} \cdot \nabla \left(  g(I,\gamma) \cdot \frac{\nabla I}{\abs{\nabla I}}  \right); ~~ g(I, \gamma) = \exp ( - {\abs{\nabla I}}^2 / \gamma^2 )
\label{eq:self-snakes}
\end{equation}
where $g(I, \gamma)$ is the curvature diffusion coefficient and $\gamma$ is its control parameter. Unlike the PM filter (\ref{eq:perona-malik}), SSCF diffuses local luminance curvature, not luminance itself. It is also edge-preserving. The number of SSCF iterations and $\gamma$ are set such that the SSCF filter eliminates any remaining sharply localized luminance maxima about the bubbles since there the luminance curvature is the greatest; at the same time, filtering must preserve bubble region contrast with respect to background. For our images we find that in some cases one may set $\gamma \rightarrow \infty$, which simplifies (\ref{eq:self-snakes}) to

\begin{equation}
    \pdv{I}{t} = \abs{\nabla I} \cdot \nabla \left(  \frac{\nabla I}{\abs{\nabla I}}  \right)
\label{eq:mean-curvature-flow}
\end{equation}
which is the mean curvature flow PDE. However, other cases require anistropic curvature diffusion and then we set $\gamma$ to the $0.5$ quantile of $| \nabla I |$ in $I_0 (\vec{r})$ as with the PM filtering. The number of performed SSCF iterations is set to 7 \cite{wolfram-self-snakes}.

The next stage is the soft color tone map (CTM) masking (SCTMM) which is a non-linear filter designed to clean up the image background by removing large-scale artefacts left over after denoising and to further separate background from bubbles while avoiding excessive erosion of the bubble regions. The large-scale structures in the background were actually one of the sources of the edge detection instability in the previous approach, especially for low-CNR images with higher bubble number density where bubble detection was often outright impossible due formed edge artefacts that could not be reliably removed.

Given a normalized original image $x$, the SCTMM background correction generates a new image $y$:

\begin{equation}
y = x * \underbrace{
\left( 
x - \underbrace{  \left( 1 - \text{CTM} (x,c) \right)  }_{\text{Soft thresholding}}
\right)
}_{\text{Soft background mask}}
\label{eq:ctm-correction}
\end{equation}

where $\text{CTM}(x,c)$ is the CTM operation and $c$ is the luminance compression factor. The CTM operation maps the colors (in this case the gray-scale values) of the image using gamma compression with a global compression factor $c$ \cite{reproduction-of-color-chapter-6}. The idea behind SCTMM (\ref{eq:ctm-correction}) is as follows. A pure $x * x$ product would have the effect of non-uniformly increasing the distances between the nearest luminance values (input $x$ is normalized) and luminance maxima and minima values would be much more distant from the mid-range luminance values, which are affected the most. If one masks or lowers the values of certain pixels within one of the images $x'$, then $x * x'$ would act as a "\textit{soft}" i.e. weighed mask (as opposed to a "hard" binary mask) that would shift the pixels with reduced values in $x'$ further towards the lower end of the luminance range, ideally making them background. Soft masking is preferred here because masking using binarization and then replacing the removed background using luminance interpolation or other methods will generally produce artificial and potentially very pronounced edges and/or reduce the contrast of actual edges.

Here it is required here that $x'$ is such that background and post-filtering artefacts are removed, and bubble contrast is enhanced while bubble features are eroded as little as possible. It was decided to opt for additive masking of the form $x'(x) = x - \textit{mask}$. An inverted $\text{CTM}(x,c)$ was chosen as \textit{mask} because, if the right $c$ value is set, $1 - \text{CTM}(x,c)$ will have high luminance for background and denoising artefacts since $\text{CTM}(x,c)$ reduces the difference in their luminance values. This way $x' = x - \left( 1 - \text{CTM} (x,c) \right)$ has, conversely, greatly reduced luminance for artefacts and background. The resulting product (\ref{eq:ctm-correction}) then has the desired properties and emphasizes the bubbles while reducing the impact of image artefacts. For the images considered here $c = 0.5$ generated better results.

Another major change is the substitution of explicit segmentation and edge detection with an implicit procedure. The SCTMM output is normalized and passed to the gradient filter (luminance gradient magnitude map of a Gaussian kernel convolution (\ref{eq:gaussian-convolution}) for an image; the Bessel derivative kernel is used) which estimates bubble edge regions (halos) in the image. The halos are segmented using double-threshold hysteresis binarization (pixel corner connections enabled) \cite{book-digital-image-processing} and then the edge estimates are obtained using the thinning transform \cite{book-digital-image-processing} (exploiting the edge gradient symmetry). Optionally, morphological erosion \cite{images-mathematical-morphology} can be applied to the halo segments before thinning, which is suggested if thinning outputs jagged edges. Finally, bubble shape masks are generated by applying the filling transform \cite{book-digital-image-processing} followed by a small-radius mean filtering. The advantage of this procedure is that it is more stable and does not require edge/area repairs for bubble edges/masks.

Afterwards, small-radius (2-3 pixels) mean filtering step followed by Otsu binarization \cite{otsu-thresholding} ensures that the edge dendrites that are occasionally left over at the filled bubble shape boundary are pruned -- this is much computationally cheaper than morphological pruning, which would also generally require multiple iterations to converge to edges that are Jordan curves immune to the pruning transform. The gradient filter regularization kernel scale is set to roughly match the bubble scale (Section \ref{sec:bubble-props}). The primary threshold for the hysteresis binarization of edge halos is by default $0.35$, though it had to be reduced to $0.25$ in some cases where the image quality was especially problematic. The secondary threshold is computed using the Otsu method. In our case we perform morphological erosion using disk structural elements with a 5-pixel radius. Finally, we remove segments that are in contact with image boundaries -- this is because only bubbles that are fully within the FOV are of interest.

\clearpage

\subsection{Multi-scale recursive interrogation filter (MRIF)}

\subsubsection{MRIF core}

\begin{figure}[htbp]
\centering
\includegraphics[width=1\textwidth]{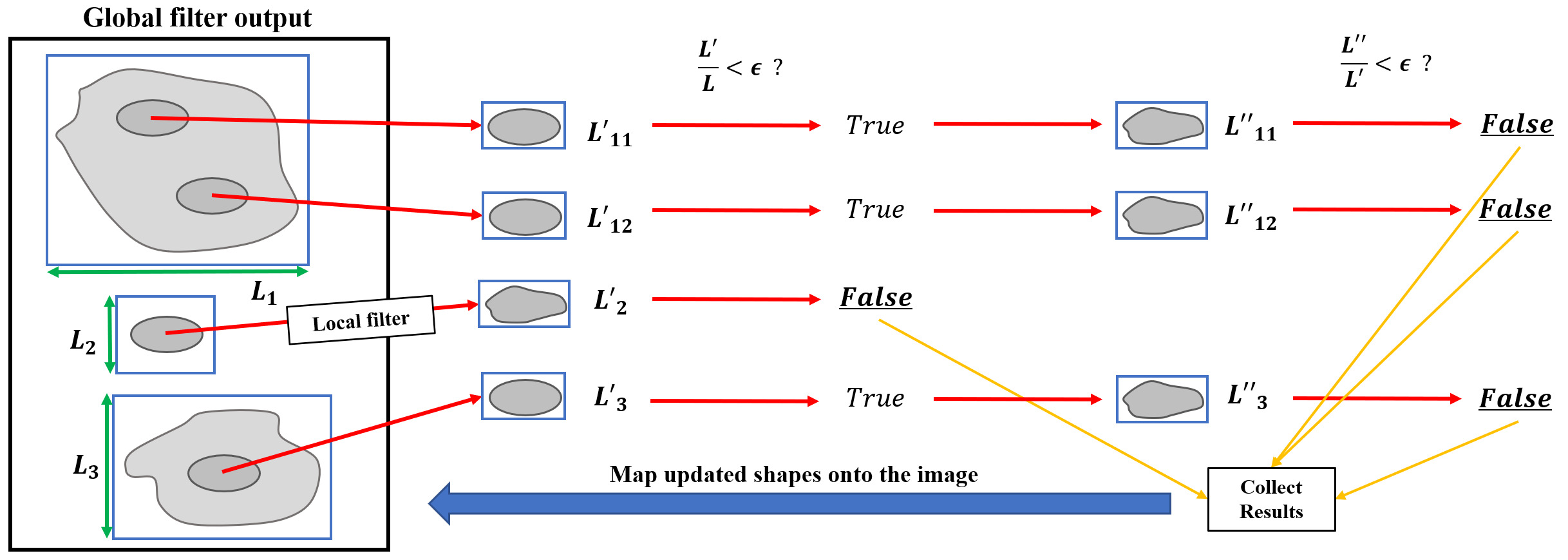}
\caption{A schematic illustration for the MRIF algorithm.}
\label{fig:mrif-illstrated}
\end{figure}

The filtering methods utilized in Algorithm \ref{alg:global-filter} are rather aggressive. Testing revealed that, while bubble detection rates are indeed significantly higher than before and bubbles are detected everywhere within the FOV, it comes at the cost of decreased shape resolution precision and a higher false positive rate. The former is very important for a more in-depth analysis of the effects of varying flow rate and MF on the behavior of bubble chains. It was decided to fine-tune the global filter such that the bubble detection rates are maximized and good first estimates of bubble shapes/sizes are obtained, and complement it with a routine that would use the first estimates to generate more precise bubble shapes and efficiently filter out false positives. To this end, we have developed an algorithm for iterative segmentation refinement -- the \textit{multi-scale recursive interrogation filter} (MRIF) outlined in Algorithm \ref{alg:mrif-core} and schematically illustrated in Figure \ref{fig:mrif-illstrated}.

\begin{algorithm}

    \nonl \textbf{Input:} \\
    \begin{itemize}[noitemsep,topsep=0pt]
    \popline  \item An image mask with first segment estimates (Algorithm \ref{alg:global-filter}, Step 12)
        \item The global SSCF filter output (Algorithm \ref{alg:global-filter}, Step 3)
    \end{itemize}
    
    \nl Define square (side length $L'$) interrogation windows (IWs) for segments based on their areas and centroids (\ref{iw-scale},\ref{virtual-iw-bounds})
    
    \nl Define the FOV image as an IW with scale $L$ based on FOV dimensions
    
    \nl \underline{\textit{For every initial segment and updated segments:}} \\
    \While{$L'/L < \varepsilon, ~~ \varepsilon > 1$ (user-defined)}{
    Map the global SSCF filter output onto the segment IW \\
    Perform local filtering (Algorithm \ref{alg:local-filter}) \\
    \eIf{updated segments were found}{
    Define new IWs (side length $L'$) for the updated segments based on their areas and centroids (\ref{iw-scale},\ref{virtual-iw-bounds})\\
    Redefine preceding IW scales as $L' \rightarrow L$
    }{
    \textbf{Break}
    }
    
   }
   
   \nl \textit{Memoize} converged IWs for all segments
   
   \nl Map the centroid coordinates of the resulting segments from converged IWs onto the original image
    
   \nl Map the updated segment masks onto the original image and build an updated global mask
    
   \nonl \textbf{Output:} \\
   \begin{itemize}[noitemsep,topsep=0pt]
    \popline  \item Updated bubble shape masks for the FOV
        \item Converged IWs for later use in Algorithm \ref{alg:false-positive-luminance-filter}
    \end{itemize}

\caption{Multi-scale recursive interrogation filter (MRIF)}
\label{alg:mrif-core}
\end{algorithm}

The key idea is to define \textit{interrogation windows} (IWs) about the initially detected bubbles to exclude irrelevant parts of the image and its intensity histogram from the analysis. This also helps to reduce the influence of any remaining artefacts left over from the global filtering. Especially for images with lower SNR and with closely packed bubbles, it may be the case that the initial segmentation is poor, i.e. two or more bubbles have been segmented as one due to the surrounding artefacts, or a bubble was connected to large-scale artifact structures, forming a large segment that obscures the true object. This means that an appropriate local filtering algorithm must be devised for IWs. However, a single local filtering pass may not be enough for various reasons, e.g. what might appear visually as a poorly segmented bubble at the scale of the current IW might actually turn out to be, after local filtering, multiple bubbles -- these would then each require another pass at a finer scale. This means that in general a series of consecutive interrogation passes could take place. Thus, MRIF performs object filtering at different scales, effectively checking that the segments have been properly resolved by the global filter and/or in the preceding local filtering iterations. A stopping criterion based on the IW size similarity between iterations makes sure that MRIF recognizes that it only makes sense to re-filter an image patch if the object is significantly smaller than the previous IW, since in this case the finer shape features may have been under-resolved.

MRIF consists of several components: an IW generator that centers the IW at the segment location and adjusts its size according to the segment size; a local filter that is responsible for filtering within IWs; a recursive routine that performs a "scale descent" and converges to the "true" segment scale starting from the first estimates; a procedure that collects the final, updates segments and maps them onto the original bubble segment mask, substituting the first estimates. The $\varepsilon$ criterion is the stopping factor that controls recursion depth, i.e. the lower IW scale threshold. In our case $\varepsilon = 2$ yielded good results. The MRIF core components are described in detail in Sections \ref{sec:interrogation-windows}, \ref{sec:local-filter} and \ref{sec:centroid-mapping}.

\subsubsection{Interrogation windows (IWs)}
\label{sec:interrogation-windows}

IWs are square windows with side length $L$ which is determined for segments as follows:
\begin{equation}
    L = \nint{
    s \cdot \sqrt{S/\pi}
    }
\label{iw-scale}
\end{equation}
where $S$ is the segment area in pixels and $s$ is a user-defined scale factor, i.e. $L$ scales with the equivalent radii of segments. Given $L$ and the segment centroid $\vec{r} = (x,y)$, an IW is defined by pixel coordinate intervals:

\begin{equation}
    \text{IW} : \bigg \{ \left[ \nint{x} - \frac{L}{2} ; \nint{x} + \frac{L}{2} \right] , \left[ \nint{y} - \frac{L}{2} ; \nint{y} + \frac{L}{2} \right] \bigg \}
\label{virtual-iw-bounds}
\end{equation}

Since $L$ for the entire image is required for the first iteration of MRIF and one should always perform local filtering at least once for every initial segment, here one can set $L = s_0 \cdot \max{ \text{dim}(z) }$ where $z$ is the SCCM-filtered FOV image (Algrithm \ref{alg:global-filter}, Step 3) and $s_0 \geq 1$ is an arbitrary scaling factor such that $L'/L > \varepsilon$ is always true for the first MRIF iteration.

Note that in general an IW may be out of the image bounds -- for this reason, two IWs are generated for a segment at every MRIF iteration: a \textit{virtual IW} defined by (\ref{virtual-iw-bounds}), and a \textit{real IW} given by

\begin{equation}
    \text{IW}' = z \cap \text{IW}
\label{iw-actual}
\end{equation}
so $\text{IW}'$ is not necessarily square. To reiterate: the current filtering scale is defined by the IW scale $L$ and the SSCF filter output is mapped onto $\text{IW}'$. For simplicity, $\text{IW}'$ will be referred to as IW from this point, as in Algorithm \ref{alg:mrif-core}, unless stated otherwise later. We observed that for our images $s \in [2;2.5]$ yielded better results.

\subsubsection{Local segment updates}
\label{sec:local-filter}

The local filter used for recursive filtering in MRIF (Algorithm \ref{alg:mrif-core}, Step 3) has elements similar to the global filter, but with important distinctions. It was designed to be less aggressive because MRIF ensures that only the crucial background context from the image is retained within an IW, e.g. the somewhat destructive SCTMM is not required. The local filter operates on an IW as described in Algorithm \ref{alg:local-filter}.

\begin{algorithm}

    \nonl \textbf{Input}: Global SSCF filter output (Algorithm \ref{alg:global-filter}) mapped onto an IW
    
    Mean filtering
    
    \nonl \underline{\textit{Implicit edge detection}}
    
    Compute edge halos using the gradient filter (Gaussian regularization + Bessel derivative kernel)
    
    Chan-Vese binarization (\ref{eq:chan-vese-functional})
    
    (Optional) Morphological erosion
    
    Thinning transform
    
    \nonl \underline{\textit{Segment filling \& cleanup}}
    
    Filling transform
    
    Mean filtering (small radius)
    
    Otsu binarization
    
    Remove border components
    
    \nonl \textbf{Output:} Updated local shape estimate
    
\caption{Local segment refinement for IWs}
\label{alg:local-filter}
\end{algorithm}

Here the mean filtering radius is a fraction of the expected bubble scale -- this is to avoid averaging out the finer shape features. The filtering radius is set based on the expected lower threshold for bubble surface perturbation wavelength. The gradient filter regularization kernel (\ref{eq:gaussian-convolution}) scale was set equal to the mean filtering radius.

Chan-Vese binarization is a variational method for object segmentation in images that does not explicitly utilize edges. It is generally more robust than edge- and histogram-based methods (e.g. Otsu, Canny \cite{otsu-thresholding, canny-edge-detection}) but is more computationally expensive \cite{getreuer-chan-vese}. However, this can be afforded relatively easily in the case of IWs that are only a fraction of the original images. Chan-Vese two-level segmentation works by assigning the following functional to an image (in this case single-channel) and minimizing it iteratively \cite{getreuer-chan-vese}:

\begin{equation}
    \min 
    \underbrace{ \mu \cdot L(C) }_{1} + 
    \underbrace{ \nu \cdot S(C) }_{2} + 
    \underbrace{ \lambda_1 \int_{D_1} \abs{I(\vec{r}) - {\langle I \rangle}_{D_1}}^2 ~ dS }_{3} + 
    \underbrace{ \lambda_2 \int_{D_2} \abs{I(\vec{r}) - {\langle I \rangle}_{D_2}}^2 ~ dS }_{4}; ~~ C = \partial D_{1,2}
\label{eq:chan-vese-functional}
\end{equation}
with respect to $C$, where $C$ is the set of segment contours, $L$ is the length of $C$, $S$ is the area enclosed by $C$, and $\mu$, $\nu$, $\lambda_1$ and $\lambda_2$ are the control parameters. The Chan-Vese process is typically initialized by defining $C$ such that the image area is covered with a checkerboard of small circular contours of adjustable size, preferably very fine \cite{getreuer-chan-vese}. Two regions within an image are defined: areas within $C$ initially given by circular contours ($D_1$) and outside $C$ ($D_2$). Minimizing (\ref{eq:chan-vese-functional}) has several effects on $C$: their combined length (term 1), area of $D_1$ (term 2), as well as the total discrepancy between the region luminance values and the region averages (terms 3 \& 4) are reduced -- the relative prevalence of these effects is dictated by $\mu$, $\nu$, $\lambda_1$ and $\lambda_2$. $C$ is defined as the zero crossing of a special level set function \cite{getreuer-chan-vese}. Optimization is performed for a certain number of iterations, generally resulting in the unification/dissolution of the initial disjoint $C$ components until, ideally, $C$ encapsulates the desired segments in the image. The Chan-Vese process also guarantees that $C$ is a set of Jordan curves.

Minimization of (\ref{eq:chan-vese-functional}) can be performed by solving the following PDE with respect to the level set function $\varphi$ for $C$ \cite{getreuer-chan-vese}:

\begin{equation}
    \pdv{\varphi}{t} = \delta(\epsilon,\varphi) \left(  
    \mu \cdot \nabla \left( 
    \frac{\nabla \varphi}{\abs{\nabla \varphi}} 
    \right) - \nu - 
    \lambda_1 { \left(I - {\langle I \rangle}_{D_1} \right)}^2 +
    \lambda_2 { \left(I - {\langle I \rangle}_{D_2} \right)}^2
    \right)
\label{eq:chan-vese-pde}
\end{equation}
with the BC

\begin{equation}
    \frac{\delta(\epsilon,\varphi)}{\abs{\nabla \varphi}} \cdot \pdv{\varphi}{\vec{n}} = 0
\label{eq:chan-vese-bc}
\end{equation}

defined for $\partial D: D = D_1 \cup D_2$. Here $\varphi = \varphi (\vec{r},t)$ with $\abs{\varphi} = 1$, $\vec{n}$ is the outward boundary normal, and $\delta(\epsilon,\varphi)$ is the level set regularization function

\begin{equation}
    \delta(\epsilon,\varphi) = \frac{\epsilon}{\pi \left( \epsilon^2 + \varphi^2 \right) }
\end{equation}

$\varphi (\vec{r},t)$ is initialized as
\begin{equation}
    \varphi ( \vec{r}, 0 ) = \sin{ \left( \frac{\pi x}{\lambda_\varphi}  \right) } \cdot \sin{ \left( \frac{\pi y}{\lambda_\varphi}  \right) }
\label{chan-vese-checkerboard}
\end{equation}

where $\lambda_\varphi$  determines the wavelength of the initialized checkerboard pattern. Equation (\ref{eq:chan-vese-pde}) with (\ref{eq:chan-vese-bc}) and (\ref{chan-vese-checkerboard}) is then solved iteratively by alternating between updates for ${\langle I \rangle}_{D_1}$ and ${\langle I \rangle}_{D_2}$, and $\varphi$.

In our case $\mu = 0.03$, $\nu = 0$, $\lambda_1 = \lambda_2 = 1$ and $\epsilon = 1$. As indicated in \cite{getreuer-chan-vese}, (\ref{chan-vese-checkerboard}) with $\lambda_\varphi = 5$ is a good initialization strategy leading to fast convergence, but we have observed that $\lambda_\varphi \in [2;5)$ yields faster convergence without perceivable quality degradation in our cases. The reason Chan-Vese instead of Otsu or hysteresis binarization is used is that the former exhibits much stabler edge halo detection in IWs and is not explicitly tied to image histograms (i.e. does not \textit{only} minimize inter-class and maximize intra-class variance for the level sets like the Otsu method) which can be very different across IWs. Once binarization is complete, one performs the same sequence of operations as with the global filter (Algorithm \ref{alg:global-filter}, Steps 7-12). For local filtering, erosion for segments is performed with 1- to 5-pixel radius disk elements and the mean filtering (cleanup) radius is 1-3 pixels.

\subsubsection{Mapping updated segments to original images}
\label{sec:centroid-mapping}

\begin{figure}[htbp]
\centering
\includegraphics[width=0.55\textwidth]{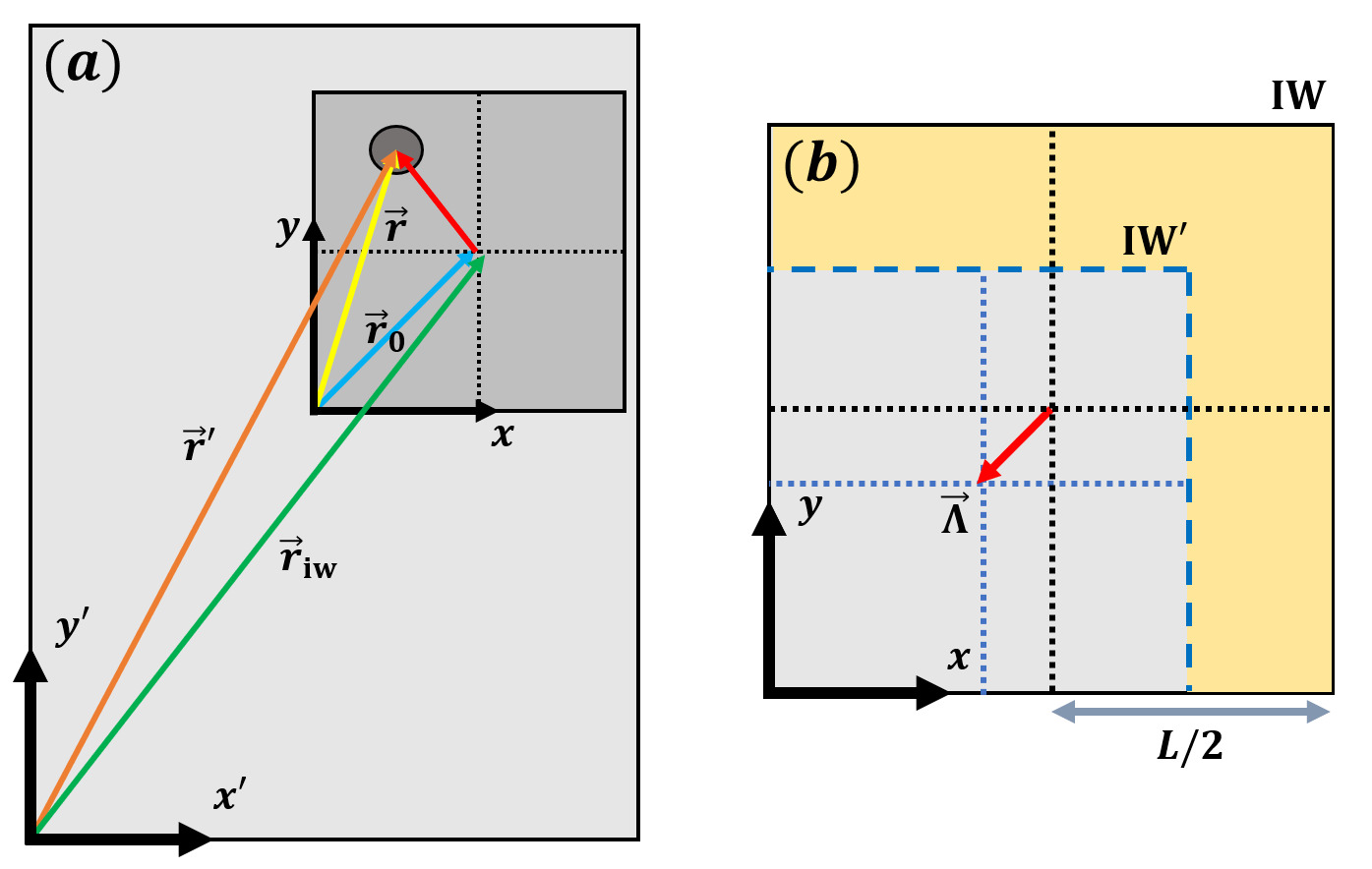}
\caption{Coordinate mapping from IWs onto the full FOV for objects detected in IWs: (a) transformation from the IW coordinate system to the image coordinate system and (b) crop correction for the IW center coordinates within the FOV. In (a) the circle represents the object detected in the IW (dark gray square). In (b) the yellow area indicates the out-of-bounds part of the virtual IW.}
\label{fig:iw-coordinate-mapping}
\end{figure}

Centroids are mapped from converged IWs onto original images (full FOV) using the following transformation:

\begin{equation}
\vec{r}~' = \vec{r}_{\text{iw}} + \left( \vec{r} - \vec{r_0}(L) \right) - \vec{\Lambda}(L, \vec{r}_{\text{iw}})
\label{eq:centroid-mapping}
\end{equation}

where  $\vec{r}~'$ is the updated centroid location in the FOV, $\vec{r}$ are the updated centroid coordinates from MRIF in the IW coordinate system, $\vec{r}_{\text{IW}}$ is the location of the virtual IW center in the FOV, $\vec{r}_0 (L)$ is the center of the virtual IW in its coordinate system and $\vec{\Lambda} (L, \vec{r}_{\text{iw}})$ is the IW crop correction for $\vec{r}_0 (L)$. The mapping is illustrated in Figure \ref{fig:iw-coordinate-mapping}. Since IW and IW' both always contain the segment (if any detected) and their centers are always within the FOV, for an IW fully within the bounds of the FOV one simply extends a radius vector $\vec{r}_{\text{iw}}$ (stored by MRIF, Algorithm \ref{alg:mrif-core}) to the IW center and then from that to the bubble centroid via $\vec{r} - \vec{r_0}(L)$ (Figure \ref{fig:iw-coordinate-mapping}a, the red vector). If the virtual IW is partially out of bounds, then the actual $\text{IW}'$ (Figure \ref{fig:iw-coordinate-mapping}b) has a different center coordinate $\vec{r_0}' = \vec{r}_0 + \vec{\Lambda}(L, \vec{r}_{\text{iw}})$ since it is displaced when the IW is cropped (\ref{iw-actual}). The crop correction given an image $z$ is as follows:

\begin{equation}
\begin{split}
    \vec{\Lambda}(L, \vec{r}_{\text{iw}}) = \vec{e}_x \cdot
    \begin{cases}
    x_\text{min}/2, & x_\text{min} < 0\\
    0, & x_\text{min} \geq 0
    \end{cases}
    + \vec{e}_x \cdot
    \begin{cases}
    0, & x_\text{max} \leq w\\
    -(x_\text{max}-w)/2, & x_\text{max} > w
    \end{cases} \\
    + \vec{e}_y \cdot
    \begin{cases}
    y_\text{min}/2, & y_\text{min} < 0\\
    0, & y_\text{min} \geq 0
    \end{cases}
    + \vec{e}_y \cdot
    \begin{cases}
    0, & y_\text{max} \leq h\\
    -(y_\text{max}-h)/2, & y_\text{max} > h
    \end{cases}
\end{split}
\label{eq:iw-crop-correction}
\end{equation}

where $x_\text{min}$, $x_\text{max}$, $y_\text{min}$ and $y_\text{min}$ are given by $L$ and $\vec{r_\text{iw}}$ via (\ref{virtual-iw-bounds}) and $\text{dim}(z) = (w,h)$. The IW crop correction is required as a separate step because the IW coordinate system origin is not necessarily within the FOV.

\subsection{Luminance map-based false positive filtering}

The MRIF procedure, in addition to its intrinsic false positive filtering capacity stemming from the recursive multi-scale analysis, is supplemented by a dedicated false positive identification algorithm that processes the segments output by MRIF. The procedure is outlined in Algorithm \ref{alg:false-positive-luminance-filter}.

\begin{algorithm}

    \nonl \textbf{Input:} 
    \begin{itemize}[noitemsep,topsep=0pt]
    \popline  \item Updated bubble shape masks for the FOV (Algorithm \ref{alg:mrif-core})
        \item Global SCTMM filter output (Algorithm \ref{alg:global-filter}, Step 4)
        \item Converged IWs (Algorithm \ref{alg:mrif-core})
    \end{itemize}
    
    \nl Map the global SCTMM filter output onto converged IWs
    
    \nl Apply the SCTMM correction (\ref{eq:ctm-correction})
    
    \nl Multiply by the segment binary mask
    
    \nl Normalize the image
    
    \nl Compute $\left< I \right> \cdot \max (I)$ for all segment regions
    
    \nl \underline{\textit{Thresholding for all segments:}}
    
    \eIf{$\left< I \right> \cdot \max (I) < \eta; ~~ \eta \in [0;1]$ (user-defined)}{
    Flag the segment as a false positive
    }{
    Nothing
    }
    
    \nl Remove the identified false positives from the bubble shape mask
    
    \textbf{Output:} FOV bubble shape masks without detected false positives
    
\caption{Luminance map-based false positive filtering for MRIF output}
\label{alg:false-positive-luminance-filter}
\end{algorithm}

Algorithm \ref{alg:false-positive-luminance-filter} exploits the observation that SCTMM applied locally to the global SSCF filter output mapped onto converged IWs should produce very strong intensity maxima and an overall higher intensity in the segment region in the case of a true bubble while the opposite should hold for false positives. The product of mean and maximum intensity is used so that neither of the two criteria alone are enough to pass the filter, since it might be the case that a region with an otherwise background level intensity might exhibit a tightly localized intensity maximum; similarly, mean intensity filtering alone is not enough since a bubble should have a strong maximum of transmission intensity about its centroid, which in itself is not as strongly correlated to the mean intensity. Another way to interpret this is that, if the maximum and mean thresholding have certain probabilities of accepting a false positive, then the max/mean product thresholding has a false positive acceptance probability at least lower than the greater of the two components. Note that Algorithm \ref{alg:false-positive-luminance-filter} uses a luminance compression factor $c = 0.5$ for SCTMM. We found that false positives are efficiently filtered (i.e. also minimizing true positive elimination) when $\eta \in [0.1;0.125]$.

After filtering the MRIF output using Algorithm \ref{alg:false-positive-luminance-filter}, all properties of interest are measured for all remaining bubble segments and logical filters can be applied for further false positive elimination. In our case logical filters check for implausible bubble coordinates, sizes and aspect ratios and remove the outliers from the dataset of measured bubble shapes. Finally, the resulting data can be post-processed and interpreted.

\section{Performance analysis}

\subsection{First segment estimates}
\label{sec:global-filter-performance}

\begin{figure}[htbp]
\centering
\includegraphics[width=\textwidth]{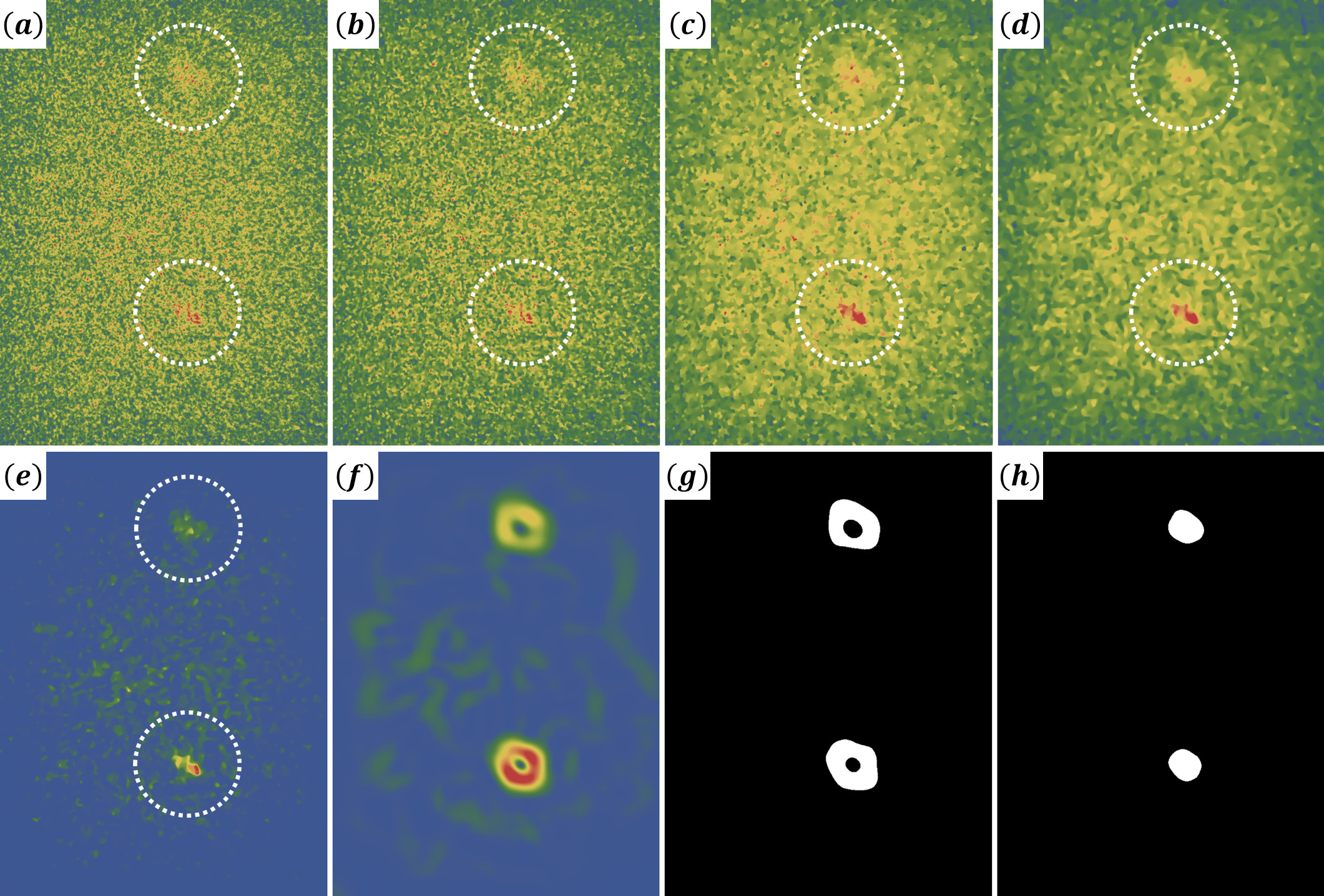}
\caption{First segment estimation stages (Algorithm \ref{alg:global-filter}): (a) original pre-processed (Algorithm \ref{alg:pre-processing}) image; (b) PM-filtered image; (c) Poisson TV-filtered image; (d) SSCF-filtered image; (e) SCTMM output; (f) gradient filter output (edge halos); (g) hysteresis-segmented edge halos (prior to erosion); (h) bubble masks obtained after erosion, thinning, filling, small-radius mean filtering, Otsu binarization and border component removal. The color scheme is as in Figure \ref{fig:og-fov-and-crop}.}
\label{fig:global-filter-steps}
\end{figure}

\begin{figure}[htbp]
\centering
\includegraphics[width=\textwidth]{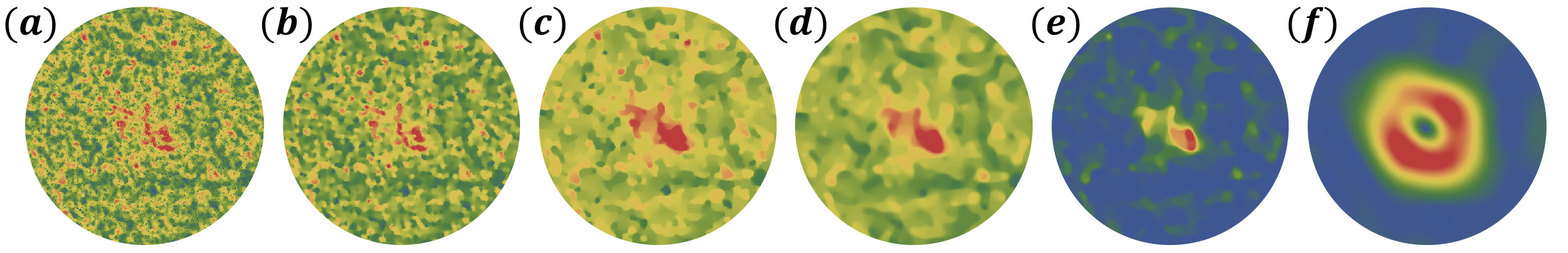}
\caption{Effects of global filtering (Algorithm \ref{alg:global-filter}) on the luminance map of the neighborhood of the lowermost bubble in Figure \ref{fig:global-filter-steps-bubble} indicated by a white dashed circle: (a) neighborhood projection of the original pre-processed image; (b) PM-filtered; (c) Poisson TV-filtered; (d) SSCF-filtered; (e) SCTMM output; (f) gradient filter output (edge halos).}
\label{fig:global-filter-steps-bubble}
\end{figure}

\begin{figure}[htbp]
\centering
\includegraphics[width=\textwidth]{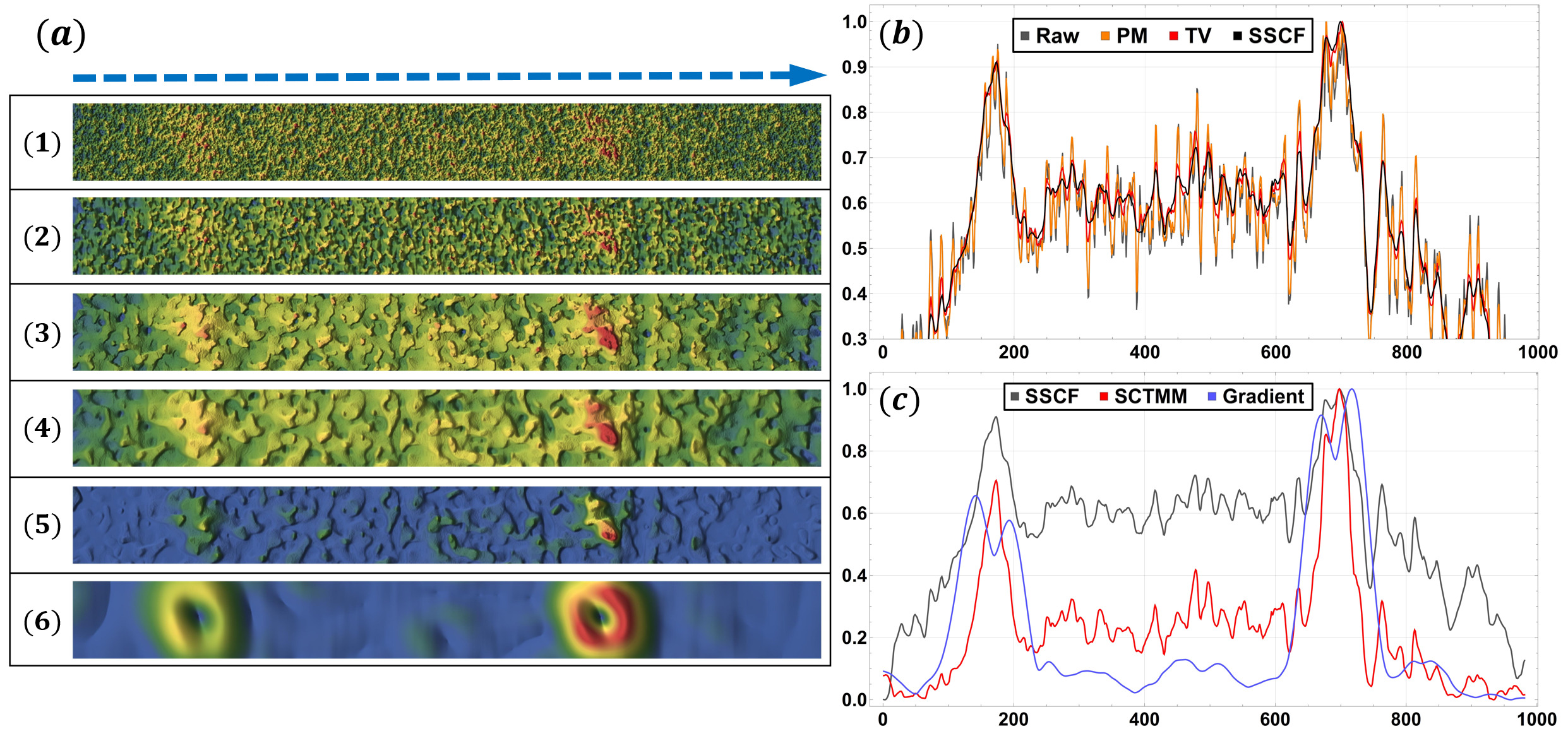}
\caption{
The effect of filtering stages on bubble signal versus noise and background: (a) pseudo-3D shaded relief plots of strips taken from Figures \ref{fig:global-filter-steps}a-f (sub-figures 1-6, respectively) containing bubbles and background in between; (b) mean luminance profiles over image strips (1-4) in the direction indicated by a blue dashed arrow in (a) for pre-processed, PM-, TV- and SSCF-filtered images; (c) mean luminance profiles for 
SSCF-, SCTMM- and gradient-filtered strips (4-6). Note that all luminance profiles in (b) and (c) have been normalized for direct comparison.}
\label{fig:global-filter-steps-stripes}
\end{figure}

Figure \ref{fig:global-filter-steps} shows the effect of subsequent operations that the global filtering routine (Algorithm \ref{alg:global-filter}) performs for a pre-processed image. The difference between Figures \ref{fig:global-filter-steps}a and \ref{fig:global-filter-steps}b is that the noise with low wavelengths (sharply localized luminance maxima and minima left over from pre-processing) have been eliminated. Next, the TV filter (Figure \ref{fig:global-filter-steps}c) consolidates the high luminance values within the bubble regions (white dashed circles), increasing the SNR for the bubbles. Also, noise is significantly damped and the wavelength of its features is increased even more. However, the TV filter does not remove the sparse larger-scale luminance maxima still seen in Figure \ref{fig:global-filter-steps}c in the background between the bubbles as efficiently as it is desired without degrading the CNR for the bubbles. This function is performed by SSCF (Figure \ref{fig:global-filter-steps}c) which specifically diffuses the leftover intensity maxima, increasing bubble SNR (and CNR due to reduced noise in the background) further. In the case of Figure \ref{fig:global-filter-steps} one has $\gamma \rightarrow \infty$ for (\ref{eq:self-snakes}). A close-up view of one of the bubble neighborhood's transformations is shown in Figure \ref{fig:global-filter-steps-bubble} where one can visually notice that bubble CNR is increased by SSCF (d) and that intensity maxima are indeed eliminated from the background. A more detailed analysis of how how noise filtering stages affect the image and the bubbles can be seen in Figure \ref{fig:global-filter-steps-stripes} where a vertical strip containing both bubbles is taken from the images in Figure \ref{fig:global-filter-steps} and their relief plots (a) and mean luminance profiles (b and c) are shown.

\clearpage

With SNR increased by the sequence of PM, TV and SSCF filters, SCTMM is now applied to increase the CNR -- this is especially clearly seen in Figure \ref{fig:global-filter-steps-stripes}c where one can see that SCTMM significantly flattens the background while preserving bubble signal intensity, as intended. This enables the gradient filter to produce bubble edge halos with an even greater CNR (note the high depth of the halo "wells" in Figure \ref{fig:global-filter-steps-stripes}a-6, which extend well below the binarization threshold, down to background luminance levels), enabling clean segmentation, as seen in Figure \ref{fig:global-filter-steps}g.

\clearpage

\subsection{MRIF}
\label{sec:mrif-performance}

\begin{figure}[htbp]
\centering
\includegraphics[width=0.65\textwidth]{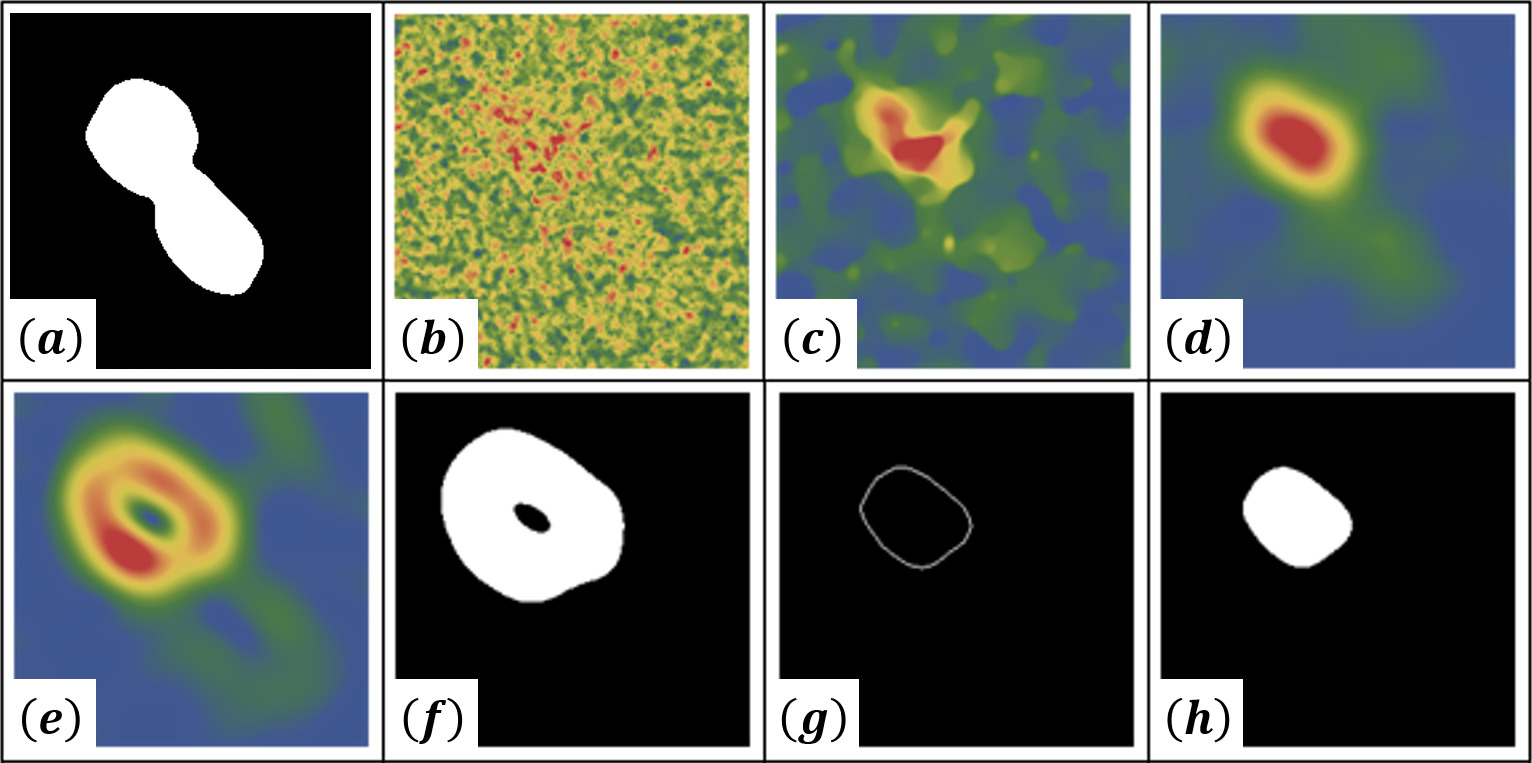}
\caption{Updating a first segment estimate using the local filter (Algorithm \ref{alg:local-filter}): (a) first estimate in an IW; (b) original pre-processed image projected onto the IW; (c) SSCF output projection onto the IW (Algorithm \ref{alg:global-filter}); (d) mean-filtered SSCF output; (e) gradient filter output; (f) Chan-Vese segmentation output; (g) edges extracted via erosion and thinning; (h) updated local segment for the IW. Segment filling and cleanup are similar to what is done in Algorithm \ref{alg:global-filter}.}
\label{fig:local-filtering-steps}
\end{figure}

The first segment estimator could have been enough for bubble shape extraction, when tuned appropriately, if not for the fact that the image considered in Figures \ref{fig:global-filter-steps}-\ref{fig:global-filter-steps-stripes} is one of the better examples in terms of noise and artefacts present, i.e. a considerable portion of the images captured in our experiments are of a much poorer quality. Not only are the obtained shapes often imprecise or deformed, they can at times be decidedly non-physical -- one such example is show in Figure \ref{fig:local-filtering-steps} where in (a) a segment is shown that looks like two bubbles that are in the process of merging.

Such events are not expected at the flow rate for which this image was acquired, therefore Figure \ref{fig:local-filtering-steps}a shows an obvious artefact. The bottom-right corner of (b) contains closely packed high-luminance spots which likely have been combined and merged with the bubble region in the upper-left corner. However, once MRIF targets the segment and the local filter is applied to the SSCF output projected onto the segment IW (c) in stages (d-h), the artefact is no longer present and a single bubble is correctly resolved.

\begin{figure}[htbp]
\centering
\includegraphics[width=0.95\textwidth]{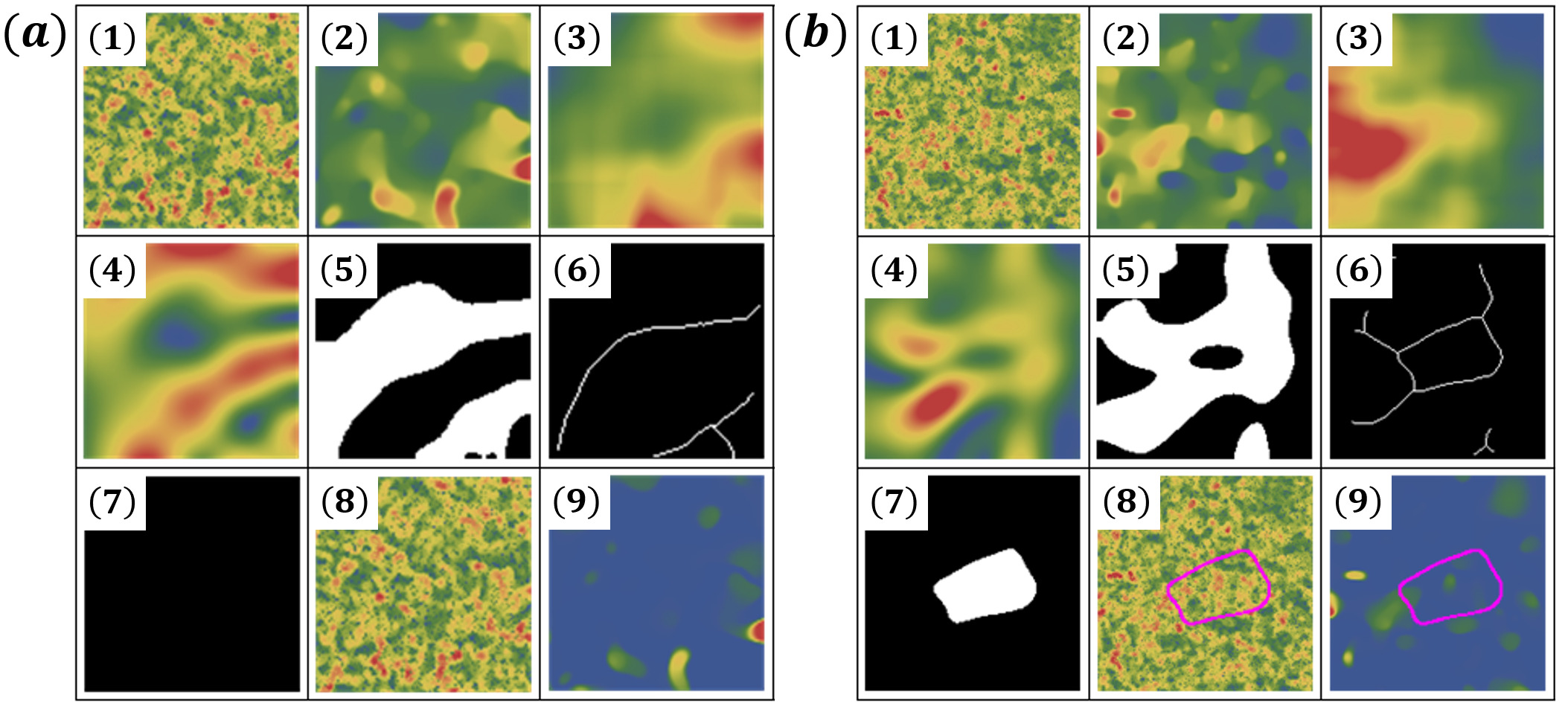}
\caption{Instances of false positives revealed by MRIF with (a) no local segments found and (b) an artefact (purple outline) that will be eliminated later by Algorithm \ref{alg:false-positive-luminance-filter} based on the segment luminance map. Sub-figures (1-9) in (a) and (b) are: (1) is the original image projected onto an IW; (2-7) are the respective local filtering stages (Figure \ref{fig:local-filtering-steps}c-h); (8) is (1) with detected edge overlays; (9) is SCTMM applied to (2).}
\label{fig:mrif-false-positives}
\end{figure}

In addition to such artefacts, since the global filter was tuned to maximize the odds of detecting bubbles in the FOV, there are cases where detected segments are false positives. Two instances of such segments interrogated by MRIF are shown in Figure \ref{fig:mrif-false-positives}. In (a) one can see that the local filter has revealed that there is indeed no segment contained within the IW.

However, interrogating the false positives in an IW might on occasion produce segments yet again, as in (b), where the gradient filter stage (b4) generated a structure that resembles a bubble edge halo. It was then segmented and, through edge cleanup, transformed into a segment that seems eligible -- but simply overlaying it over the original image projected onto the IW, one can see that this is not the case. However, MRIF effectively performs a two-factor false-positive check, and in such cases the luminance map-based filter (Algorithm \ref{alg:false-positive-luminance-filter}) serves as a backup. Once the global SCTMM output is projected onto the IW and SCTMM is applied to the resulting image (b9), one can see that the segment overlay contains only background, and thus this false positive will be eliminated since it has $\left< I \right> \cdot \max (I) < \eta$.

\begin{figure}[h!]
\centering
\includegraphics[width=0.80\textwidth]{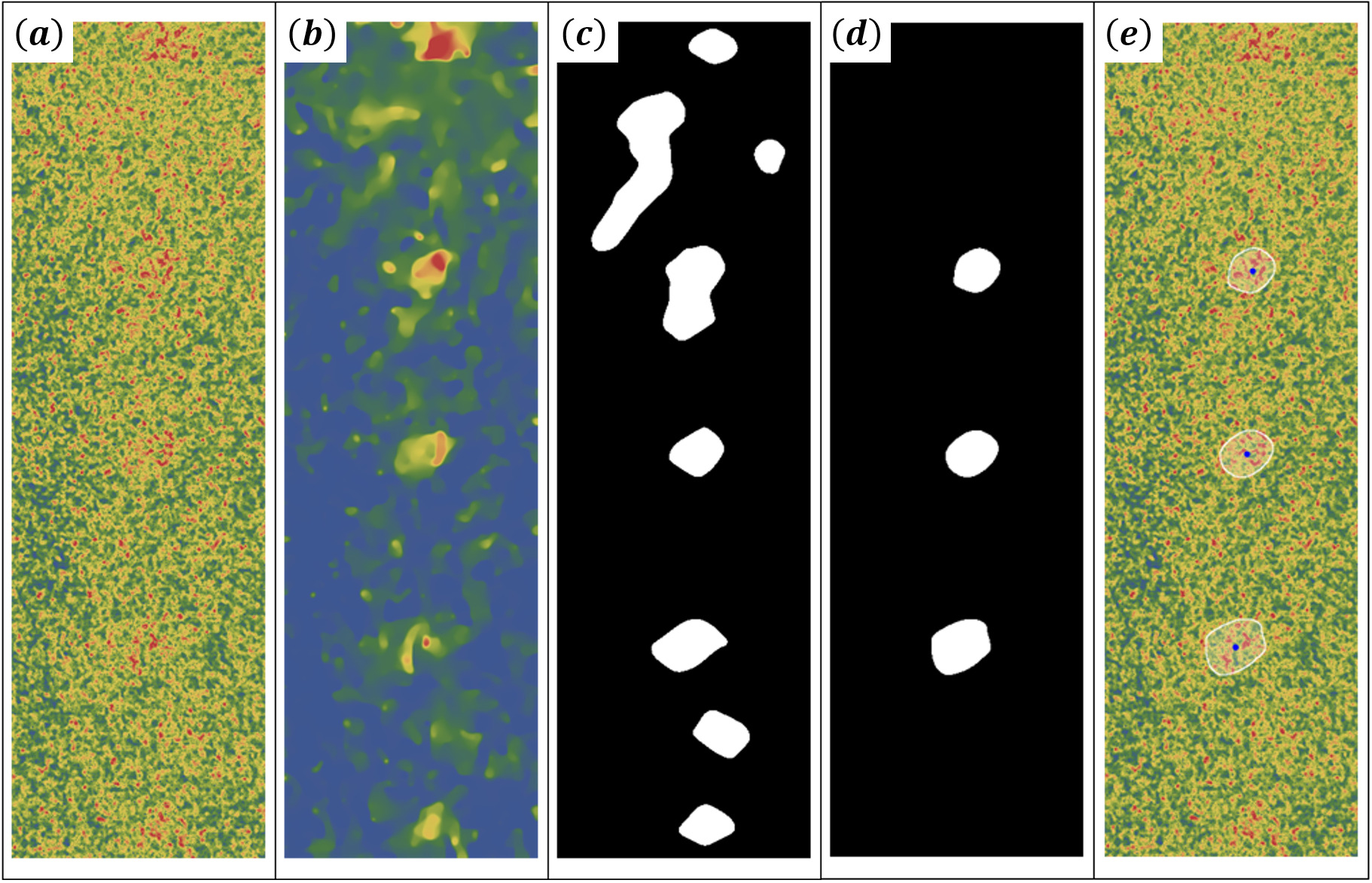}
\caption{An example of segmentation improvement by MRIF: (a) horizontally cropped pre-processed image -- note the very low CNR in the upper half of the image; (b) global SCTMM filtering result; (c) first segment estimates; (d) updated segments output by MRIF; (e) output segment overlays for (a). Note that the top- and bottom-most segments from (c) are not present in (d) and (e) -- this is the correct behaviour, since one can visually see in (a) and (b) that the corresponding bubbles are partially outside of the FOV, and thus are not eligible for analysis.}
\label{fig:mrif-example-1}
\end{figure}

An example containing several instances of false positives, under-resolved shapes and bubble regions merged with noise patterns is shown in Figure \ref{fig:mrif-example-1}. Notice that the image quality, even visually, is much worse than in Figures \ref{fig:global-filter-steps}-\ref{fig:global-filter-steps-stripes}. The artefacts in the upper part of the image stem from lower CNR in (b), whereas one of the bottom artefacts comes from a noise structure in the background that resembles a bubble. However, as seen in (d) and (e), MRIF successfully removes all false-positives and artefacts while improving the shape estimates.

\begin{figure}[h!]
\centering
\includegraphics[width=0.95\textwidth]{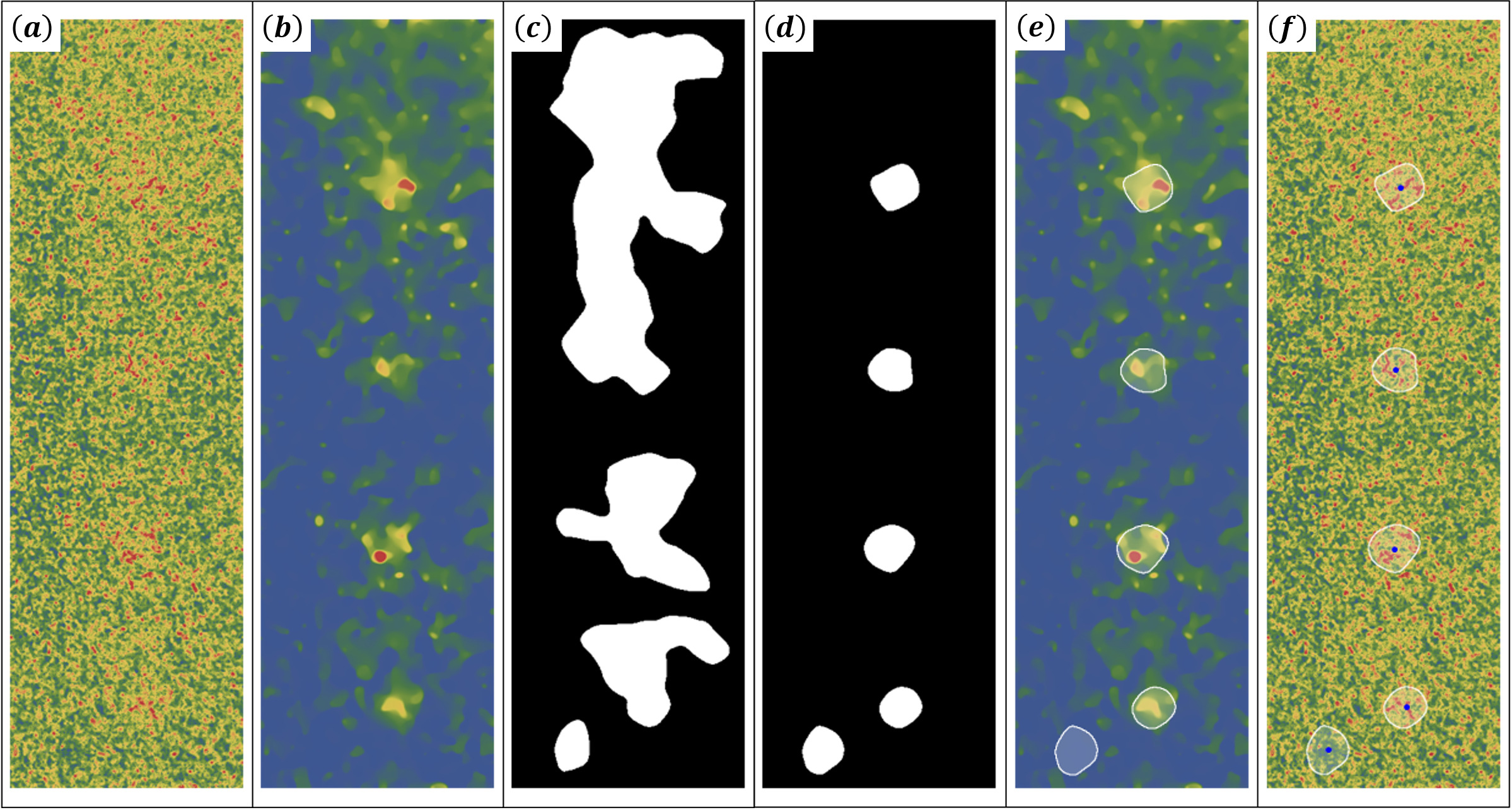}
\caption{Another example of MRIF resolving initially occluded and incorrectly detected segments. Sub-figures (a-d) and (f) correspond to Figures \ref{fig:mrif-example-1}a-e, respectively, and (e) shows the MRIF output overlays for the global filter output. The false positive in the bottom-left corner of (d-f) that survived MRIF interrogation will be eliminated by Algorithm \ref{alg:false-positive-luminance-filter}, since it corresponds to background (e).}
\label{fig:mrif-example-2}
\end{figure}

An even more difficult case is seen in Figure \ref{fig:mrif-example-2} where very large segments appear. The largest one in the upper part of (c) has occluded two of the four bubbles visible in (a-b). This is also an instance where the crop correction (\ref{eq:iw-crop-correction}) is significant for remapping the updated segments onto the FOV (\ref{eq:centroid-mapping}). MRIF successfully resolves bubbles from the first estimates, as seen in (d-f). Notice the segment in the bottom-left part of (e) and (f) -- its luminance map contains background only, so it will be later eliminated by Algorithm \ref{alg:false-positive-luminance-filter}.

\begin{figure}[htbp]
\centering
\includegraphics[width=1\textwidth]{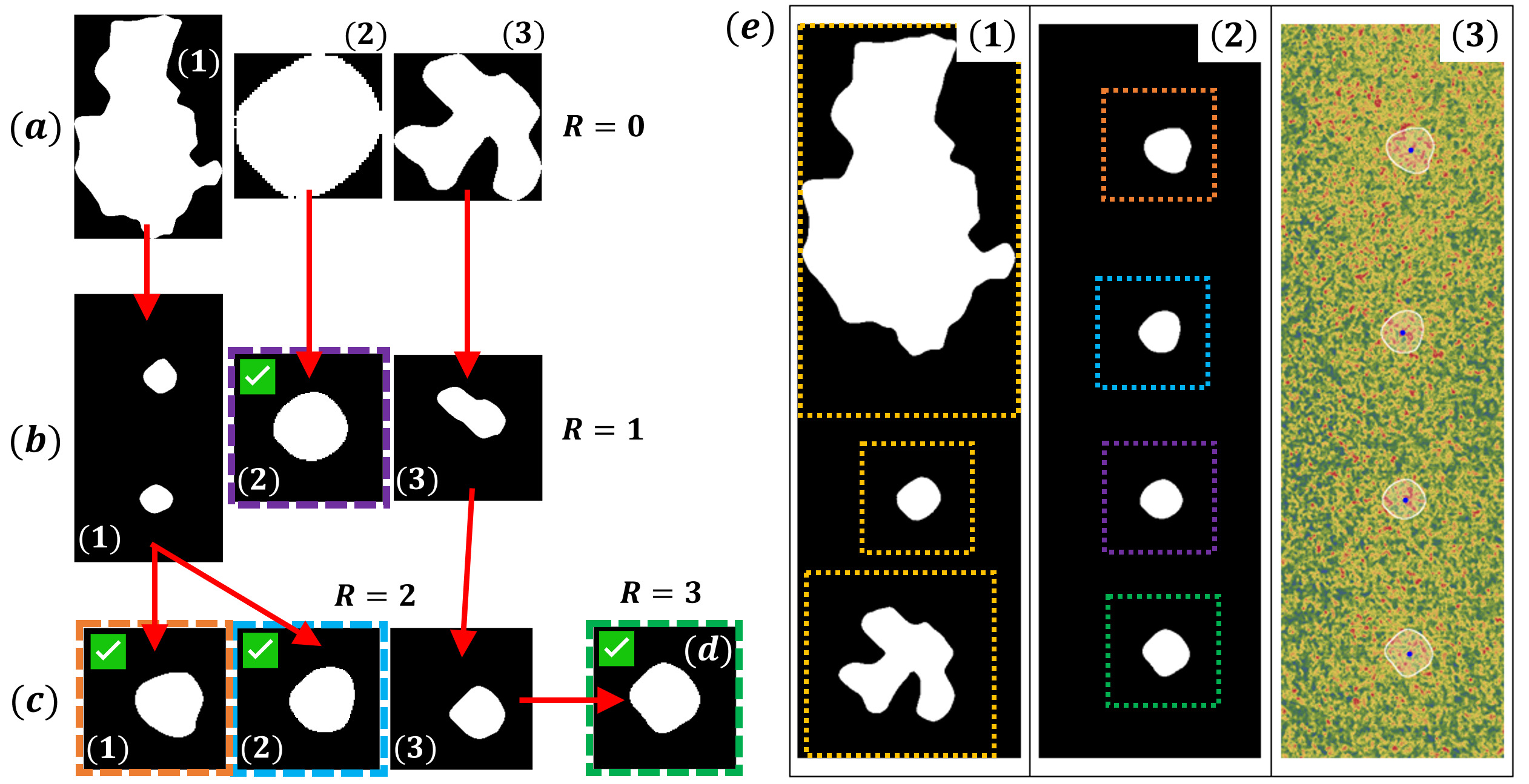}
\caption{An illustration of the iterative interrogation process for an image with one of the worst overall SNR values, where $R$ is the MRIF recursion depth. Sub-figures (a-d) show $R$ running 0 through 3 for the segments detected at each depth. Sub-figure (e) displays (1) the initial segments, (2) MRIF output and (3) output overlays for the original image. Converged segments are flagged with green ticked boxes. Colored frames in (e2) correspond to (b2), (c1-2) and (d) via respective colors. Orange frames in (e1) indicate the IWs at $R=1$: note the high aspect ratio of the topmost IW -- its virtual counterpart is significantly out-of-bounds and therefore considerable crop correction (\ref{eq:iw-crop-correction}) is assigned to properly map (c1-2) onto the FOV.}
\label{fig:mrif-recursion}
\end{figure}

To see how MRIF iteratively resolves cases like the above two examples, consider Figure \ref{fig:mrif-recursion} where the updates for the first segment estimates are shown for MRIF iterations. One can see in (e3) that the SNR and CNR are even worse than in the cases shown in Figures \ref{fig:mrif-example-1} and \ref{fig:mrif-example-2}. The largest segment seen in (e1) and (a1) is first resolved into two bubbles (b1) and then each bubble is interrogated once more, obtaining more precise shapes. The bottom-most segment in (e1) requires the most MRIF iterations -- the first two, (b3) and (c3), remove portions of the artefact that had obscured the bubble, and the last iteration (d) updates the resolved shape. The resulting bubble shapes are then mapped onto the FOV as indicated in (b-e).

\clearpage

\clearpage

Here it is important to reiterate that the performance of MRIF strongly depends on the user-defined IW scaling factor $s$ (\ref{iw-scale}) and the critical IW length scale ratio $\varepsilon$ for the latest and the (potential) next recursion iterations. Specifically, the $\chi = s/\varepsilon$ ratio is of interest -- we suggest $\chi \geq 1$ in general. An optimal ratio, which, as we observe, should be the same for all image sequences acquired under similar conditions, enables MRIF to efficiently "strip" the first segment estimates of artefacts and perform one final update for the resolved bubbles, as in Figures \ref{fig:mrif-recursion}a-3 - \ref{fig:mrif-recursion}d. Here $\chi = 1.25$ for examples shown in Figures \ref{fig:mrif-example-1}-\ref{fig:mrif-recursion}, with $s = 2.5$ and $\varepsilon = 2$. Before adjusting $\chi$, we would recommend that the user determine the $s$ value which gives the best performance for the local filter, also adjusting the settings for the latter -- this will determine a starting value for $\epsilon$ before further optimization.

\subsection{Results for existing \& new data}
\label{sec:results-existing-and-new-data}

\begin{figure}[h!]
\centering
\includegraphics[width=0.85\textwidth]{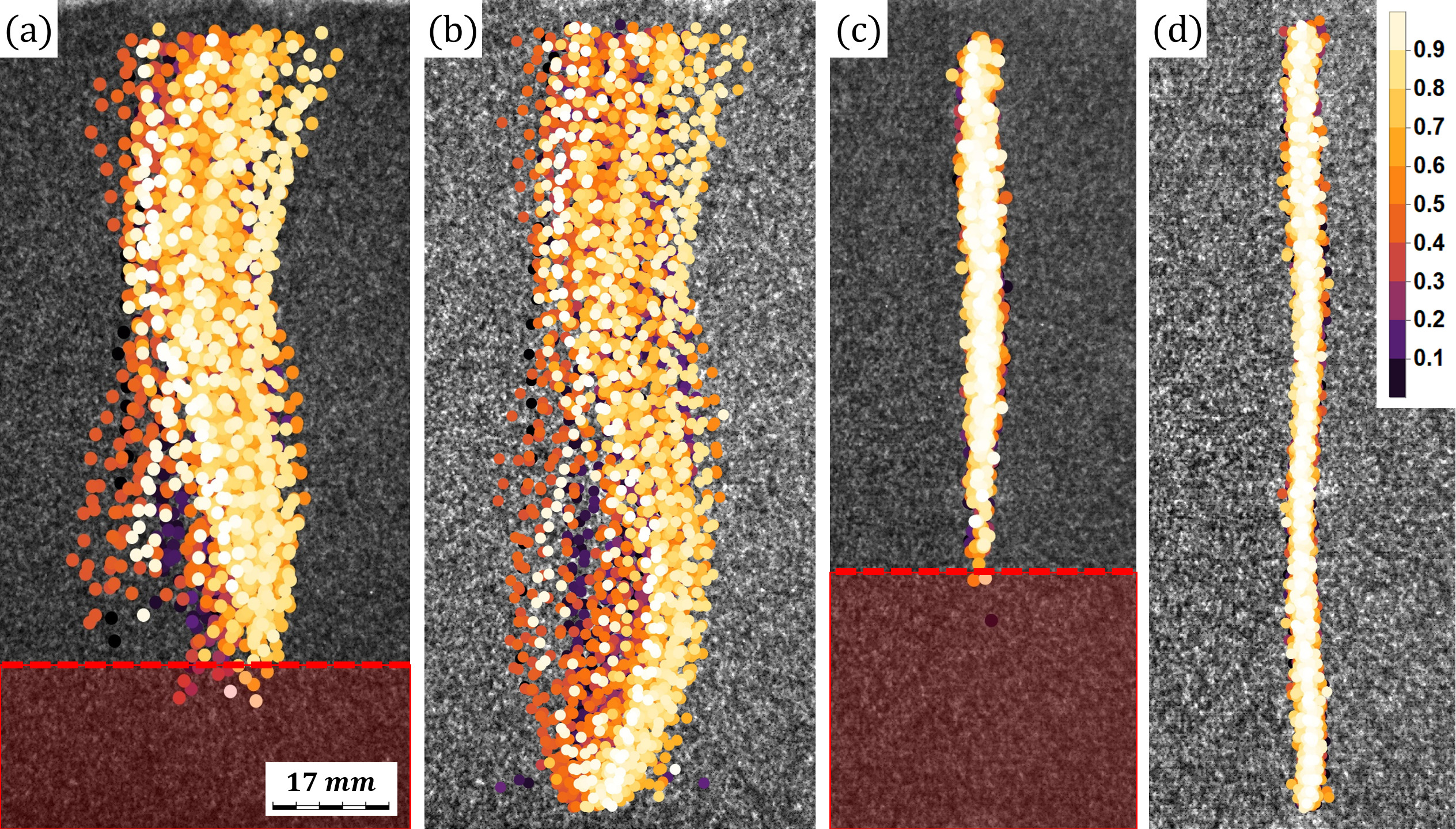}
\caption{The locations of bubbles detected in the FOV for a sequence of 3000 images (30 seconds) at a $100~sccm$ flow rate for the model system from \cite{birjukovsArgonBubbleFlow2020, birjukovsPhaseBoundaryDynamics2020} (Section \ref{sec:image-acquisition}, horizontal inlet): no applied MF, (a) previous and (b) current image processing code; applied $\sim 265~ mT$ horizontal MF, (c) previous and (d) current image processing code. Bubble locations are marked with dots color-coded in the chronological order of appearance. Note the color legend in (d): imaging starts at 0 and ends at 1. The red-tinted areas indicate the blind zones of the previously used image processing code. All sub-figures are to scale.}
\label{fig:pipeline-versus-detections}
\end{figure}

\begin{figure}[h!]
\centering
\includegraphics[width=1\textwidth]{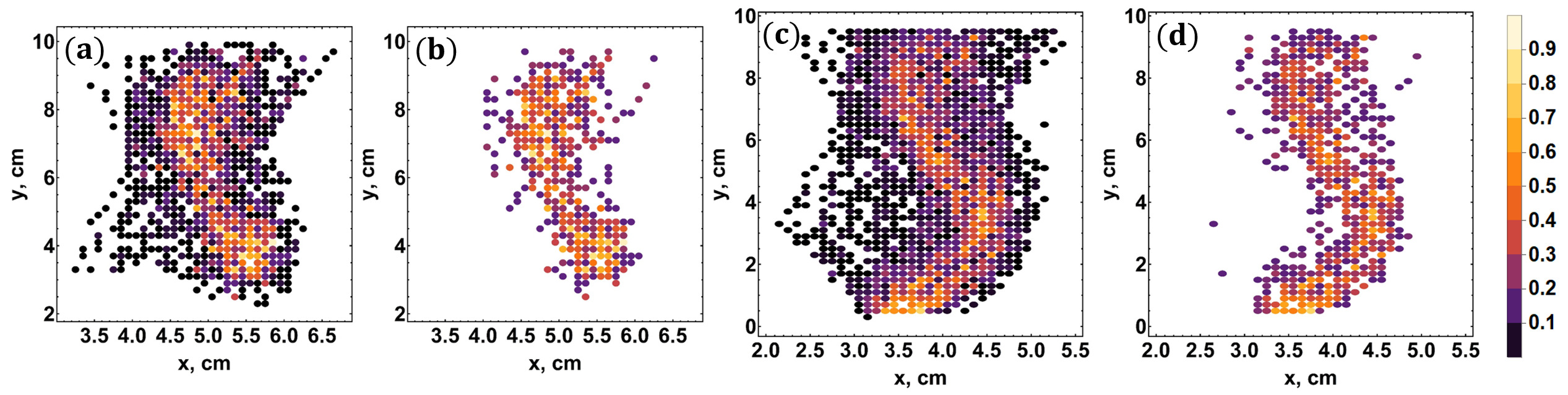}
\caption{Normalized bubble detection density histograms with $(dx,dy) = (2,4)~ mm$ bins color-coded by detection counts: (a-b) the case in Figure \ref{fig:pipeline-versus-detections}a with (a) all bins and (b) bins with 3+ detections shown; (c-d) the case in Figure \ref{fig:pipeline-versus-detections}b with (c) all bins and (d) bins with 4+ detections shown. Note the color legend to the right of (d).}
\label{fig:pipeline-versus-detection-density}
\end{figure}

\begin{figure}[htbp]
\centering
\includegraphics[width=1\textwidth]{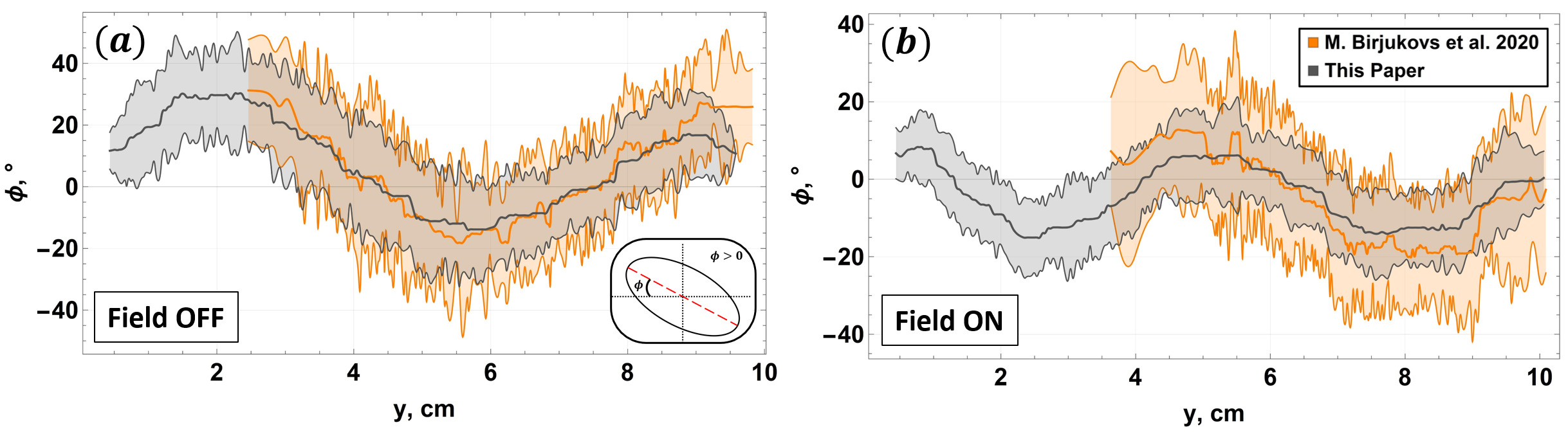}
\caption{Bubble tilt angle versus elevation (averaged curves and error bands) over the FOV bottom for (a) the case with no applied MF (Figures \ref{fig:pipeline-versus-detections}a and b) and (b) applied horizontal $\sim 265~mT$ MF (Figures \ref{fig:pipeline-versus-detections}c and d), both cases at a $100~sccm$ flow rate. Orange indicates the previous paper \cite{birjukovsPhaseBoundaryDynamics2020} and the current results are shown in gray. The tilt angle definition is shown in the bottom-right corner of (a).}
\label{fig:pipeline-versus-tilt-angle}
\end{figure}

Aside from the examples shown in Sections \ref{sec:global-filter-performance} and \ref{sec:mrif-performance}, it is also of interest how the developed approach performs for entire image sequences in terms of bubble detection density in the FOV and the physicality of obtained results, i.e. bubble trajectory and shape properties. Figures \ref{fig:pipeline-versus-detections}-\ref{fig:pipeline-versus-tilt-angle} demonstrate the differences in performance for the preceding image processing pipeline \cite{birjukovsArgonBubbleFlow2020, birjukovsPhaseBoundaryDynamics2020} and the methodology presented here.

One can clearly see in Figure \ref{fig:pipeline-versus-detections} that the new version of the image processing code outperforms the previous version by completely avoiding the blind zones in the lower part of the FOV for both image sequences. Notice also that bubble tracks visible in (b) are much more coherent than in (a). Figure \ref{fig:pipeline-versus-detection-density}, in turn, shows that with the new methods one can now clearly resolve the classical S-shaped mean trajectory cluster formed by zigzag trajectories, as seen in (c) and (d), as opposed to (a) and (b) where a significant portion of the events is missing. The deflection bias in the $x>0$ direction in (c) and (d) is determined by the horizontal inlet releasing gas in that direction.

Another point of interest are the tilt angle dynamics resolved in \cite{birjukovsPhaseBoundaryDynamics2020} versus the current results -- this is showcased in Figure \ref{fig:pipeline-versus-tilt-angle}. Three things are important to note here: first, as a consequence of the blind zone elimination, the new curves extend all the way through the FOV; second, the average trends yielded by both approaches indicate that the previously used code indeed resolved the dynamics without unacceptable inaccuracy; third, the error bands are considerably narrower about the averaged curves for the present results. The latter is especially true for the case with applied MF shown in (b) where the SNR was much lower then in the image sequence corresponding to (a). This indicates that the new approach indeed yields significant improvement not just in bubble detection, but also in shape boundary resolution. The experimentally obtained results in \cite{birjukovsPhaseBoundaryDynamics2020} were in a rather good agreement with performed simulations, meaning one thus has indirect validation of the presented approach.

\begin{figure}[htbp]
\centering
\includegraphics[width=0.85\textwidth]{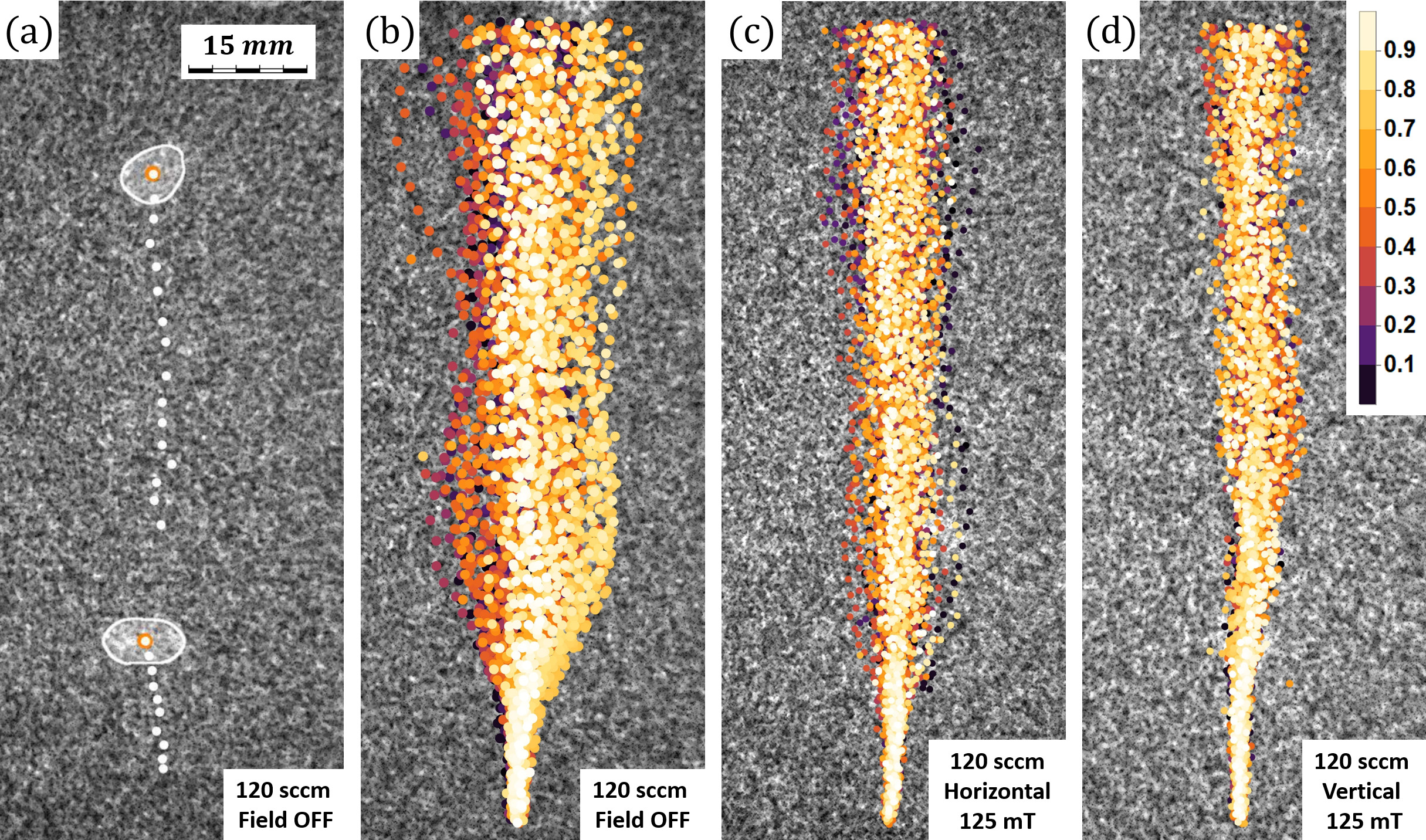}
\caption{The locations of bubbles detected in the FOV for a sequence of 3000 images (30 seconds) at a $120~sccm$ flow rate for the new model system (Section \ref{sec:image-acquisition}, vertical inlet): (a) an example of detected bubbles: white contours are shapes, orange dots are the current positions and white dots are the preceding detections; (b-d) all detected bubble positions with (b) no applied MF, (c) horizontal $\sim 125~mT$ MF and (d) vertical $\sim 125~mT$ MF. Bubble locations in (b-d) are marked as in Figure \ref{fig:pipeline-versus-detections}. All sub-figures are to scale.}
\label{fig:trajectories-field-versus}
\end{figure}

The developed approach was also applied to the newly acquired data to ensure consistency in the code output across different experimental campaigns -- one instance of the new results is shown in Figure \ref{fig:trajectories-field-versus}. Again, the bubbles are resolved over the entire FOV for all three cases shown. An in-depth physical analysis of the bubble dynamics is beyond the scope of this paper and is reserved for a follow-up article.

\clearpage

An important note on Figure \ref{fig:pipeline-versus-tilt-angle}: the averaged curves and the error bands were computed from the image processing code output as outlined in Algorithm \ref{alg:data-post-processing}. The quantile spline envelopes (QSEs) were computed following \cite{antonov-qse} and using the code (\textit{Wolfram Mathematica} package) available on \textit{GitHub}: \href{https://github.com/antononcube/MathematicaForPrediction/blob/master/QuantileRegression.m}{Anton Antonov
(\textit{antononcube}): MathematicaForPrediction/QuantileRegression.m}. For all the datasets represented in Figure \ref{fig:pipeline-versus-tilt-angle}, we used $q=0.9$ with 3-rd order splines and $\nint{2.5\% \cdot N}$ spline knots for QSEs where $N$ is the number of points in the dataset; we set $N_\text{QSE} = \nint{14\% \cdot N}$ and $\delta_\text{b} = \nint{0.5\% \cdot N}$ (point density-adaptive physical bin size); the TV regularization parameters \cite{total-variation-rof-model} were $1$ for QSEs and $0.25$ for the binned data.

\begin{algorithm}

    \nonl \textbf{Input}: Tilt angle (or any other bubble property) values over elevation (or any other independent variable)
    
    Construct the upper and lower quantile spline envelope (QSE) functions for the input data with the quantile $q$
    
    Sample the QSEs evenly in intervals of $N_\text{QSE}$ points over the independent variable range 
    
    Smooth the QSEs using Gaussian TV filtering
    
    Designate the data points above the upper and below the lower lower QSEs as outliers and remove from the dataset
    
    Bin the remaining data points over the independent variable range into bins with size of $\delta_\text{b}$ points
    
    Compute means and standard deviations for the resulting bins
    
    Smooth the resulting mean value sequence using Gaussian TV filtering
    
    \nonl \textbf{Output:} An averaged curve for the dataset with error bands
    
\caption{Post-processing for the output of the image processing code.}
\label{alg:data-post-processing}
\end{algorithm}

Finally, even though one cannot check how many bubbles the image processing code actually failed to detect without manual inspection (not feasible), one can evaluate the amount of detection events that are ruled out as false positives at the various stages of code execution for a sequence of images. The results for the five image sequences considered above (Figures \ref{fig:pipeline-versus-detections} and \ref{fig:trajectories-field-versus}) are presented in Table \ref{tab:false-positive-elimination-rates}.

\begin{table}[!h]
\begin{center}
\begin{tabular}{| c | c | c | c | c | c |}
\hline
\textbf{Image sequence} & \textbf{MRIF} & \textbf{Algorithm \ref{alg:false-positive-luminance-filter}} & \textbf{Property filters} & \textbf{Total} & \textbf{Remainder} \\ 
\hline
Figure \ref{fig:pipeline-versus-detections}b & 1.75 & 2.63 & 1.21 & 5.49 & 94.5\\ \hline
Figure \ref{fig:pipeline-versus-detections}d & 1.51 & 4.63 & 6.80 & 12.5 & 87.5\\ \hline
Figure \ref{fig:trajectories-field-versus}b & 1.53 & 2.05 & 1.25 & 4.76 & 95.2\\ \hline
Figure \ref{fig:trajectories-field-versus}c & 0.71 & 0.51 & 0.18 & 1.39 & 98.6\\ \hline
Figure \ref{fig:trajectories-field-versus}d & 1.62 & 0.89 & 0.32 & 2.81 & 97.2\\ \hline
\end{tabular}
\end{center}
\caption{False positive elimination rates ($\%$) over three stages -- MRIF, the luminance map-based filter (Algorithm \ref{alg:false-positive-luminance-filter}) and the object property filter (with respect to the input that each filter received), and the percentage of detection events input to MRIF that were eliminated in total.}
\label{tab:false-positive-elimination-rates}
\end{table}

Notice that in the most difficult case of the five (Figure \ref{fig:pipeline-versus-detections}d) most of the work is done by the luminance map-based filter and the object property filter. However, the intrinsic filtering capacity of MRIF is significant because it filters out the detection events that very likely would have passed both of the two following stages.

\section{Experimental validation}

\subsection{Stationary reference body}
\label{sec:statonary-reference-body}

\begin{figure}[htbp]
\centering
\includegraphics[width=0.80\textwidth]{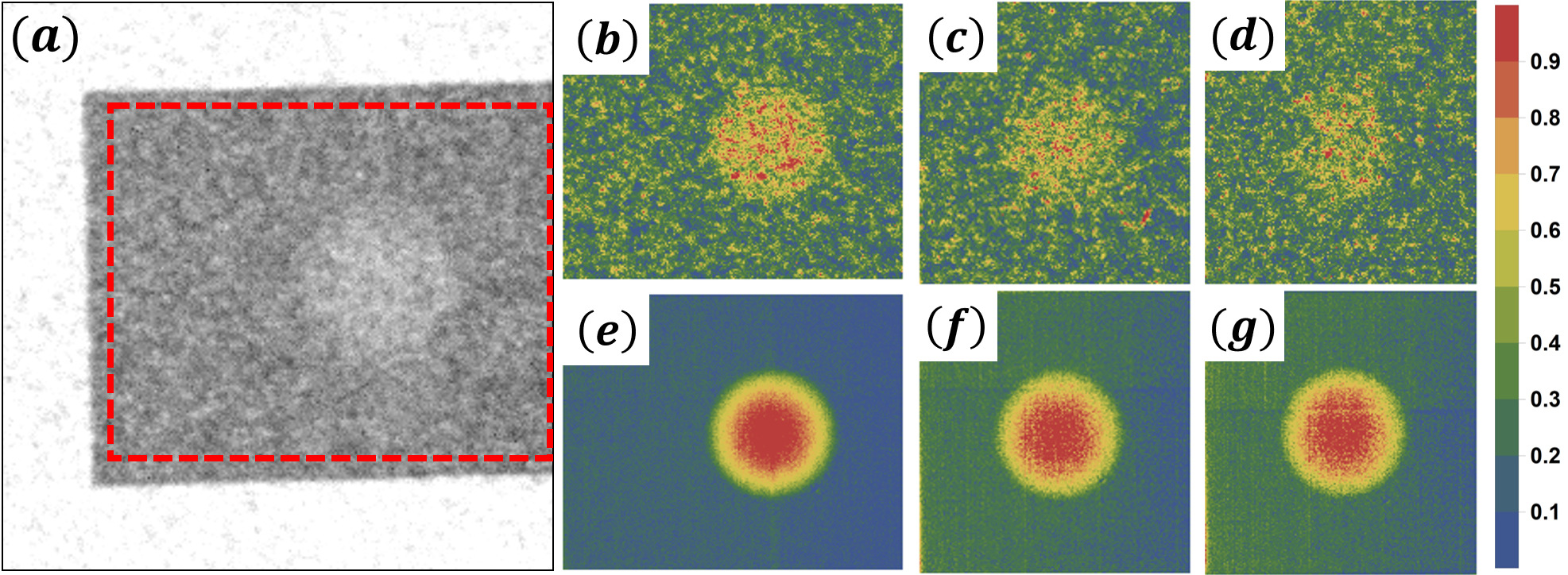}
\caption{Static reference body: (a) an example radiograph; (b-c) normalized images cropped as in (a) for neutron flux transmission through the shorter axis, the longer axis, and the latter with an extra distance to the scintillator, respectively; (e-g) normalized mean luminance maps for the respective image sequences. Note the color bar to the right.}
\label{fig:reference-body-spheres}
\end{figure}

The developed image processing algorithm is first validated by applying it to the images of a stationary reference body described in Section \ref{sec:image-acquisition}. Three imaging cases are considered here: neutron flux transmission through the shorter body axis, the longer axis, and the latter with an extra distance from the body to the scintillator. Thus, the SNR of the neutron-transparent spherical cavity within the body progressively decreases for these cases. This is illustrated in Figure \ref{fig:reference-body-spheres}.

A neutron radiography image of the reference body (slightly inclined) is shown in (a) where one can see the rectangular brass frame (darker) and a circular projection of the spherical void (brighter), as well as the surrounding background due to air. Neutron transmission in the case of (a) is along the shorter of the body axes. In all three imaging cases the images are cropped as indicated in (a). Examples of cropped images of the spherical void within the body with exposure very similar ($\sim 1.3 \times$) to that for the bubble images (100 FPS, $\sim 1.3 \times$ neutron flux) are shown in (b-d) for the three cases listed above, respectively. The corresponding mean luminance maps shown in (e-g) were obtained by averaging over $\sim 5.8$K, $\sim 12$K and $\sim 9.5$K images (entire recorded sequences), respectively --  these averaged images are used to obtain reference shapes. The shapes detected by the image processing code in images like (b-d) are then compared to the reference shapes to compute shape detection error metrics. Note that all images seen in Figure \ref{fig:reference-body-spheres} and used for validation are obtained from raw images by pre-processing via Algorithm \ref{alg:pre-processing}, as with bubble flow images.

\begin{figure}[htbp]
\centering
\includegraphics[width=0.9\textwidth]{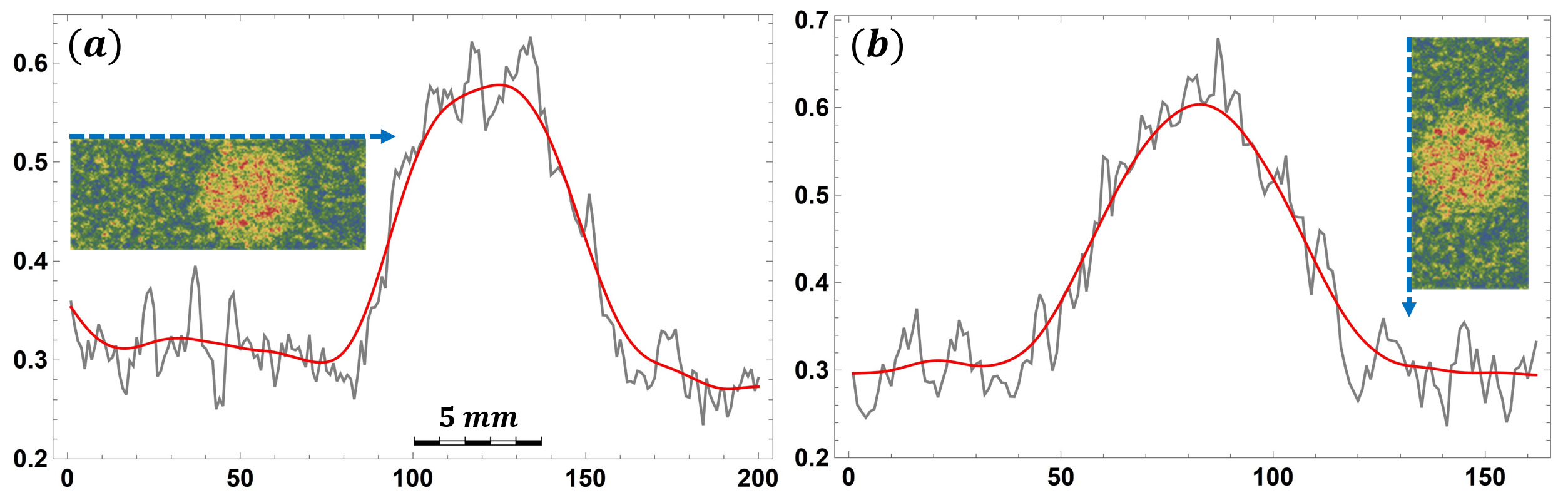}
\caption{Neighborhood analysis for the reference spherical void, transmission along the shorter axis: normalized width mean of luminance (gray) over the length of (a) horizontal and (b) vertical patches roughly fit to the sphere dimensions, spanning the cropped body images. Note the pixel-to-$mm$ scale bar in (a). Patch scan directions are indicated by the dashed blue arrows. The image patches are not to scale and their luminance maps are normalized as in Figure \ref{fig:reference-body-spheres}. The red curve is the
total variation filtered (Gaussian, regularization parameter equal to 1 \cite{total-variation-rof-model}) gray curve.}
\label{fig:sphere-profile-short}
\end{figure}

\begin{figure}[htbp]
\centering
\includegraphics[width=0.9\textwidth]{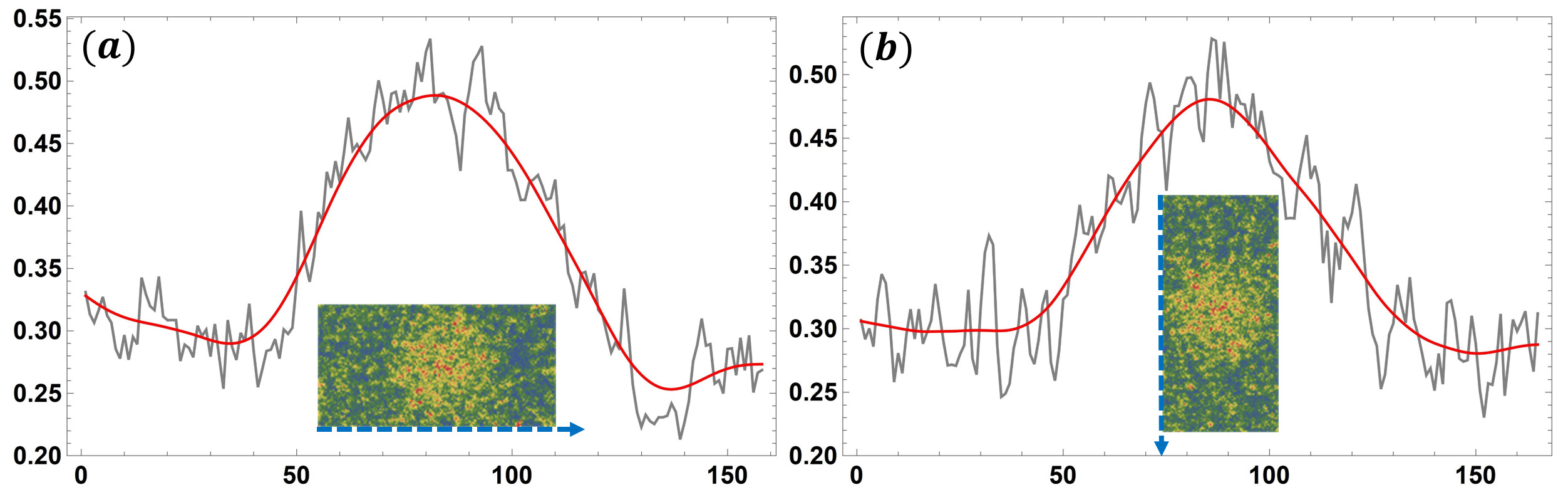}
\caption{Neighborhood analysis for the reference spherical void, transmission along the longer axis: normalized width mean of luminance (gray) over the length of (a) horizontal and (b) vertical patches roughly fit to the sphere dimensions, spanning the cropped body images.}
\label{fig:sphere-profile-long}
\end{figure}

\begin{figure}[htbp]
\centering
\includegraphics[width=0.9\textwidth]{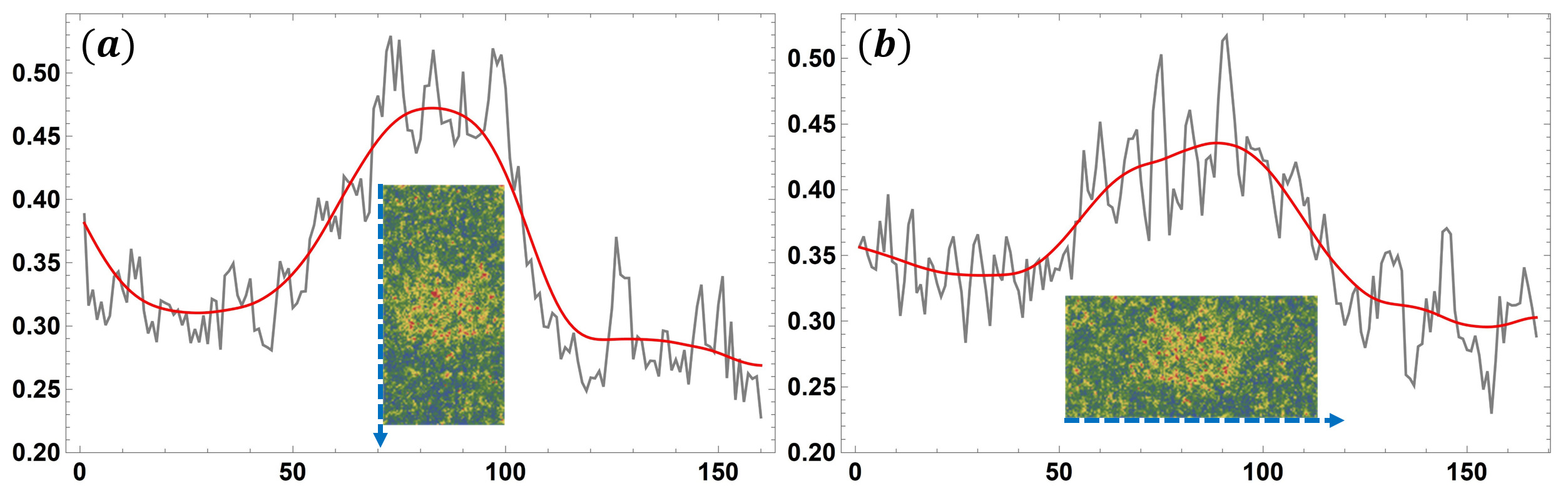}
\caption{Neighborhood analysis for the reference spherical void, transmission along the longer axis with a extra distance to the scintillator: normalized width mean of luminance (gray) over the length of (a) horizontal and (b) vertical patches roughly fit to the sphere dimensions, spanning the cropped body images.}
\label{fig:sphere-profile-long-extra}
\end{figure}

Note that the images shown in Figure \ref{fig:reference-body-spheres} have the image side length to sphere diameter ratios that are very similar to what one has for MRIF IWs. To compare the reference body images to the bubble images in terms of image quality, consider Figures \ref{fig:sphere-profile-short}-\ref{fig:sphere-profile-long-extra} versus Figure \ref{fig:bubble-profiles}.

Notice that the case shown in Figure \ref{fig:sphere-profile-long-extra} is very similar to the bubble neighborhood case in Figure \ref{fig:bubble-profiles}. In fact, Figure \ref{fig:sphere-profile-long-extra} exhibits arguably even worse CNR and SNR, meaning that, despite different materials (gallium and argon versus brass and air) and slightly different neutron flux, the reference measurements are representative of the flow imaging conditions and can be used for direct validation of the new image processing code.

To obtain the reference shapes from the images in Figure \ref{fig:reference-body-spheres}e-g, one first applies the Gaussian TV filter \cite{total-variation-rof-model}

\begin{equation}
    \pdv{I}{t} = 
    \nabla \left( \frac{\nabla I}{ \abs{\nabla I } } \right) + 
    \frac{1}{\mu} \left( I_0 - I \right)
\label{eq:total-variation-rof-model}
\end{equation}

where the notation is as in (\ref{eq:total-variation}) (note that the left-most term is different), $\mu = 0.25$ is the regularization parameter and 15 TV iterations are performed \cite{wolfram-total-variation}. Afterwards, double-Otsu hysteresis binarization is performed followed by image border component removal and mean filtering (5-pixel radius).

To ensure fair verification, we apply both the global (Algorithm \ref{alg:global-filter}) and the local (Algorithm \ref{alg:local-filter}) filters to the reference void images with parameters identical to those used for bubble images -- these are provided in Sections \ref{sec:first-segment-estimates} and \ref{sec:local-filter}.

Once all images are processed and masks given by the global and the local filter are obtained, we compute the following shape detection error metrics for both filters:

\begin{itemize}[noitemsep,topsep=0pt]
    \item $S_\Delta$ -- the area of the difference between the detected and the reference masks
    \item $\delta S$ -- the absolute difference in the areas of the detected and the reference masks
    \item $\delta r$ -- the absolute difference of the detected and the reference mask effective radii
    \item $\delta c$ -- the absolute difference in the circularity (the ratio of the equivalent disk circumference to the shape polygonal length) of the of the detected and the reference mask
    \item $(\delta x, \delta y)$ -- the absolute difference in centroid coordinates between the detected and the reference masks
\end{itemize}
where all metrics are normalized to the respective properties of the reference masks, except $(\delta x, \delta y)$ is normalized to the reference radius.

Here the $\delta c$ and $\delta r$ metrics serve primarily as "red flags" against gross inaccuracies in the detection of the circular void projection shapes. Under normal circumstances, one should have $\delta c < 3$-$5\%$ and the $\delta r$ histogram should conform to that of $\delta S$. The other three metrics are used directly for shape detection error quantification where the most important one is $S_\Delta$.

\subsection{Moving reference body}

\begin{figure}[htbp]
\centering
\includegraphics[width=0.9\textwidth]{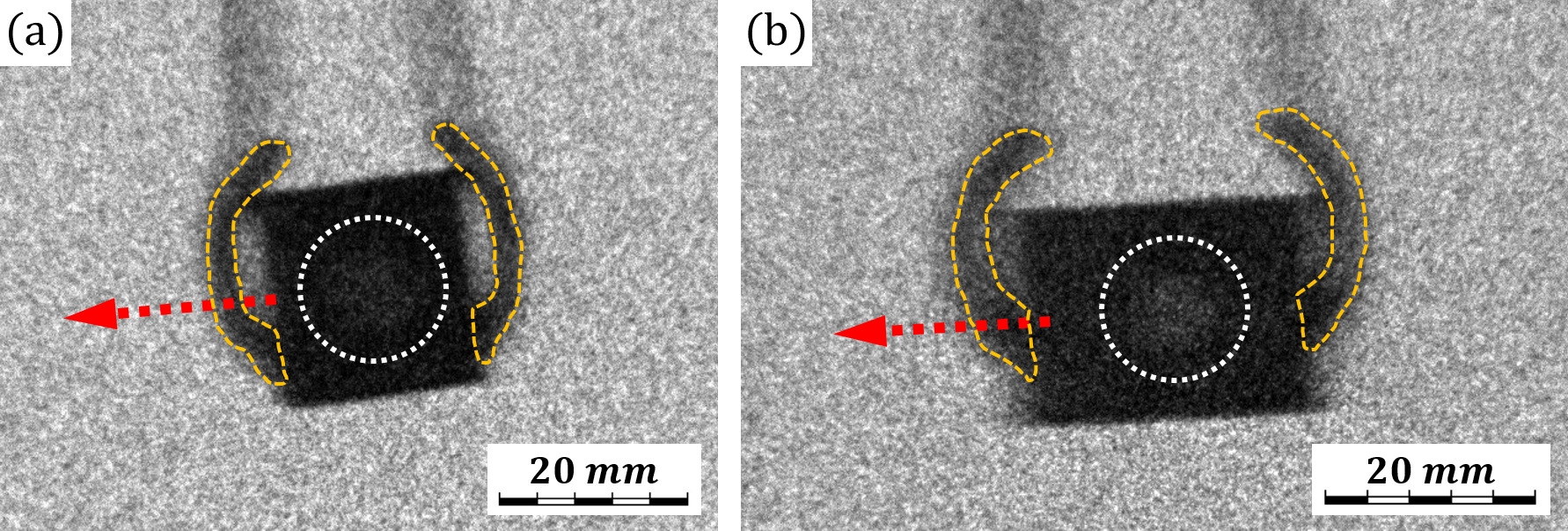}
\caption{Neutron radiography image of the pendulum used for the reference experiments with the moving reference body: neutron flux transmission along the (a) longer and (b) shorter axes. The red dotted arrows indicate the direction of motion and the white dotted circles highlight the locations of the spherical void. The pendulum arm is a standard lab holder and the reference body is held by clamps (dashed orange lines). Note the motion blur visible at the body boundaries and the holder arms above the clamps.}
\label{fig:kekw-pendulum}
\end{figure}

The principles behind imaging a moving reference body are as above, except the body is now attached to a pendulum (Figure \ref{fig:kekw-pendulum}) that periodically oscillates, and thus the body travels back and forth through the FOV. The motion is mostly horizontal and is initially strongly damped until the pendulum amplitude reaches a state where its oscillations exhibit a very slow decay. This enables us to determine the dynamics of the error metrics outlined in Section \ref{sec:statonary-reference-body} and assess the effects of motion blur as the body and the void within it decelerate.

Here, before the global and local filters can be applied to the cropped reference body images, one must first segment the body within the FOV (Figure \ref{fig:kekw-pendulum}), crop the masked image to an IW about the mask centroid, repair the IW images as necessary, and then apply the filters. This procedure is somewhat more involved that the one in Section \ref{sec:statonary-reference-body} and is outlined in Algorithm \ref{alg:void-region-extraction-repair}.

The Poisson TV filter (\ref{eq:total-variation}) reduces the noise in the image and the image luminance map inversion makes the darker body area foreground and the surrounding air (brighter) background. The SCTMM filter (\ref{eq:ctm-correction}) then increases the body CNR and the body is segmented using the Chan-Vese process (\ref{eq:chan-vese-functional}). Morphological opening (disk structural elements, 15-pixel radius) and erosion (disk structural elements, 5-pixel radius) help separate the body from the clamps (Figure \ref{fig:kekw-pendulum}), as well as to remove artefacts, if any. Border component removal is done because only body segments fully within the FOV are eligible for analysis. Note that here we use $c=0.5$ for SCTMM and $\beta = 1$ for the TV filter.

\clearpage

\begin{algorithm}

    \nonl \textbf{Input}: A normalized pre-processed (Algorithm \ref{alg:pre-processing}) image as in Figure \ref{fig:kekw-pendulum}
    
    \nonl \underline{\textit{Image filtering}}
    
    Poisson TV filtering (\ref{eq:total-variation})
    
    Luminance map inversion
    
    SCTMM filtering (\ref{eq:ctm-correction})
    
    Image normalization
    
    \nonl \underline{\textit{Reference body segmentation}}
    
    Chan-Vese binarization (\ref{eq:chan-vese-functional})
    
    Morphological opening (disk elements)
    
    Morphological erosion (disk elements)
    
    Delete border components
    
    \nonl \underline{\textit{Void region extraction}}
    
    Mask area thresholding
    
    Define an IW about the body centroid based on the body segment area (\ref{iw-scale},\ref{virtual-iw-bounds})
    
    Project the input image onto the IW masking the pixels outside the body segment

    \nonl \underline{\textit{Void region restoration}}
    
    Body background extrapolation to masked IW regions using texture synthesis-based inpainting

    Image normalization

    (Optional) Remove high-luminance outliers and normalize the image
    
    \nonl \textbf{Output:} A repaired IW containing the spherical void surrounded by the body background
    
\caption{Void region (IW) extraction and repair from the pendulum images.}
\label{alg:void-region-extraction-repair}
\end{algorithm}

\begin{figure}[htbp]
\centering
\includegraphics[width=1\textwidth]{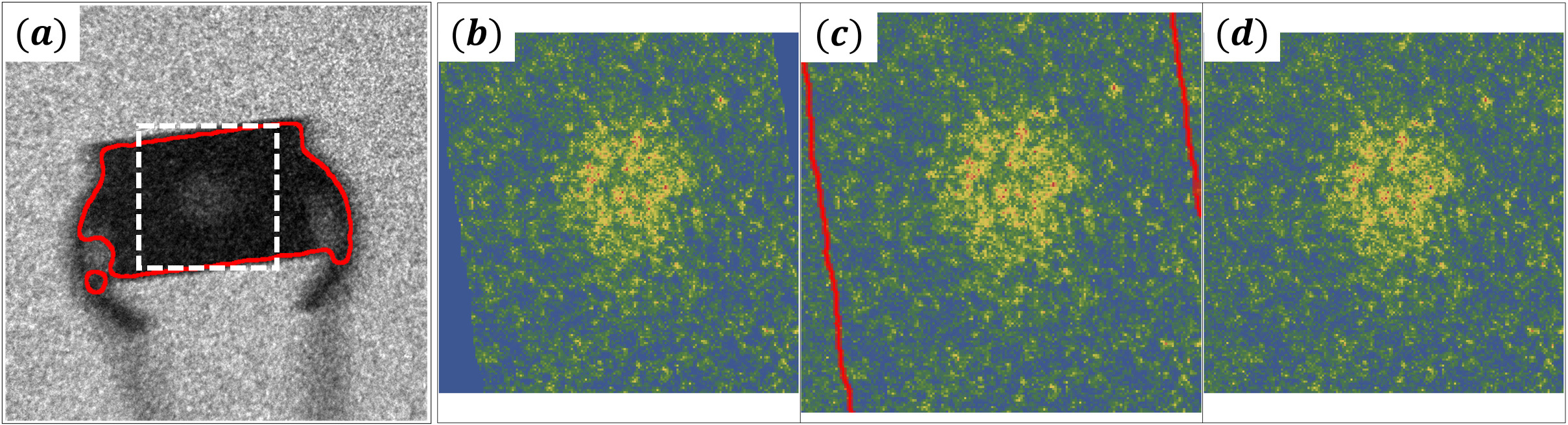}
\caption{(a) Detected reference body segments (red contours) with an IW (white dashed frame) centered about the body position after segment area thresholding; (b) original image projected onto the IW; (c) image regions designated for inpainting (red borders); (d) restored void region.}
\label{fig:void-region-extraction-repair}
\end{figure}

After area thresholding removes mask components that have abnormally small areas (usually left over clamp segment fragments), an IW is defined about the body centroid via (\ref{iw-scale}) and (\ref{virtual-iw-bounds}) with $s = 1.25$, and the input image with masked non-segment pixels is projected onto the IW. An example is illustrated in Figure \ref{fig:void-region-extraction-repair}. Note that (a), which shows the detected segments, contains a small artefact to the bottom left of the body segment -- such segments are removed by area thresholding. An IW is then defined about the centroid of the remaining body segment and a void region is extracted, which is shown in (b). Note, however, that the IW contains a portions of the masked background from (a), which can interfere with the global and local filters.

It is therefore necessary to extrapolate the body background about the void into such regions. These regions are detected by inverting the IW luminance map and running histogram-based segmentation with a single threshold of $0.999$. The masked regions are then segmented from the IW, as shown in (c). Texture synthesis-based inpainting is then performed with a maximum $N_\text{neigh}=150$ neighboring pixels used for texture comparison and a maximum of $N_\text{samp}=300$ sampling instances for texture fitting \cite{wolfram-mathematica-inpaint}.

\clearpage

In rather rare cases inpainting introduces pixels with strongly outlying luminance values, which are eliminated as follows. An Otsu threshold $I_c$ is computed for the input image, and then the image is binarized using the $k_I \cdot I_c$ threshold, where $k_I$ is the threshold scaling factor. With the right $k_I$ value, this operation segments pixels that differ by more than a certain luminance value from the bulk of the histogram. The segmented pixels are then masked in the input image. We found that $k_I = 4$ does not affect the images without significant outliers (ones that strongly affect global and local filter output) while effectively cleaning up severe outliers without modifying the bulk image histogram. An example output of Algorithm \ref{alg:void-region-extraction-repair} is shown in Figure \ref{fig:void-region-extraction-repair}d. Afterwards, the resulting repaired void regions are used as input for the global and local filters. The settings for the global and local filters are the same as in the cases with stationary reference bodies.

In the cases where neutron flux transmission was along the longer body axis (Figure \ref{fig:kekw-pendulum}a) it was more difficult to segment the body without the pendulum clamps with Algorithm \ref{alg:void-region-extraction-repair} as outlined above. Therefore, minor adjustments were made:

\begin{itemize}[noitemsep,topsep=0pt]
    \item Otsu binarization was used instead of the Chan-Vese process
    \item Morphological opening disk element radius was increased to $50$ (Step 6)
    \item Morphological dilation using disk structural elements with a 5-pixel radius was performed after Step 7
    \item Body masks were oriented to minimize the area of masked background in the IW before performing Step 11
\end{itemize}
and other Steps and parameters of Algorithm \ref{alg:void-region-extraction-repair} remained unchanged. This procedure was necessary because clamp removal from the body segments using larger-scale morphological opening resulted in body segments smaller then the body by a considerable margin, which was compensated for by morphological dilation to recover the eroded area. Body reorientation was required because the masked image margins to be filled using texture synthesis often constituted a significant fraction of the IW area, resulting in artefacts. Body orientation was detected by fitting a minimum area oriented bounding box and determining its angle $\phi$ (definition as in Figure \ref{fig:pipeline-versus-tilt-angle}a) with respect to the horizontal axis. The masked body image was then rotated by $- \phi$ radians and the body region was obtained and repaired as usual. In two cases we had to use lower values, $N_\text{neigh}=50$ and $N_\text{samp}=200$, to avoid sampling the void regions and producing high-luminance synthetic background.

In addition, another issue exists: filtering is performed for IWs with slightly different sizes and void positions. Therefore, one cannot directly overlay detected void shapes over a reference mask. One also cannot obtain a reference mask by averaging the images from the recorded sequence as with a stationary body. As a solution, we use the reference void segment detected from Figure \ref{fig:reference-body-spheres}e (best SNR and CNR) to generate a reference circle that the detected shapes will be compared against. Using the radius determined for the reference void segment, optimal circles $C_\text{ref}$ are fitted to the detected segments:

\begin{equation}
    C_\text{ref} = \bigg \{ \{ \vec{c}_0, R_\text{ref} \} \bigg| \min_{\vec{c}_0} \sum_k ( \norm{\vec{r}_k - \vec{c}_0} - R_\text{ref} )^2 \bigg \}
\label{eq:fit-reference-mask}
\end{equation}

where $\vec{r}_k$ are the segment boundary pixel positions, $\vec{c}_0$ is the optimized circle position and $R_\text{ref}$ is the reference void radius measured from Figure \ref{fig:reference-body-spheres}f. This approach to reference mask placement works well only if the global and local filters are known not to systematically produce shapes with significant errors in centroid position with respect to the true void position, which has been verified using the image sequences with stationary bodies.

\clearpage

Given this, it is expected that (\ref{eq:fit-reference-mask}) affects $S_\Delta$ to a degree, and $\delta S$, $\delta r$ and $\delta c$ are unaffected by definition. However, the centroid determination error is considerably diminished, but this is an acceptable trade-off. In fact, with (\ref{eq:fit-reference-mask}) $(\delta x, \delta y)$ becomes a measure of circularity correlated to $\delta c$ and it is arguably more useful to treat it as such. In the case of the moving reference body $(\delta x, \delta y)$ can also be used to quantify shape detection errors induced by motion blur.

\subsection{Error analysis}

\begin{figure}[htbp]
\centering
\includegraphics[width=0.95\textwidth]{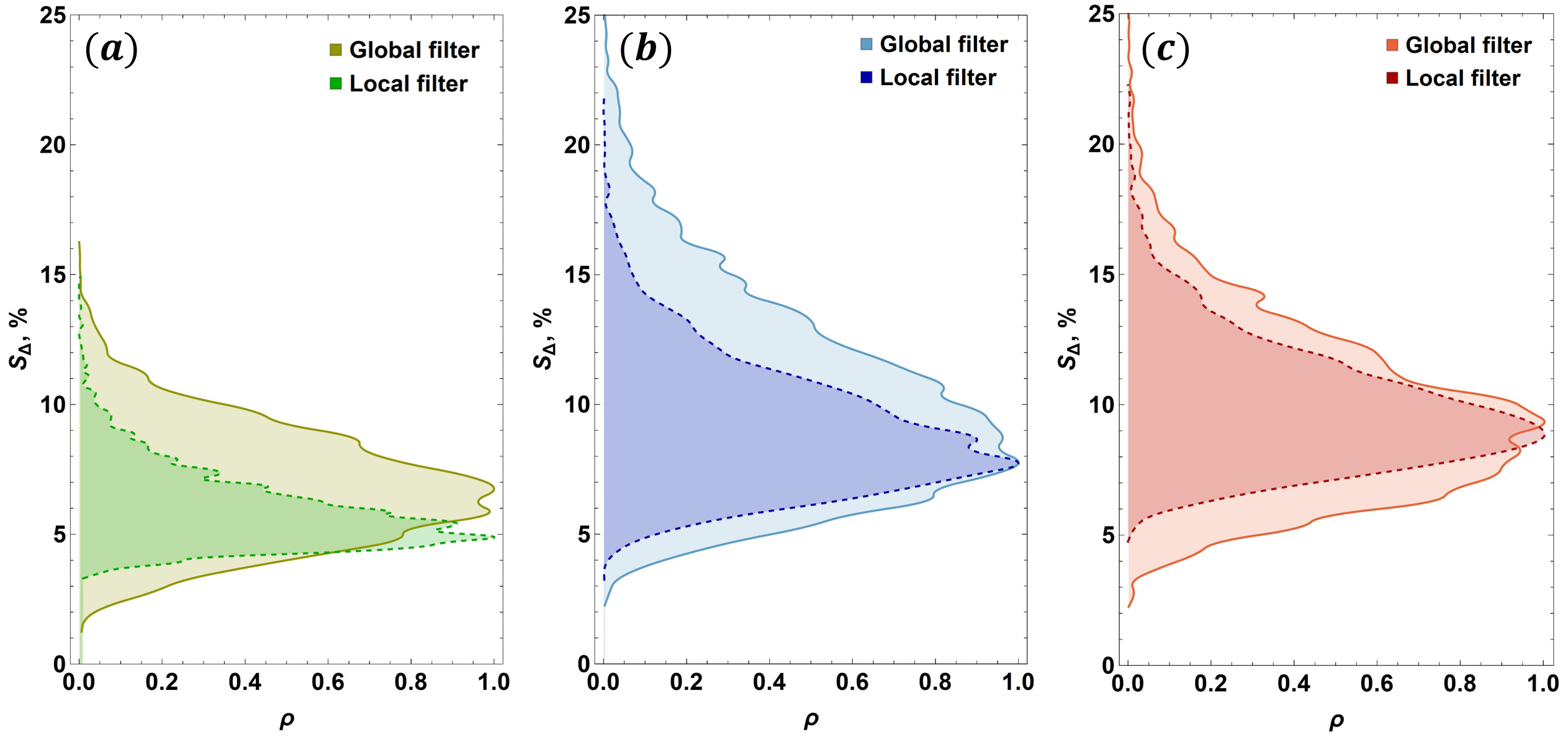}
\caption{Static reference body: smooth normalized $S_\Delta$ histograms for neutron flux transmission along the (a) smaller ($20~mm$) axis, (b) larger ($30~mm$) axis and (c) larger axis with an extra $1~cm$ distance between the scintillator and the body ($0~mm$ by default). Note the color legends in the upper right corners of (a-c) indicating results for the global and local filters. Here $\rho$ is the normalized event density. Histogram bins were determined using the Scott method and 2-order interpolation was applied to bin density values.}
\label{fig:stationary-difference-area-errors}
\end{figure}

\begin{figure}[htbp]
\centering
\includegraphics[width=0.95\textwidth]{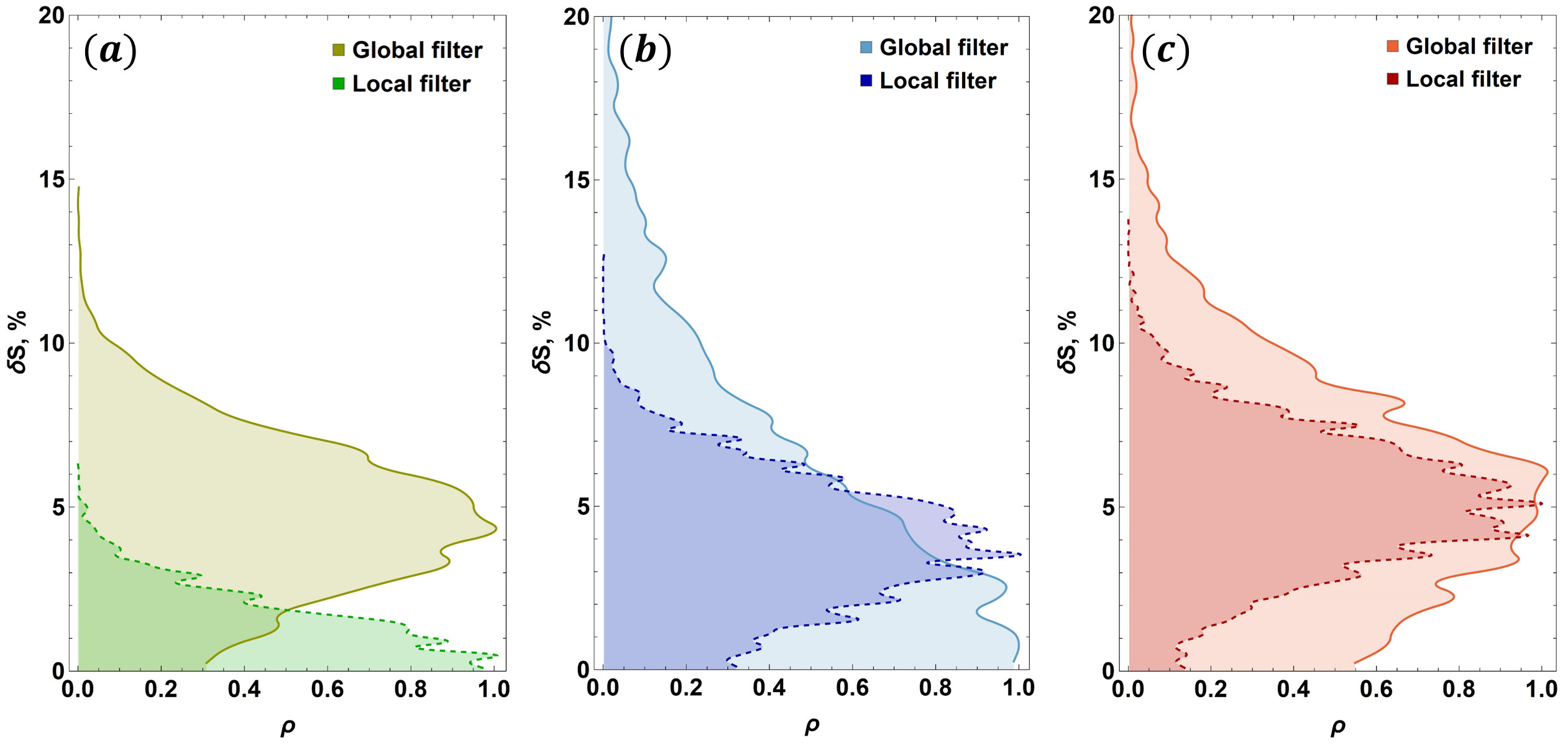}
\caption{Static reference body: smooth normalized $\delta S$ histograms for neutron flux transmission along the (a) smaller axis, (b) larger axis and (c) larger axis with an extra $1~cm$ distance between the scintillator and the body.}
\label{fig:stationary-area-difference-errors}
\end{figure}

\begin{figure}[htbp]
\centering
\includegraphics[width=0.95\textwidth]{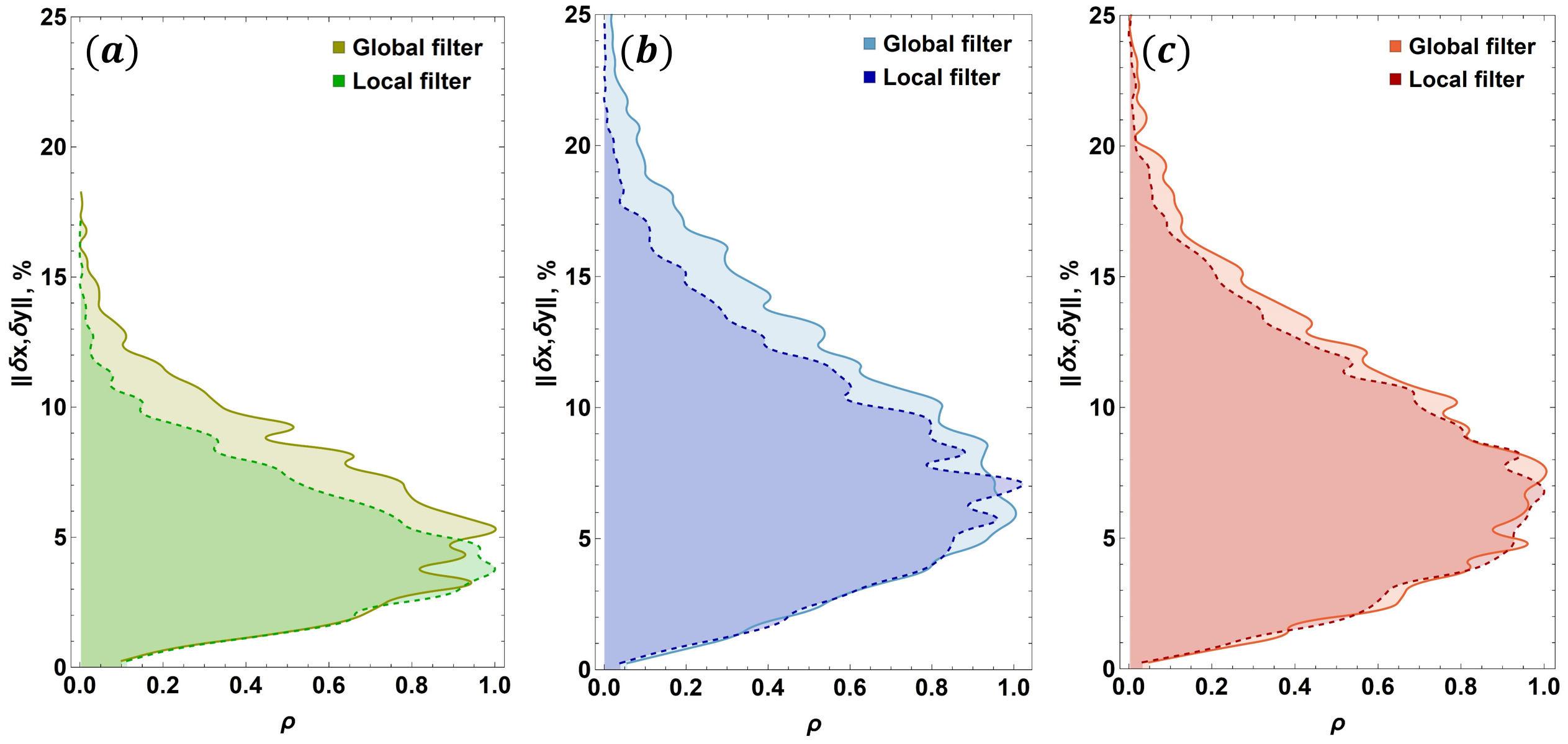}
\caption{Static reference body: smooth normalized $\norm{\delta x, \delta y}$ histograms for neutron flux transmission along the (a) smaller axis, (b) larger axis and (c) larger axis with an extra $1~cm$ distance between the scintillator and the body. Note that here $\delta x \approx \delta y$ for all instances, which is why they are not plotted individually.}
\label{fig:stationary-position-difference-errors}
\end{figure}

With the images from the reference experiments processed and the error metrics calculated, one can now assess the performance of the developed code more strictly than in Section \ref{sec:results-existing-and-new-data}. Starting with the imaging for a static reference body, the results are presented in Figures \ref{fig:stationary-difference-area-errors}-\ref{fig:stationary-position-difference-errors}. One can see in the $S_\Delta$ histograms (Figure \ref{fig:stationary-difference-area-errors}) that the global filter error distribution peaks and means are shifted towards higher $S_\Delta$ values going from neutron flux transmission along the shorter (a) to longer (b) axis, and value dispersion is also increased with considerably more instances of $S_\Delta > 10 \%$. When the body is moved $1~cm$ away from the scintillator (c), the global filter peak is shifted further towards higher values but the dispersion does not change significantly. The maximum values (not shown) are higher as well with respect to (b). Similar tendencies can be observed for the local filter errors, but the overall error values are considerably decreased and distribution peaks are shifted back to lower $S_\Delta$ values. The $S_\Delta$ dispersion and maximum values are also diminished across the board. Thus, while not radical, the improvements due to local filtering are still very clear.

The $\delta S$ distributions shown in Figure \ref{fig:stationary-area-difference-errors} are different in that there are many more instances of $\delta S < 5 \%$. Note that, while in (a) where there are relatively less $\delta S < 5 \%$ values than in (b,c) for the global filter, the local filter improves the results significantly -- dispersion is minimized and the distribution mean is shifted below $\delta S = 2 \%$. This is in contrast to what is seen in (b,c): in (b), the $\delta S$ peak is shifted towards larger values by the local filter, and in (c) it stays at roughly the same position. However, again, the local filter drastically minimized $\delta S$ dispersion and maxima. The reason why one observes relatively less $\delta S < 5 \%$ values in (b,c) than in (a) is that the local filter performance depends on how well the first three stages of the global filter improve SNR and CNR. That is to say, most of the instances with $\delta S \gtrsim 7 \%$ are converted to values about the peaks of the local filter error distribution because the local filter in most cases cannot achieve improvement to below than $\delta S \lesssim 3 \%$. Therefore it stands to reason that less such values are seen in (c) than in (b). As with $S_\Delta$ (Figure \ref{fig:stationary-difference-area-errors}), one can see that the local filter performs progressively worse from (a) to (c), as expected.

\clearpage

In the static body case one would expect $\delta x \approx \delta y$, which is indeed the case, since there is no bias due to motion blur. Figure \ref{fig:stationary-position-difference-errors} therefore shows the norm of the position error vector. Notice that here the local filter improves the error distributions only sightly, reducing the relative amount of higher $\norm{\delta x, \delta y}$ values in all cases. It is important to point out that $\delta r$ distributions conform to $\delta S$ plots and the $\delta r$ metric values are lower by a factor of 2, while $\delta c$ is negligible across the board. The error values seen in Figures \ref{fig:stationary-difference-area-errors}-\ref{fig:stationary-position-difference-errors} are within acceptable ranges.

\begin{figure}[htbp]
\centering
\includegraphics[width=0.95\textwidth]{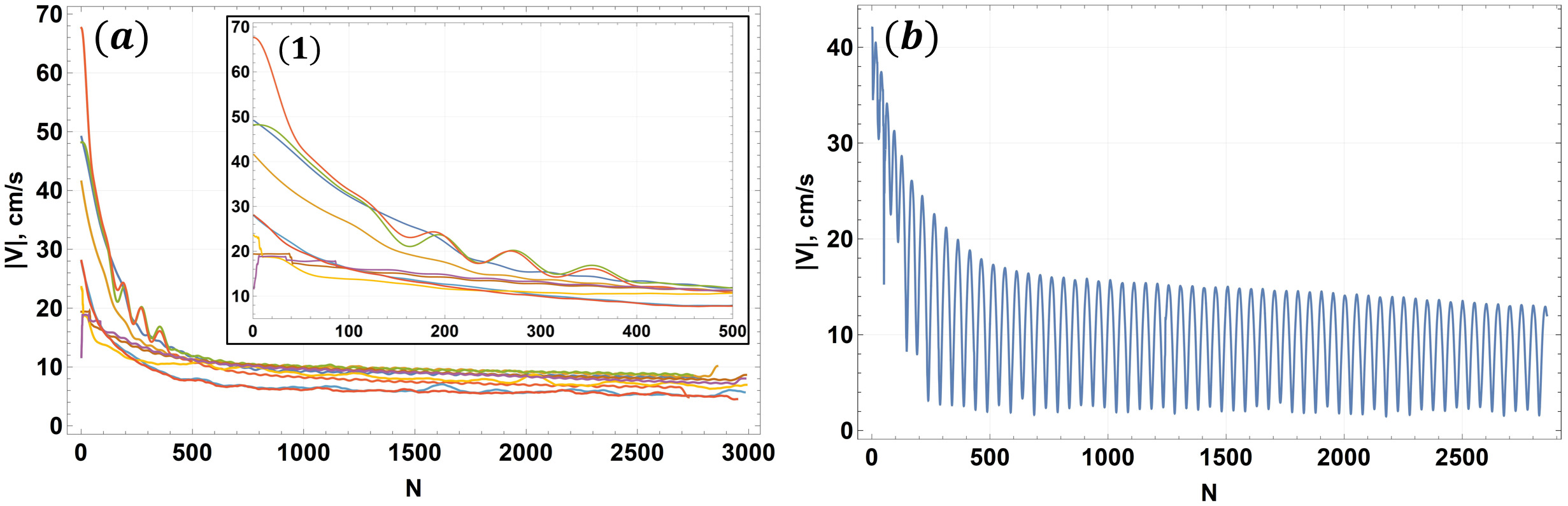}
\caption{Frame pair velocimetry (magnitude) for the moving body (Figure \ref{fig:kekw-pendulum}): (a) velocity over sequential frames for all recorded image sequences with oscillations filtered out for visual clarity; (b) velocity for one of the image sequences with oscillations shown. (a1) shows the first 500 frames in (a). Gaussian TV filtering was used for (a) and (b) with the regularization parameters set to $5$ (150 iterations) and $0.5$, respectively.}
\label{fig:pendulum-all-velocity-curves}
\end{figure}

\begin{figure}[H]
\centering
\includegraphics[width=1\textwidth]{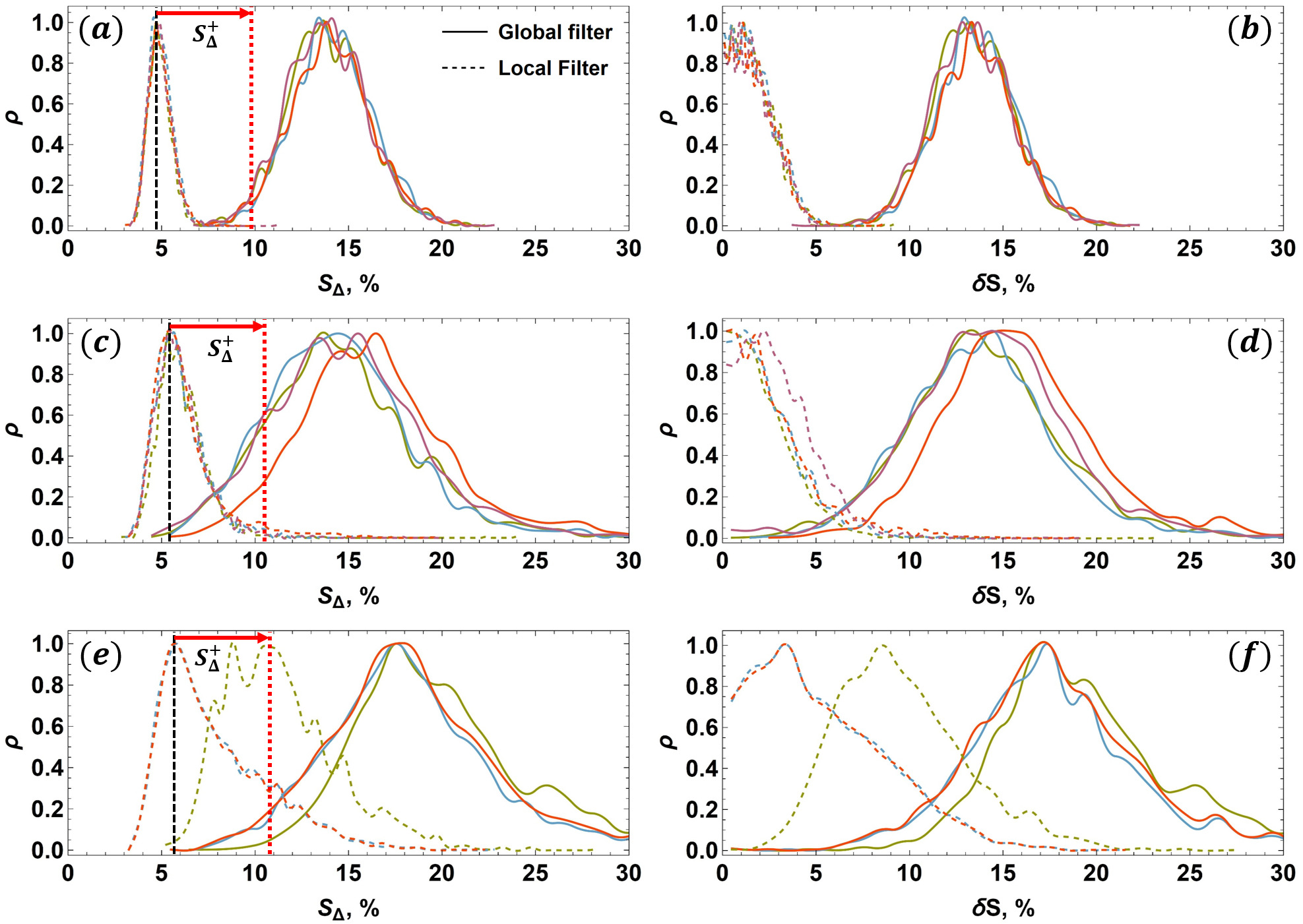}
\caption{Smooth normalized $S_\Delta$ (a,c,e) and $\delta S$ (b,d,f) histograms for the cases with a moving body (pendulum): neutron flux transmission along the (a,b) smaller axis, (c-d) larger axis and (e-f) the latter with the addition of synthetic image artefacts. Note the legend in the upper-right corner of (a) indicating the results for the global and local filtering. Colors indicate different imaging instances. $S_\Delta^{+}$ is the correction for the reference mask fitting bias (red arrows).}
\label{fig:all-moving-body-area-metrics}
\end{figure}

\begin{figure}[htbp]
\centering
\includegraphics[width=1\textwidth]{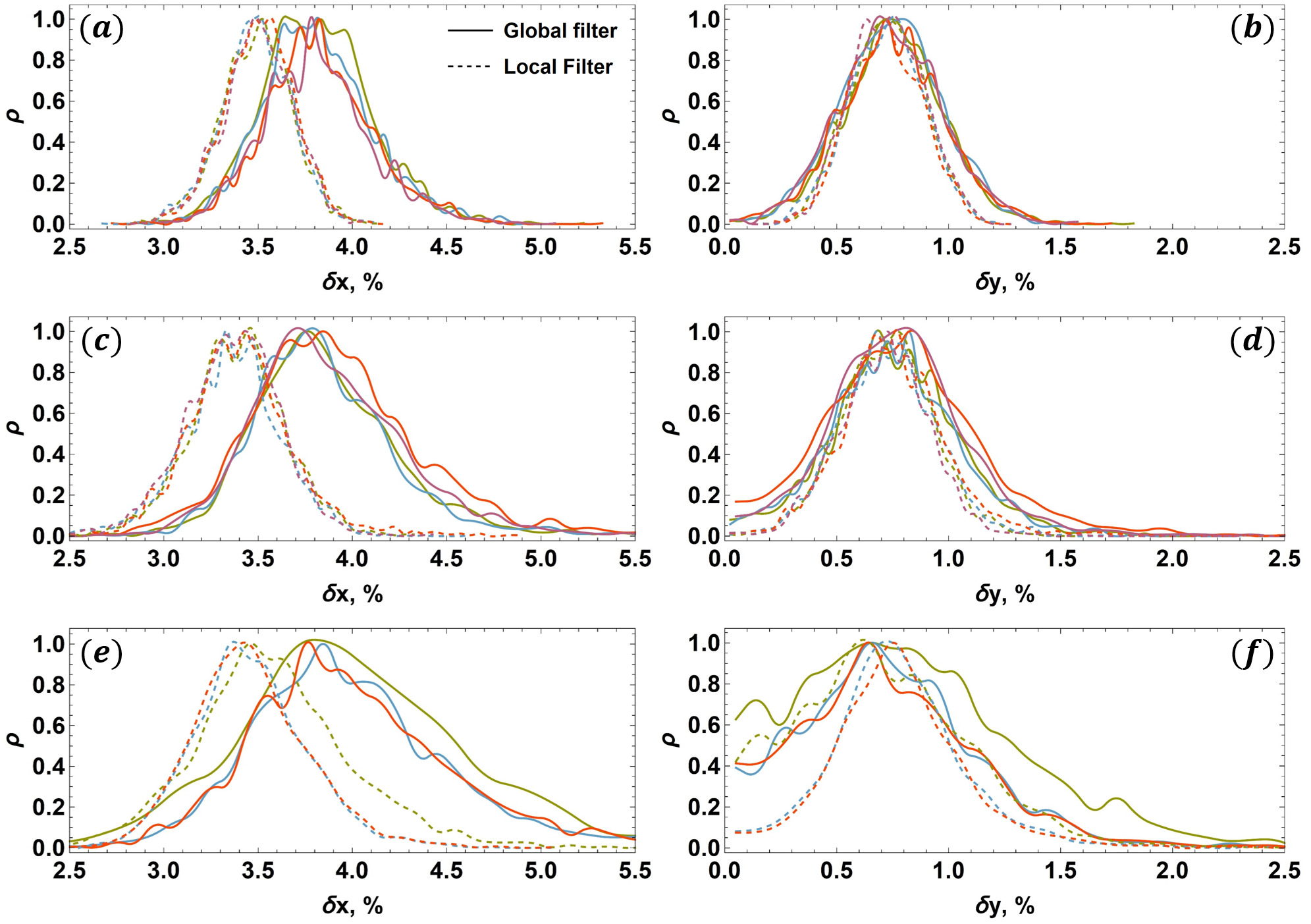}
\caption{Smooth normalized $\delta x$ (a,c,e) and $\delta y$ (b,d,f) histograms for the cases with the moving body: neutron flux transmission along the (a,b) smaller axis, (c-d) larger axis and (e-f) the latter with the addition of synthetic image artefacts.
}
\label{fig:all-moving-body-position-metrics}
\end{figure}

\begin{figure}[htbp]
\centering
\includegraphics[width=1\textwidth]{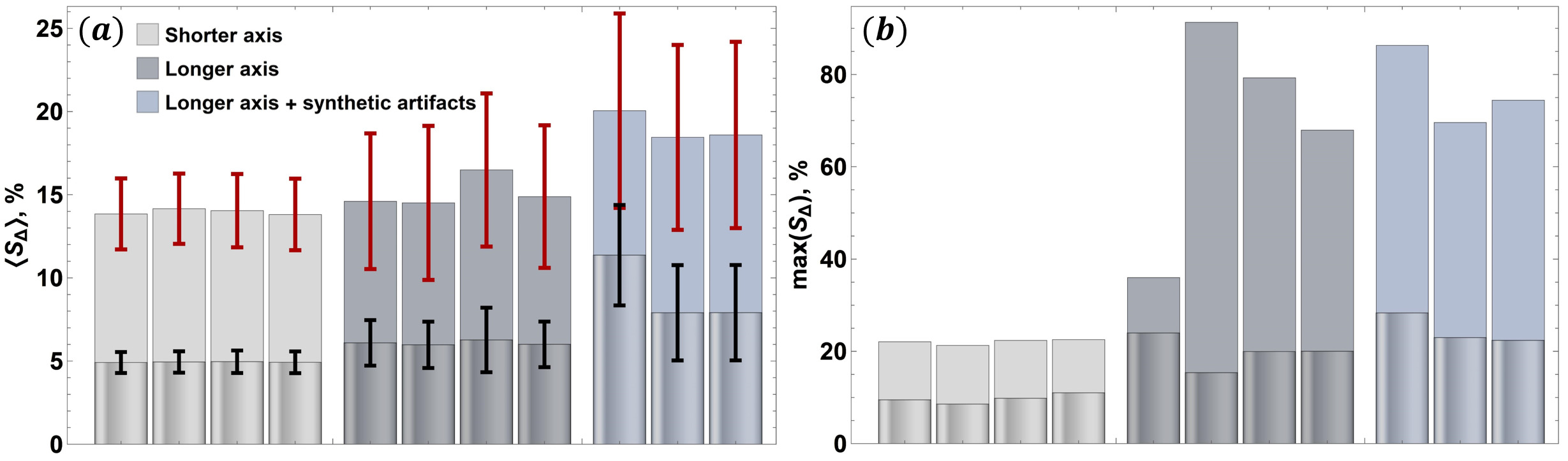}
\caption{(a) Mean and (b) maximum $S_\Delta$ values for the cases with a moving reference body. The results for the global and local filtering are represented with matte and glossy bars, respectively. The standard deviations for the mean values in (a) are indicated by the red (global) and black (local) error bars. Note the color legend in the upper-left corner of (a) indicating the three different test groups considered in Figures \ref{fig:all-moving-body-area-metrics} and \ref{fig:all-moving-body-position-metrics}.}
\label{fig:mean-max-difference-areas}
\end{figure}

\begin{figure}[htbp]
\centering
\includegraphics[width=1\textwidth]{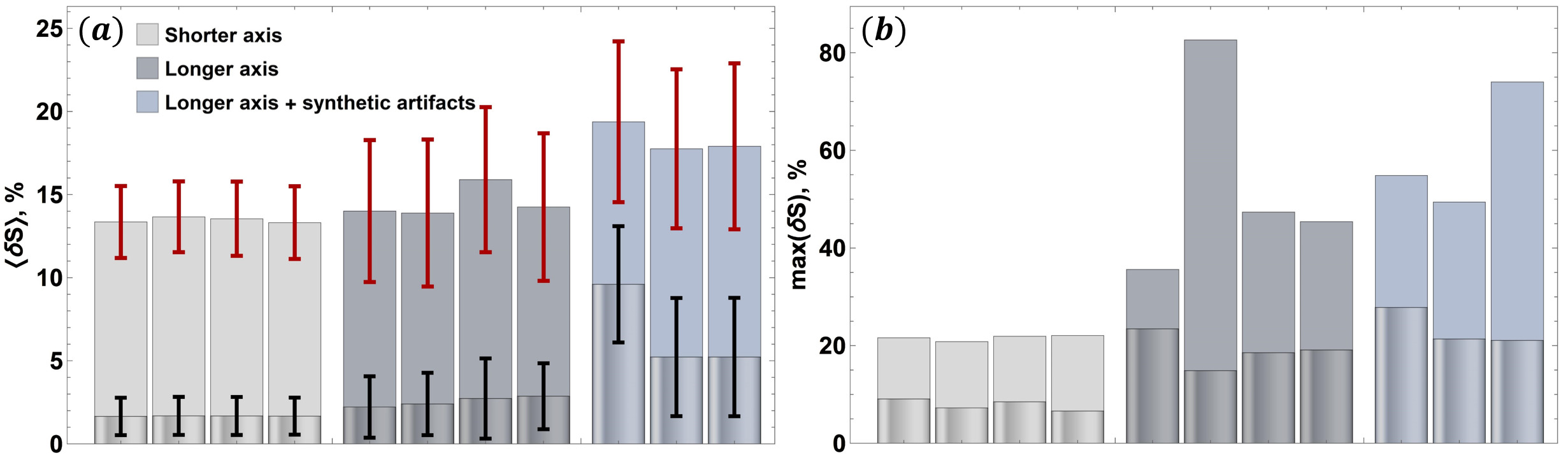}
\caption{(a) Mean and (b) maximum $\delta S$ values for the cases with a moving reference body. The results for the global and local filtering are represented with matte and glossy bars, respectively.}
\label{fig:mean-max-area-differences}
\end{figure}

\begin{figure}[htbp]
\centering
\includegraphics[width=1\textwidth]{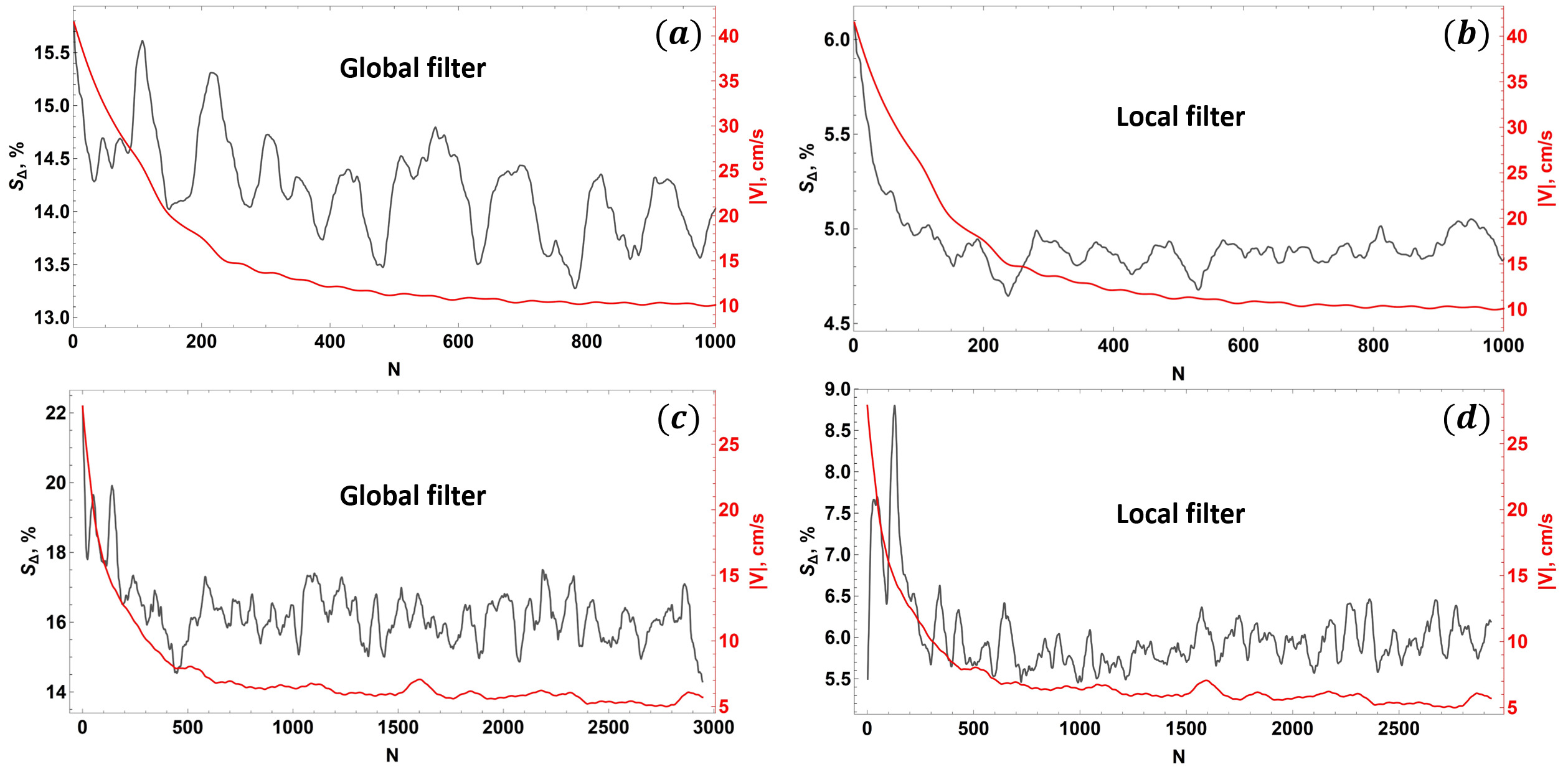}
\caption{An illustration of the effect of body motion on the shape detection errors: examples from two image sequences. Neutron transmission along (a-b) the shorter and (c-d) longer axis of the reference body. (a,c) show the $S_\Delta$ dynamics over consecutive frames for the global filter and (b,d) show the results for the local filter. (a,b) show the first 1000 frames for visual clarity. Note that 3000 images were analyzed in total for both shown cases. The error time series show here are obtained from raw data by removing the outliers above and below the Gaussian TV-filtered (regularization parameter 1) $q=0.9$ QSEs (3-rd order splines, $\nint{90\% \cdot N}$ spline knots) and applying the Gaussian TV filter (regularization parameter 2) to the remaining data points. The velocity curves are as in Figure \ref{fig:pendulum-all-velocity-curves}a.}
\label{fig:motion-blur-errors}
\end{figure}

Moving on to the moving reference body imaging, the results of these tests are shown in Figures \ref{fig:pendulum-all-velocity-curves}-\ref{fig:motion-blur-errors}. Figure \ref{fig:pendulum-all-velocity-curves} shows the pendulum (and reference void) velocity dynamics over consecutive frames for all imaging series. The $[20;40]~ cm/s$ range of expected bubble velocities (Section \ref{sec:bubble-props}) is covered by the performed measurements, as seen in (a) and (a1). Importantly, in addition to shorter and longer axis transmission tests, we also used purposefully inappropriate texture synthesis and outlier removal settings in Algorithm \ref{alg:void-region-extraction-repair} to generate three sets of data with synthetic image edge artefacts (luminance similar to the void regions) and single pixels with luminance exceeding the image maximum by an order of magnitude (greatly reduced image CNR) to see how the filters perform. Consider Figure \ref{fig:all-moving-body-area-metrics} where $S_\Delta$ and $\delta S$ distributions are shown for the three test groups.

\clearpage

As before, the global filter performance in terms of both $S_\Delta$ and $\delta S$ progressively decreases as the transmission axis length is increased and then image artefacts are introduced to the data from the longer axis transmission measurements. Notice that the $S_\Delta$ and $\delta S$ values in Figure \ref{fig:all-moving-body-area-metrics} are in all cases greater than in the static body cases (Figures \ref{fig:stationary-difference-area-errors}-\ref{fig:stationary-position-difference-errors}). This is clear enough to be seen visually, as histogram peaks in (a,b) are just under $\lesssim 15\%$ errors for the global filter output, while only being just under $\lesssim 8\%$ in the worst case for $S_\Delta$ in Figure \ref{fig:stationary-difference-area-errors}c. However, this is exactly where the local filter makes a radical difference -- observe in Figures \ref{fig:all-moving-body-area-metrics}a, c and e that the dispersion, maxima, minima and means of $S_\Delta$ are greatly reduced. Similarly good performance is seen in (b,d,f) in terms of $\delta S$. Notice that the changes in the local filter error distributions as image quality gets worse from (a,b) to (e,f) are consistent with what is seen for static body imaging in Figures \ref{fig:stationary-difference-area-errors} and \ref{fig:stationary-area-difference-errors}. Another important point here is that two of the shorter axis transmission instances (Figures \ref{fig:all-moving-body-area-metrics}a and b) and one from the longer axis groups (Figures \ref{fig:all-moving-body-area-metrics}c and d) are with an extra $1~cm$ body-to-scintillator distance ($2~mm$ by default) -- however, in this case the differences are not significant enough to warrant attention. This likely stems from the fact a large fraction of errors is due to motion blur, especially during the pendulum deceleration stage, which is within the first 300-500 imaging frames (Figure \ref{fig:pendulum-all-velocity-curves}). It is also very clear from Figures \ref{fig:all-moving-body-area-metrics}e and f that introducing additional image artefacts considerably degrades the global filter performance. The local filter, again, greatly improves the quality of detected shapes, but, of course, also produces greater errors as opposed to what is seen in (c) and (d).

One must remember that $S_\Delta$ is reduced by (\ref{eq:fit-reference-mask}). To estimate the reduction, consider that if the centers of two circles with equal radii are displaced by a factor $k \leq 2$ of the radius of either, the relative area of their difference is $S_\Delta (k) = 1 - (2/ \pi) \arccos (k/2) + (k/ \pi) \sqrt{1-k^2/4}$. Assuming the worst-case scenario where mask fitting via (\ref{eq:fit-reference-mask}) "improves" the position of the detected shape by the peak $\norm{\delta x, \delta y}$ value in \ref{fig:stationary-position-difference-errors}c, $k \sim 8 \%$, the $S_\Delta$ underestimation comes out to $S_\Delta^{+} \sim 5\%$. To verify this, we also observe how $S_\Delta$ changes for shapes with significant deformities detected (rarely) for stationary body images when ground truth masks (Figures \ref{fig:reference-body-spheres}e-f) are replaced with fits using (\ref{eq:fit-reference-mask}) -- this is in agreement with the above idealized estimate rather well, yielding $S_\Delta$ differences $\in (4;5\%)$. If this adjustment is applied to Figures \ref{fig:all-moving-body-area-metrics}a, c and e, the observations are consistent with the results for the stationary reference body, and the local filter $S_\Delta$ values come out to about $\sim 10\%$ at distribution peaks (versus $\sim 8\%$ for Figure \ref{fig:stationary-difference-area-errors}c), excluding one of the cases in Figure \ref{fig:all-moving-body-area-metrics}e (light green curves). To reiterate, $\delta S$, $\delta r$ and $\delta c$ are unaffected by (\ref{eq:fit-reference-mask}) by their definitions. Here the $\delta r$ distributions again conform to $\delta S$. $\delta c$ is negligible for all cases and, as seen in Figure \ref{fig:all-moving-body-position-metrics}, $\norm{\delta x, \delta y}$ is such that the asphericity of the detected shapes in within acceptable bounds.

Turning to Figure \ref{fig:all-moving-body-position-metrics} and recalling that with (\ref{eq:fit-reference-mask}) $\delta x$ and $\delta y$ are correlated to the sphericity of the detected shapes, notice that the errors are anisotropic in all cases. The greater component $\delta x$ corresponds to the main direction of motion. Importantly, this anisotropy is also observed when inspecting the shape/reference difference masks for the local and global filters, where the largest contributions to $S_\Delta$ are at the shape sides in the $-x$ and $x$ directions which are the pendulum (Figure \ref{fig:kekw-pendulum}) oscillation directions.

Since $S_\Delta$ and $\delta S$ maxima for the global filters were not included in Figure \ref{fig:all-moving-body-area-metrics} to maintain visual clarity, and it is hard to quantify dispersion visually, Figures \ref{fig:mean-max-difference-areas} and \ref{fig:mean-max-area-differences} show these quantities explicitly for all three test groups and for both global and local filters. While shape detections with $S_\Delta$ and $\delta S$ values close to the values indicated in Figures \ref{fig:mean-max-difference-areas}b and \ref{fig:mean-max-area-differences}b are very rare, it is important that the local filter can significantly improve the quality of detected shapes even in these instances.

The effects of motion blur can be assessed more quantitatively and directly -- consider Figure \ref{fig:motion-blur-errors} where $S_\Delta$ is plotted as a time series for two of the processed image sequences. One can see that higher body motion velocity indeed corresponds to greater errors, which decline over time as the body decelerates. Note that, between Figures \ref{fig:motion-blur-errors}a and b, it is evident that the intrinsic error signal due to the global filter quickly obscures the error contribution due to motion blur. However, it also seems that the global filter is affected by the motion blur more than the local filter -- as seen in (b), the $S_\Delta$ rather quickly relaxes to a quasi-stationary value at about $N \sim 250$, while in (a) the error decay persists until $N \sim 500$. Similar dynamics can be observed in (c) and (d) as well.

We assess the reference void detection failure rates in the image sequences used for the code validation -- the results are summarized in Table \ref{tab:detection-rates-reference-body}.

\begin{table}[!h]
\begin{center}
\begin{tabular}{| c | c | c | c | c | c | c | c | c | c | c | c | c | c | c |}
\hline
\textbf{Filter} & \multicolumn{3}{ c |}{\textbf{Static}} & \multicolumn{4}{ c |}{\textbf{Motion: SA}} & \multicolumn{4}{ c |}{\textbf{Motion: LA}} & \multicolumn{3}{ c |}{\textbf{Motion: LA + artefacts}}\\ 
\cline{2-15}
& \textbf{1} & \textbf{2} & \textbf{3} & \textbf{1} & \textbf{2} & \textbf{3} & \textbf{4} & \textbf{1} & \textbf{2} & \textbf{3} & \textbf{4} & \textbf{1} & \textbf{2} & \textbf{3}\\
\hline
Global & 0 & 0 & 0 & 0 & 0 & 0 & 0 & 0.33 & 0.37 & 1.24 & 1.10 & 3.08 & 2.51 & 2.85 \\ \hline
Local & 0 & 0  & 0  & 0 & 0 & 0 & 0 & 0.44 & 2.54 & 1.71 & 1.27 & 0.87 & 1.83 & 1.80 \\ \hline
\end{tabular}
\end{center}
\caption{Reference void failure rates ($\%$) for the validation experiments. Static: tests (1-3) are with neutron flux transmission along the shorter axis (SA) of the body, the longer axis (LA), and the latter with an extra distance to the scintillator. For moving body imaging, tests with an extra body-to-scintillator distance are (3-4) for SA and (4) for LA. Tests with LA and synthetic artefacts are based on data from LA runs (2) and (4).}
\label{tab:detection-rates-reference-body}
\end{table}

There are no detection failures for static body imaging and for a moving body with neutron flux transmission along the shorter axis, and very small percentages are found for the longer axis case. The cases with synthetic artefacts have significantly higher failure rates for the global filter. However, all cases exhibit values $< 5\%$, which indicates acceptable performance.

Finally, Table \ref{tab:motion-blur-error-summary} indicates peak errors induced by motion blur at near the maximum pendulum velocity for the local filter for all the imaging instances, including $S_\Delta^{+} \sim 5\%$.

\clearpage

\begin{table}[!h]
\begin{center}
\begin{tabular}{| c | c | c | c | c | c | c | c | c | c | c | c |}
\hline
  & \multicolumn{4}{ c |}{\textbf{Motion: SA}} & \multicolumn{4}{ c |}{\textbf{Motion: LA}} & \multicolumn{3}{ c |}{\textbf{Motion: LA + artefacts}}\\ 
\cline{2-12}
& \textbf{1} & \textbf{2} & \textbf{3} & \textbf{4} & \textbf{1} & \textbf{2} & \textbf{3} & \textbf{4} & \textbf{1} & \textbf{2} & \textbf{3}\\
\hline
$\max |V|$, $cm/s$  & 49.1 & 41.5 & 48.1 & 50.2 & 19.0 & 19.2 & 27.3 & 28.0  & 19.2 & 28.0 & 28.0 \\ \hline
$\max S_\Delta$, $\%$ & 11.5 & 11.1  & 11.1 & 10.8 & 12.2 & 13.0 & 13.8 & 13.2 & 19.6 & 18.4 & 18.5 \\ \hline
\end{tabular}
\end{center}
\caption{Peak $S_\Delta$ values (including $S_\Delta^{+}$) for the local filter at near maximum body velocity for all validation experiments.}
\label{tab:motion-blur-error-summary}
\end{table}

Note that the peak errors are within $14\%$ for tests without synthetic artefacts, and within $20\%$ for instances with artefacts added to images.

\section{Further improvements \& extensions}
\label{sec:improvements}

While the demonstrated image processing performance is satisfactory for the problems that the code was developed for, its degree of parallelization could still be increased, especially for MRIF, and we expect that significant speedup can be attained for several components.

The developed code has been tested using the following hardware:

\begin{enumerate}[noitemsep,topsep=0pt]
    
    \item \href{https://ark.intel.com/content/www/us/en/ark/products/97128/intel-core-i7-7700-processor-8m-cache-up-to-4-20-ghz.html}{Intel Core i7-7700} (4 cores/8 threads) with 64 Gb $2400$ MHz DDR4 RAM
    
    \item \href{https://ark.intel.com/content/www/us/en/ark/products/126686/intel-core-i7-8700-processor-12m-cache-up-to-4-60-ghz.html}{Intel Core i7-8700} (6 cores/12 threads) with 64 Gb $2400$ MHz DDR4 RAM
    
    \item \href{https://ark.intel.com/content/www/us/en/ark/products/186605/intel-core-i9-9900k-processor-16m-cache-up-to-5-00-ghz.html}{Intel Core i9-9900K} (8 cores/16 threads) with 64 Gb $2666$ MHz DDR4 RAM
    
    \item \href{https://ark.intel.com/content/www/us/en/ark/products/198016/intel-xeon-w-2255-processor-19-25m-cache-3-70-ghz.html}{Intel Xeon W-2255} (10 cores/20 threads) with 192 Gb $2933$ MHz DDR4 ECC RAM
    
    \item \href{https://ark.intel.com/content/www/us/en/ark/products/198017/intel-core-i9-10980xe-extreme-edition-processor-24-75m-cache-3-00-ghz.html}{Intel Core i9-10980XE} (18 cores/36 threads) with 256 Gb $2933$ MHz DDR4 RAM
    
\end{enumerate}
and we found that memory utilization for parallel execution of Algorithm \ref{alg:global-filter} using hyperthreading (all images are processed independently) and all available threads for a sequence of 3000 images (properties given in Section \ref{sec:image-props}) requires almost all of the memory for the first and the last two machines, while the second and third machines ran out of memory and the image batch size had to be decreased. While this is not a critical issue, the mean execution time per 1000 images reduces significantly between machines as the thread count increases, so we expect the reduction in memory utilization to be worthwhile. For context, the first machine fully processes 3K images and outputs results in $\sim 2$ hours, while the fourth machine finishes in $\sim 1.3$ hours. A more detailed performance report will be provided in a follow-up publication for a greater number of processed image sequences.

It is also planned to implement a feedback loop that will enable coupling with our recently developed object tracking algorithm MHT-X \cite{zvejnieks-mhtx-arxiv} for iterative reinforced object detection and tracking. Finally, we will also apply the developed methods to image sequences with even smaller bubble-bubble distances where bubble collisions also occur.

\section{Conclusions}
\label{sec:conclusions}

To summarize, we have demonstrated the new version of our image processing methodology for resolving gas bubbles travelling through liquid metal imaged using dynamic neutron radiography. The showcased components of our code, such as the multi-scale recursive interrogation filter (MRIF) and the underlying global and local image filters, as well as soft color tone map masking, proved effective for detecting bubbles and extracting their dynamic shapes from images with low SNR and CNR. Output quality was further improved by the implemented luminance map-based false positive filter that bolstered the MRIF's intrinsic false positive filtering function.

It as shown by direct comparison that the new image processing code clearly outperforms the previous version used in \cite{birjukovsArgonBubbleFlow2020, birjukovsPhaseBoundaryDynamics2020}, while the outputs of both are still consistent. In addition, we have validated the new methods experimentally by imaging a reference body, both stationary and in motion, with a precisely machined spherical cavity. Results indicate that that local filtering performed by MRIF largely limits the shape detection errors: relative shape mismatch area and shape area difference with respect to reference shapes are within acceptable bounds of $\sim 14 \%$ ($\sim 20 \%$ with synthetic artefacts) and $\sim 10\%$ ($\sim 14 \%$ with synthetic artefacts), respectively (accounting for motion blur and the worst-case underestimation correction), while the asphericity of the detected shapes is rather negligible. As such, we find that applying the current methodology to the neutron radiography images obtained for our model systems with bubble chain flow is safe in that physically meaningful results with manageable errors can be expected. Note that we have also used the present image processing code to benchmark our object tracking code MHT-X \cite{zvejnieks-mhtx-arxiv}.

In follow-up articles, we are going to process the data acquired in the previous and latest neutron imaging campaigns using the methods presented in this paper and our MHT-X code, and showcase the effects of applied horizontal and vertical magnetic field with different strengths on bubble chain flow in a rectangular liquid gallium vessel. Bubble trajectories (length, curvature, oscillation frequencies, envelopes, etc.), velocity (both overall spectra and dynamics, including acceleration) and shapes (aspect ratio, tilt angle, etc., and dynamics thereof) will be assessed and compared for different flow conditions -- in addition to magnetic field configurations, a range of gas flow rates will be considered. We will also attempt to perform dynamic mode decomposition for the bubble shapes extracted from neutron radiography images and compare the dynamics against simulations -- this will be done using the methods recently developed in \cite{klevs2021dynamic}.

Finally, we expect that the developed image processing pipeline and/or separate elements thereof should be applicable beyond the current application and context, which we also plan to demonstrate in follow-up papers. In the meantime, the image processing code is available on \textit{GitHub}: \href{https://github.com/Mihails-Birjukovs/Low_C-SNR_Bubble_Detection}{Mihails-Birjukovs/Low\_C-SNR\_Bubble\_Detection}. The code will be improved as outlined in Section \ref{sec:improvements}.

\section{Acknowledgements}

The authors acknowledge the support due to the ERDF project ”Development of numerical modelling approaches to study complex multiphysical interactions in electromagnetic liquid metal technologies” (No. 5 1.1.1.1/18/A/108). The work is also supported by the Paul Scherrer Institut (PSI) and a DAAD Short-Term Grant (2021, 57552336). The authors would also like to express gratitude to Jevgenijs Telicko (UL) and Peteris Zvejnieks (UL) for assistance with the experiments, as well as to Imants Bucenieks (UL) who assembled the designed magnetic field systems.

\printbibliography[title={References}]

\end{document}